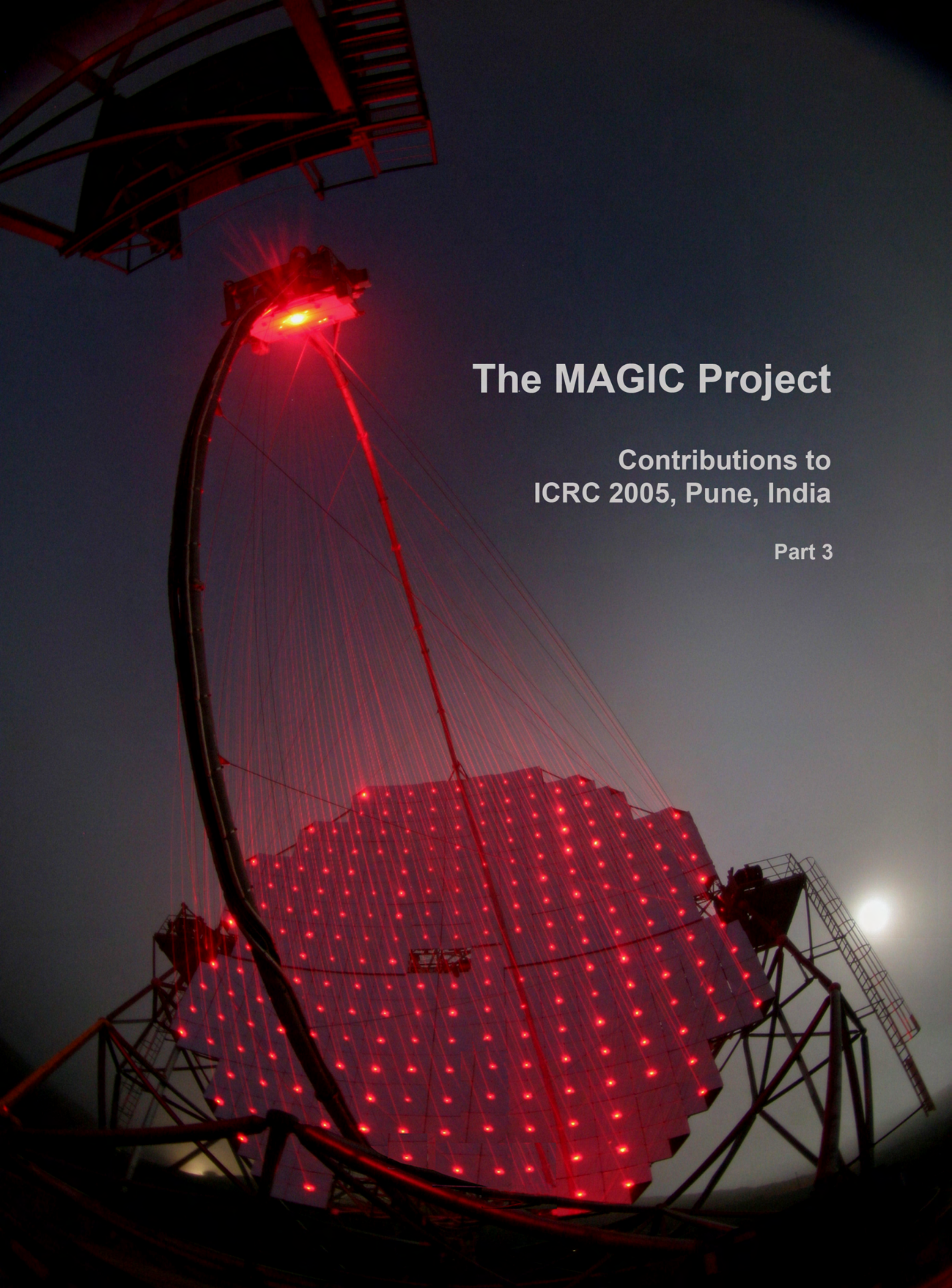

# The MAGIC Project

## Contributions to ICRC 2005, Pune, India

### Part 3

## Contents

**Contributions:**

### Part 3: MAGIC Detector and Analysis Details

# MAGIC: List of collaboration members

J. Albert i Fort [a], E. Aliu [b], H. Anderhub [g], P. Antoranz [k], A. Armada [b], M. Asensio [k], C. Baixeras [c], J.A. Barrio [k], H. Bartko [d], D. Bastieri [e], W. Bednarek [f], K. Berger [a], C. Bigongiari [e], A. Biland [g], E. Bisesi [h], O. Blanch [b], R.K. Bock [d], T. Bretz [a], I. Britvitch [g], M. Camara [k], A. Chilingarian [i], S. Ciprini [j], J.A. Coarasa [d], S. Commichau [g], J.L. Contreras [k], J. Cortina [b], V. Danielyan [i], F. Dazzi [e], A. De Angelis [h], B. De Lotto [h], E. Domingo [b], D. Dorner [a], M. Doro [b], O. Epler [l], M. Errando [b], D. Ferenc [m], E. Fernandez [b], R. Firpo [b], J. Flix [b], M.V. Fonseca [k], L. Font [c], N. Galante [n], M. Garczarczyk [d], M. Gaug [b], J. Gebauer [d], R. Giannitrapani [h], M. Giller [f], F. Goebel [d], D. Hakobyan [i], M. Hayashida [d], T. Hengstebeck [l], D. Höhne [a], J. Hose [d], P. Jacon [f], O.C. de Jager [o], O. Kalekin [l], D. Kranich [m], A. Laille [m], T. Lenisa [h], P. Liebing [d], E. Lindfors [j], F. Longo [h], M. Lopez [k], J. Lopez [b], E. Lorenz [d, g], F. Lucarelli [k], P. Majumdar [d], G. Maneva [q], K. Mannheim [a], M. Mariotti [e], M. Martinez [b], K. Mase [d], D. Mazin [d], C. Merck [d], M. Merck [a], M. Meucci [n], M. Meyer [a], J.M. Miranda [k], R. Mirzoyan [d], S. Mizobuchi [d], A. Moralejo [e], E. Ona-Wilhelmi [b], R. Orduna [c], N. Otte [d], I.Oya [k], D. Paneque [d], R. Paoletti [n], M. Pasanen [j], D. Pascoli [e], F. Pauss [g], N. Pavel [l], R. Pegna [n], L. Peruzzo [e], A. Piccioli [n], M. Pin [h], E. Prandini [e], R. de los Reyes [k], J. Rico [b], W. Rhode [p], B. Riegel [a], M. Rissi [g], A. Robert [c], G. Rossato [e], S. Rügamer [a], A. Saggion [e], A. Sanchez [e], P. Sartori [e], V. Scalzotto [e], R. Schmitt [a], T. Schweizer [l], M. Shayduk [l], K. Shinozaki [d], N. Sidro [b], A. Sillanpää [j], D. Sobczynska [f], A. Stamerra [n], L. Stark [g], L. Takalo [j], P. Temnikov [q], D. Tescaro [e], M. Teshima [d], N. Tonello [d], A. Torres [c], N. Turini [n], H. Vankov [q], V. Vitale [d], S. Volkov [l], R. Wagner [d], T. Wibig [f], W. Wittek [d], J. Zapatero [c]

[a] Universität Würzburg, Germany
[b] Institut de Fisica d'Altes Energies, Barcelona, Spain
[c] Universitat Autonoma de Barcelona, Spain
[d] Max-Planck-Institut für Physik, München, Germany
[e] Dipartimento di Fisica, Università di Padova, and INFN Padova, Italy
[f] Division of Experimental Physics, University of Lodz, Poland
[g] Institute for Particle Physics, ETH Zürich, Switzerland
[h] Dipartimento di Fisica, Università di Udine, and INFN Trieste, Italy
[i] Yerevan Physics Institute, Cosmic Ray Division, Yerevan, Armenia
[j] Tuorla Observatory, Pikkiö, Finland
[k] Universidad Complutense, Madrid, Spain
[l] Institut für Physik, Humboldt-Universität Berlin, Germany
[m] University of California, Davis, USA
[n] Dipartimento di Fisica, Università di Siena, and INFN Pisa, Italy

[o] Space Research Unit, Potchefstroom University, South Africa
[p] Fachbereich Physik, Universität Dortmund, Germany
[q] Institute for Nuclear Research and Nuclear Energy, Sofia, Bulgaria

# Part 3: MAGIC Detector and Analysis Details

# The Mirrors for the MAGIC Telescopes

D. Bastieri,[2] D. Agguiaro,[2] J. Arnold,[4] C. Bigongiari,[2] F. Dazzi,[2] M. Doro,[2] N. Galante,[3] M. Garczarczyk,[1] E. Lorenz,[1] D. Maniero,[2] M. Mariotti,[2] R. Mirzoyan,[1] A. Moralejo,[2] D. Pascoli,[2] A. Pepato,[2] L. Peruzzo,[2] M. Rebeschini,[2] A. Saggion,[2] P. Sartori,[2] V. Scalzotto,[2] D. Tescaro,[2] and N. Tonello[1] for the MAGIC Collaboration⋆.

*(1) Max–Planck–Institut für Physik, Föhringer Ring 6, D–80805 München, Germany.*
*(2) Università and INFN Padova, Via Marzolo 8, I–35131 Padova, Italy.*
*(3) Università di Siena and INFN Sezione di Pisa, Via F. Buonarroti 2, I–56127 Pisa, Italy.*
*(4) LT Ultra GmbH, Wiesenstrasse 9, D-88634 Aftholderberg, Germany.*
⋆ *Updated collaborator list at* http://magic.mppmu.mpg.de/collaboration/members/index.html.
Presenter: D. Bastieri (*bastieri@pd.infn.it*), ita-bastieri-D-abs1-og27-poster

The MAGIC Telescope has the largest reflecting surface, among other Cherenkov detectors, in order to have an energy threshold well below 100 GeV. In this contribution, we describe the technology used for the production and the optical qualities of the surface currently mounted onto the telescope. MAGIC features now 964 square mirrors, 50 cm by side, each of spherical shape. We present also a new technology that, producing pre-shaped panels, will finally yield lighter mirrors and can even be applied to make bigger mirrors of 1 m by side.

## 1. Introduction

The MAGIC telescope[1] features a huge reflecting surface[2] of 236 $m^2$ and overall parabolic shape, which allows detected photons to keep the correct timing information. The surface was segmented into 964 smaller elements (50 cm × 50 cm), each machined to spherical shape with the curvature radius that better fits the required parabolic shape. Each element is an aluminium honeycomb core *sandwiched* between two outer Al-layers using laminating adhesives. The sandwich, called *raw blank*, is later worked and polished with milling machines. Details can be found in sec. 2, while the optical properties of the mirrors can also be found in sec. 3.

## 2. The Mirrors

MAGIC mirrors are mainly a 5-mm thick AlMgSi1.0 plate, pre-machined to spherical shape and polished with a milling tool equipped with a diamond tip of *large* ($\sim$ 1 m) curvature radius. Plates are glued to an Al-honeycomb inside a thin Al-box making up the *raw blank* of $\sim$ 4 kg of weight. The reflecting surface is subdivided into zones with varying curvature radii ($34.125 \div 36.625$ m) to match the parabolic shape of the dish. After diamond milling, front plates are coated with a hard, transparent protective layer against scratches and aging. Mirrors are then grouped in 3 or 4 onto panels and each panel can be moved and aligned by an active mirror control system. Each mirror is also equipped with a heating system to prevent ice and dew formation.

### 2.1 Raw blank production

Raw blanks assembled in Padova are composed of a 1-mm thick Al 3003 box, 2.5 cm high, containing the Al 5052 honeycomb of 2.0 cm of thickness and the heating printed circuit board (PCB). Four small plates, 5-mm thick, are embedded into the honeycomb and glued to the outer box. They host four screws each, to fix the finished mirror to a panel. The box is then closed with the aluminium plate. Final assembly of the raw blank parts is done using three layers of 3M glue foils between box, honeycomb, PCB and front plate (see fig. 1 *left*). The gluing procedure consists in a 4-hour cycle. During this time the raw blanks, closed in an evacuated bag, are heated to 120° and stand to 3 atm of pressure. Up to 12 mirrors can be made in the same gluing cycle.

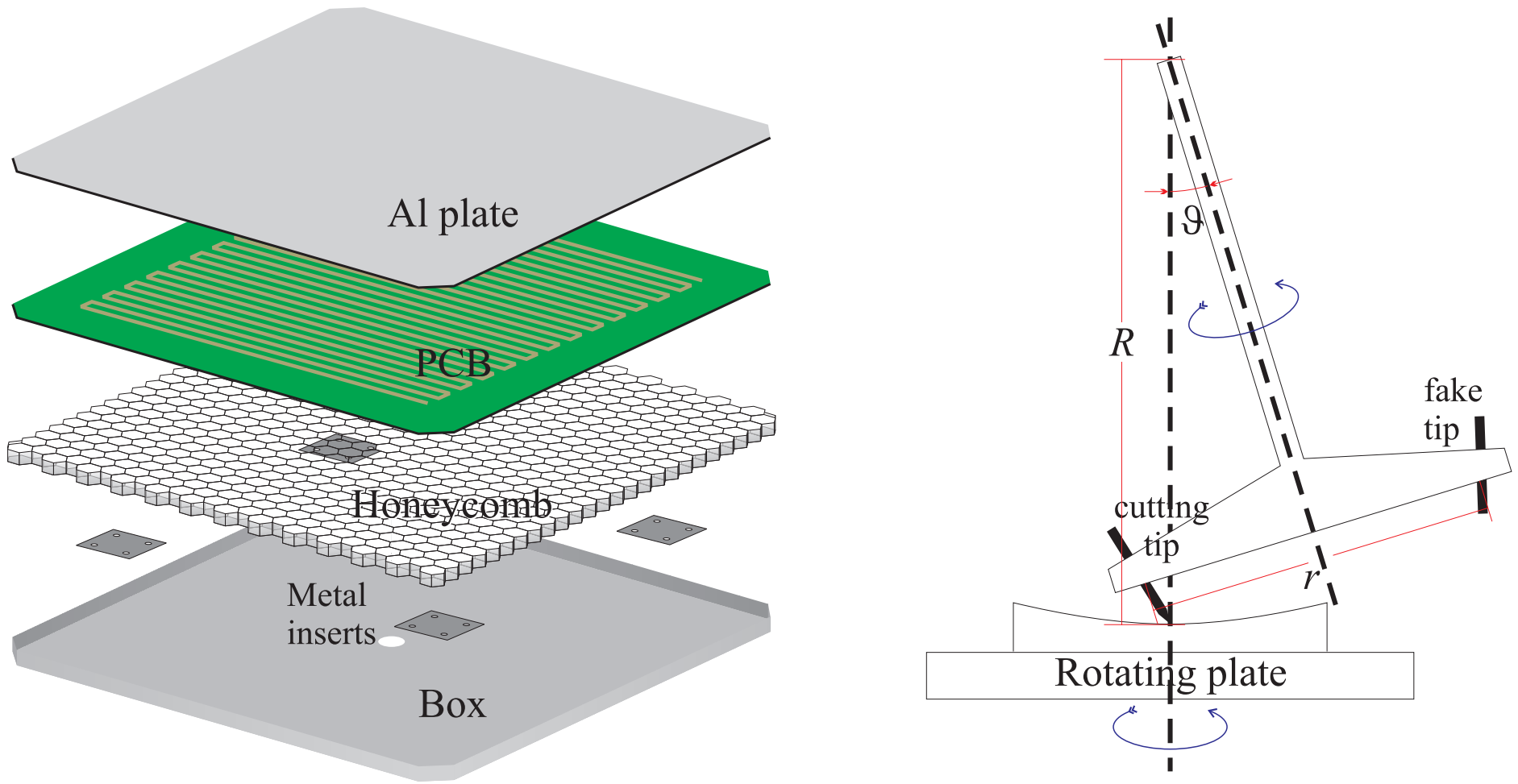

**Figure 1.** Exploded view of the *raw-blank* structure (*left*). Sketch of the milling tool and the rotating table (*right*).

About a quarter of the actual mirrors were machined from raw blanks made by MPI, which used a very similar assembling method. Their behaviour is quite similar to the mirrors made using raw blanks produced in Padova.

### 2.2 Premilling and diamond polishing

Production starts with a rough, but quick, *premilling* of the raw blanks with an accuracy of better than $\frac{1}{10}$ mm. The machining is done by fixing the raw blank onto a rotating plate and using a "T"-shaped tool, the *fly-cutter*, as shown in fig. 1 (*right*). The final shape is a spherical surface of radius $R = \frac{r}{\sin\vartheta}$ where $r$ is the radius of the milling tool and $\vartheta$ is the angle between the rotation axis of the milling tool and that of the plate.

The diamond milling of the surface is done by the LT Ultra company (Aftholderberg, Germany). After diamond milling, the roughness of the surface is well below 10 nm *rms*, as can be seen in fig. 2 (*left*) for a typical profile analysed with a commercial surface roughness tester. From the same picture one can also see one *step* of the milling machine, that can follow the desired profile at a level of the micrometer.

After coating, the overall reflectivity of the mirrors is between 85% and 90% in the visible band (see fig. 2 *right*). Once all mirrors are mounted in place, the actual reflectivity of the whole surface can be measured. Different measurement techniques agree, estimating the light collection efficiency to be $77\% \pm 4\%$[1].

### 2.3 Pre-shaped raw blanks

To further reduce production times, we investigated the feasibility of assembling pre-shaped raw blanks. Using pre-shaped raw-blanks, two major issues could be improved:

- the thickness of the Al slab, needed for the milling, was reduced from 5 to $1 \div 2$ mm;
- *premilling* could be completely skipped.

Pre-shaped mirrors are assembled, as the old one, in an *autoclave* environment, but are sandwiched between two curved moulds, that shape the final raw blank with the requested curvature radius, between 34 and 36 m.

Let us remind that, in this range, the *sagittae* vary $\sim 0.2$ mm. Therefore, we can produce all pre-shaped raw-blanks with just one *gross* curvature radius and let the diamond milling machine refine them by removing just a minimal amount of material. This results in a faster, and less expensive, overall procedure.

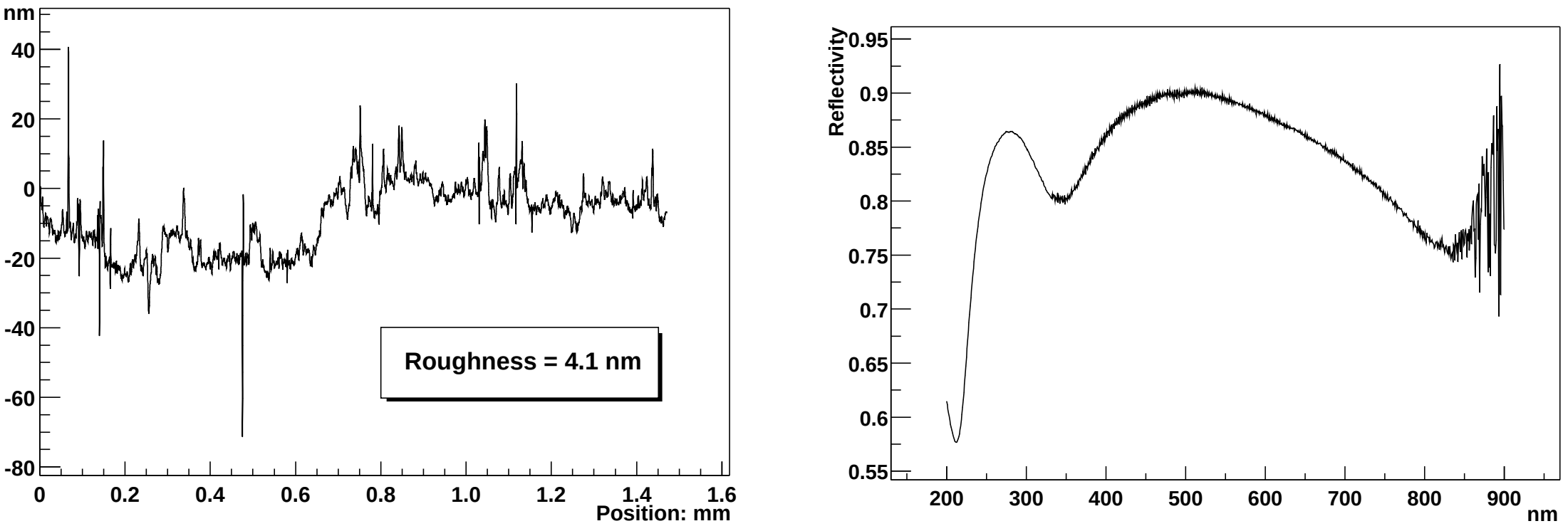

**Figure 2.** Roughness (*left*) and optical reflectivity (*right*) of a sample taken from a MAGIC mirror after milling and coating. In the roughness diagram, the actual profile (measured in nanometers) is plotted against the position (in millimeters).

### 2.4 Larger raw blanks and mirrors

Working with thinner, but pre-curved, Al-plates also makes the assembling and machining of larger mirrors easier: in fact, a $1\,\mathrm{m} \times 1\,\mathrm{m}$ spherical mirror of $\sim 34\,\mathrm{m}$ of curvature radius requests that $\sim 4\,\mathrm{mm}$ of material would have to be removed from its centre if it were assembled with a flat plate, whereas virtually no material at all is removed from pre-shaped mirrors. Moreover, as MAGIC currently uses panels hosting four fixed mirrors each for active optics, increasing the mirror size also eliminates the necessity to use back-panels, as the mirrors themselves could be controlled with minor refinements to the actual active optics device.

Larger mirrors have nevertheless some drawbacks. In fact, MAGIC is made up with many small spherical mirrors that best fit the desired overall parabolic shape: increasing mirror size makes the fit harder, at least for the outer mirrors, where the requested paraboloid differs more from a sphere. Astigmatic mirrors can adapt better to parabolic shapes, but their production can be quite difficult, and for MAGIC-II[3], if machining of astigmatic mirrors does not prove to be feasible via the diamond milling technique, it could be envisaged the construction of a mixed-size surface, with 1-m mirrors in the inner rings and 50-cm ones outside.

## 3. Optical quality checks of the mirrors

To check the optical quality, we use an ultrabright blue LED that is reflected by the mirror under study onto a white screen: the reflected image, the *spot*, is analysed with a CCD camera. The centre of the screen and the LED are at a distance of $\sim 40\,\mathrm{cm}$, and are symmetric with respect to the mirror axis. The distance between the mirror and the LED (and between the mirror and the screen) is equal to the nominal curvature radius (or twice the focal length) of the mirror itself, in such a way that a point image is reflected again into a point image.

For the quality check we compute the $R_{90}$, that is the radius of the circle, taken from the centre of gravity of the spot, containing 90% of the total, reflected light. As the picture is taken at twice the focal, when focusing light-rays coming from *infinity* the spot is actually half the size of the measured one. Looking at fig. 3, the result is that 90% of the light from a parallel beam will be focused, on average, within a circle of 1 cm of diameter, or less than half of MAGIC pixel size (PMTs of 1"∅).

The effective radius of curvature is defined operatively as the distance between the spot and the mirror where the $R_{90}$ is minimum. It is the effective radius of curvature that is taken into account for the correct positioning of the mirror onto the parabolic dish, having to match the local mean curvature radius of the paraboloid.

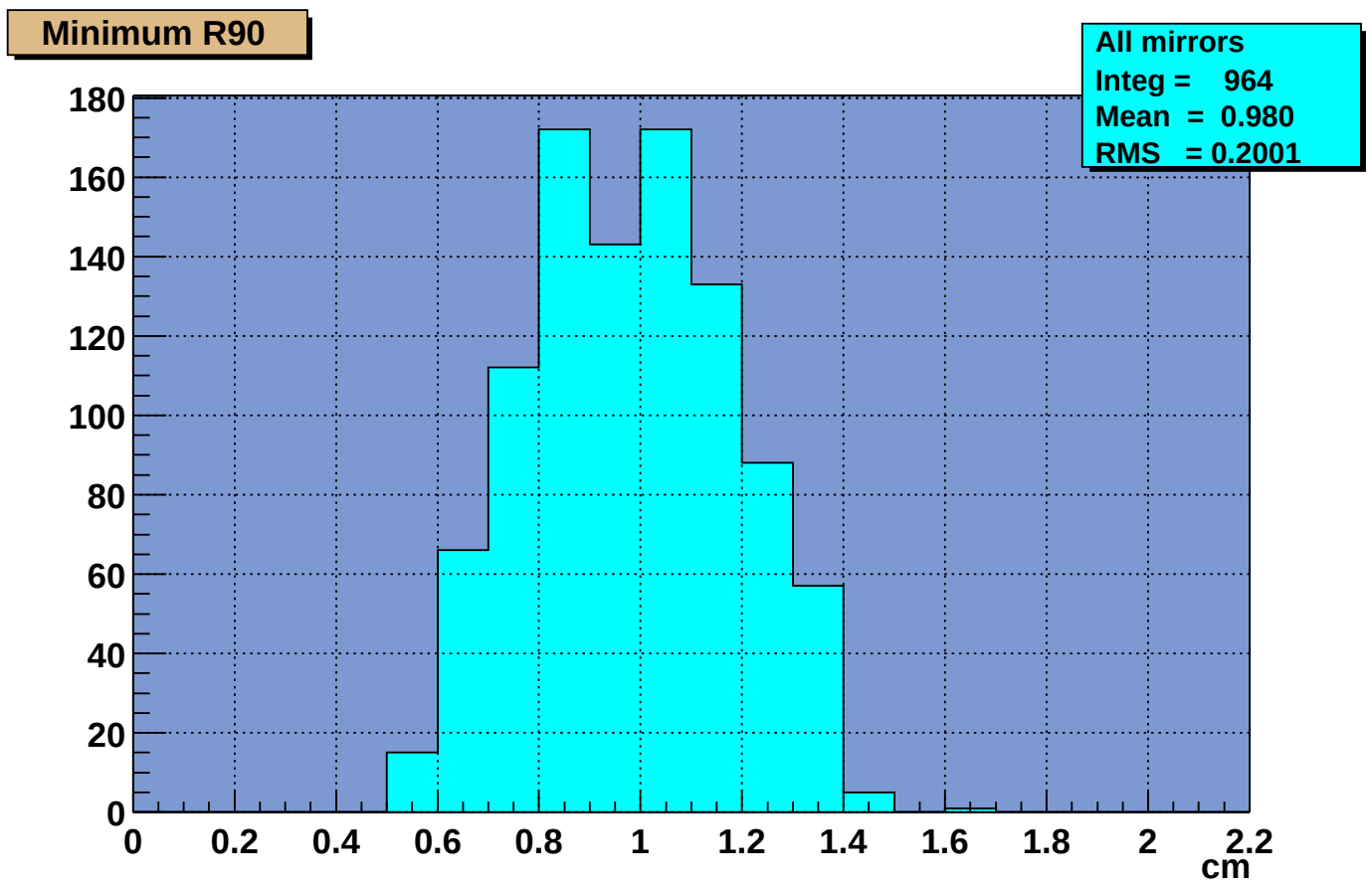

**Figure 3.** Distribution of $R_{90}$ for the actual MAGIC mirrors.

## 4. Conclusions

The reflecting surface of the MAGIC telescope is operating in open air already for some years. During this period, few percents of it were damaged by water infiltrations inside the sandwich structure. Due to the extremely bad weather conditions of last winter, the continuos changing of state between water and ice had the effect of detaching some of the aluminium plates from the raw blanks structure. Maintenance and substitution of damaged mirrors is now on-going. Moreover, an improved sealing has been now devised in order to prevent water creeping inside the mirror structure.

On the other hand, even in these extreme weather conditions, with also strong wind blowing and *calima* (Sahara sand particles with $5 \div 10\,\mu\mathrm{m}\,\varnothing$), the hard surface seems to resist quite well: samples coming from mirrors that had to be substituted evidenced no change in local reflectivity.

Few mirrors made from pre-shaped raw-blanks are already installed on MAGIC and survived safely the last exceptionally-hard winter. In the near future it is foreseen to install also some larger 1-m mirror, in order to test the technology and adopt it for the construction of MAGIC II.

## 5. Acknowledgements

The production of the reflecting surface of the MAGIC Telescope was made possible by the financial contributions of the Italian INFN and the German BMBF, to whom goes our grateful acknowledgement. We would also like to thank the IAC for the excellent working conditions provided at El Roque de los Muchachos.

# Comparison of On-Off and Wobble mode observations for MAGIC

T.Bretz[a], D.Dorner[a], B.Riegel[a], D.Höhne[a], K.Berger[a] for the MAGIC Collaboration
*(a) Institut für theoretische Physik und Astrophysik, Universität Würzburg, Würzburg, Germany*
*(b) Updated collaborator list at:* http://magic.mppmu.mpg.de/collaboration/members/index.html

Presenter: T.Bretz (tbretz@astro.uni-wuerzburg.de), ger-bretz-T-abs1-og23-oral

One of the main goals of the MAGIC telescope is to reduce the energy threshold accessible to the imaging Cherenkov technique. Due to the fact that the background suppression becomes increasingly difficult for lower energies it is important to reduce the possible systematic error of the background determination. An observation mode in which the source is not in the center of the camera allows to have a symmetrical sky position, for which the background can be determined. This ensures having similar systematics like nearly the same zenith angle and the same weather conditions. A comparison with the classical On-Off observation will be presented.

**Introduction** When observing a gamma ray source candidate with an IACT, standard observations are done such that the source is located in the center of the camera. To determine the background *behind* the source an independent sky region, under observation conditions as similar as possible, is observed or the background is extrapolated from an off-source region in the same measurement (eg. the region between $(\vartheta/\circ)^2 = 0.2$ and $(\vartheta/\circ)^2 = 0.4$ in 1, right). In this case having the source in the most privileged position in the camera there is the possibility of having a fake excess of events being a feature of the background rather than a gamma ray source. Instead of observing the source in the center of the camera, the telescope can be pointed to a sky position slightly off-source (for the MAGIC telescope typically $0.4°$). The background can then be extracted from a so called *anti-source* position symmetrical w.r.t the camera center. Due to the ALT-AZ mount of the telescope source and anti-source will rotate around the center of the camera which might smear out systematics due to camera defects. The distance between them is chosen such that the background measurement is not influenced too much by the source itself. This observation mode is called *Wobble Mode* because several off-axis positions around the source are observed in an alternating fashion. Typically two position symmetric w.r.t the source are convenient making source and anti-source more equivalent (eg. zenith angle). It was extensively and successfully used by the HEGRA telescopes and is currently used by the H.E.S.S. telescopes.
In this paper a comparison of observations made in wobble mode and in the on-off mode is presented.

**On-Off observation mode** In on-off observation mode the source is located in the camera center. For background determination a similar sky region must be observed. Because it cannot be done with the same instrument at the same time, it must be done under different conditions (weather, night-sky-background, etc.). Thus scaling of the background measurement is necessary to achieve an agreement of the background levels of the on- and off-source observations. This scaling factor can only be determined by comparing parts of the data not influenced by the signal, eg. events which have their origin not at the source position. Having good statistics this scaling factor can be determined accurately enough to get plausible results, however still introducing an additional systematic error. Having detected a source, studies on its spectrum are essential to extract the interesting physics about the source. Therefore the data sample is divided into subsamples of different energy as determined by an energy estimation algorithm [3]. To be able to calculate the gamma ray flux for each of these energy bins independently it is necessary, that the scaling of the background in all bins produces reasonable results. Having slightly different conditions for signal and background measurements, e.g. a little higher humidity or the changed star light at another sky region, the scale factor for all energy bins might not be unique anymore.
Furthermore the construction of MAGIC's camera makes the center of the camera a privileged position. Due to the gap between two outer sectors of the camera (a sketch is shown in 3) and because only the fine-pixel part of the camera triggers the readout, events are aligned towards the center.

**Wobble observation mode** In wobble observations a position symmetric w.r.t. the camera center for the source is available. This anti-source position is used to determine the background to be subtracted from the signal. This offers the possibility of continuous monitoring of a source without having to interrupt the observations for taking off data. For reasons of symmetry almost identical conditions for both, signal- and background-observation, are available. Consequently in the first order no scaling has to be applied to the background measurement. Due to inhomogeneities (eg. single broken PMTs) in the camera acceptance it might be necessary to apply a small correction. The rotation of the source around the camera center smears out possible inhomogeneities and further decreases a necessary small correction factor. It also improves the quality of the background measurement because it does not correspond to a unique sky position anymore. However, there is some danger that the background measurement is not statistically independent from the signal of a strong source. Therefore a special anti-source cut has to be applied, i.e. any event analyses can not be assigned to both positions. Already simple technique (classical $\alpha$ method) suppress this bias well. More advanced analysis', assigning a single source position to each shower image like Whipple's Disp-method [5], suppress this bias almost completely. This is done by applying source-position dependent cuts, also as a veto to the anti-source position and vice versa. Figure 1 shows a $\vartheta^2$ signal- and background-plot [2] taken from wobble mode Crab observations without applying any scaling. Figure 2 shows the size distribution (solid line) of Monte Carlo gamma events assigned to the source position by surviving such cuts (plus gamma-/background-separation cuts) and the distribution (dashed line) of the same events which were wrongly assigned to the anti-source position only. It can be seen that the there are less than $1\%$ wrongly assigned events in the lowest size bins which is negligible compared to statistical errors. For lower sizes this effect increases further, but it can be suppressed by more sophisticated cuts or using the Monte Carlo distribution for correction.

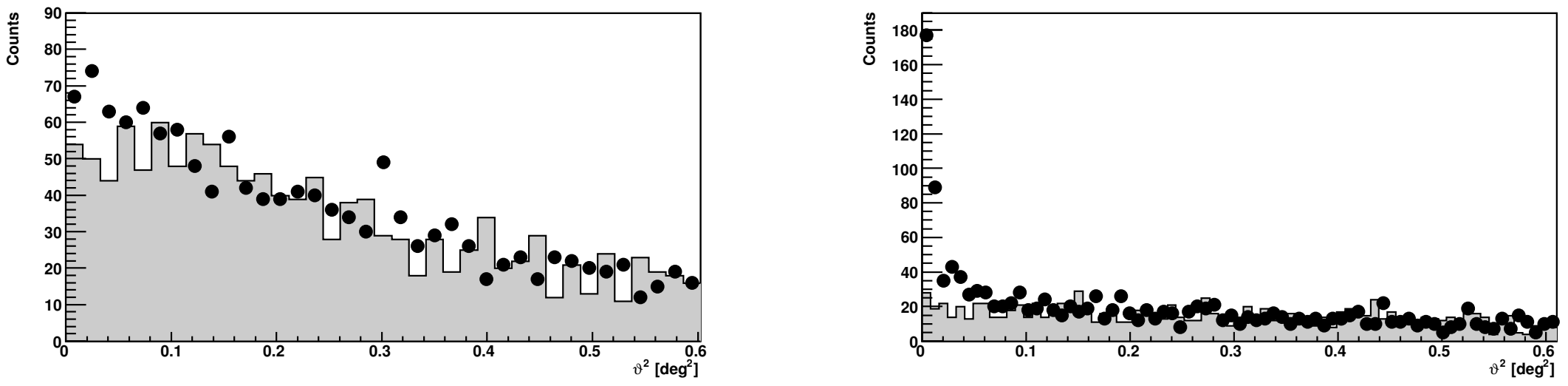

**Figure 1.** . Signal- (black dots) and background (gray shaded) as taken from wobble mode Crab observations (loose cuts) and without applying any scaling. Left: low size bin (around $90GeV$). Right: medium size bin (around $600GeV$).

**Discussion** An example for a measurement can be seen in 1. For simplicity the binning was not chosen in estimated energy but in size, the total measured intensity of the shower image (which has an almost linear correlation to the energy of the primary particle). The plots show the $\vartheta^2$-signal (black dots) of a wobble mode measurement of Crab and the measured background (signal of the anti-source). It can be seen that both measurements fit quite well and only small scaling corrections might be necessary, introducing only a small additional systematic error. Furthermore the scaling factor also depends on the way of its determination (eg. unifying the integral between $(\vartheta/\circ)^2 = 0.3$ and $(\vartheta/\circ)^2 = 0.5$), introducing another small systematic error. Extrapolating an accurate background from the large $\vartheta$-region into the low $\vartheta$-region, in the low-energy case as shown in the plot, seems problematic. Different functions may fit the background well at large $\vartheta$ or $\alpha$, while giving quite different predictions for the background in the signal region. well. Having identical conditions for signal and background measurement in wobble mode means also an important reduction in the work necessary to ensure the compatibility of both measurements, eg. making sure that the image parameter distributions fit well which is influenced by environmental conditions like weather, zenith angle, etc. Because the compatibility of signal and background measurement is given by construction the results are more robust.

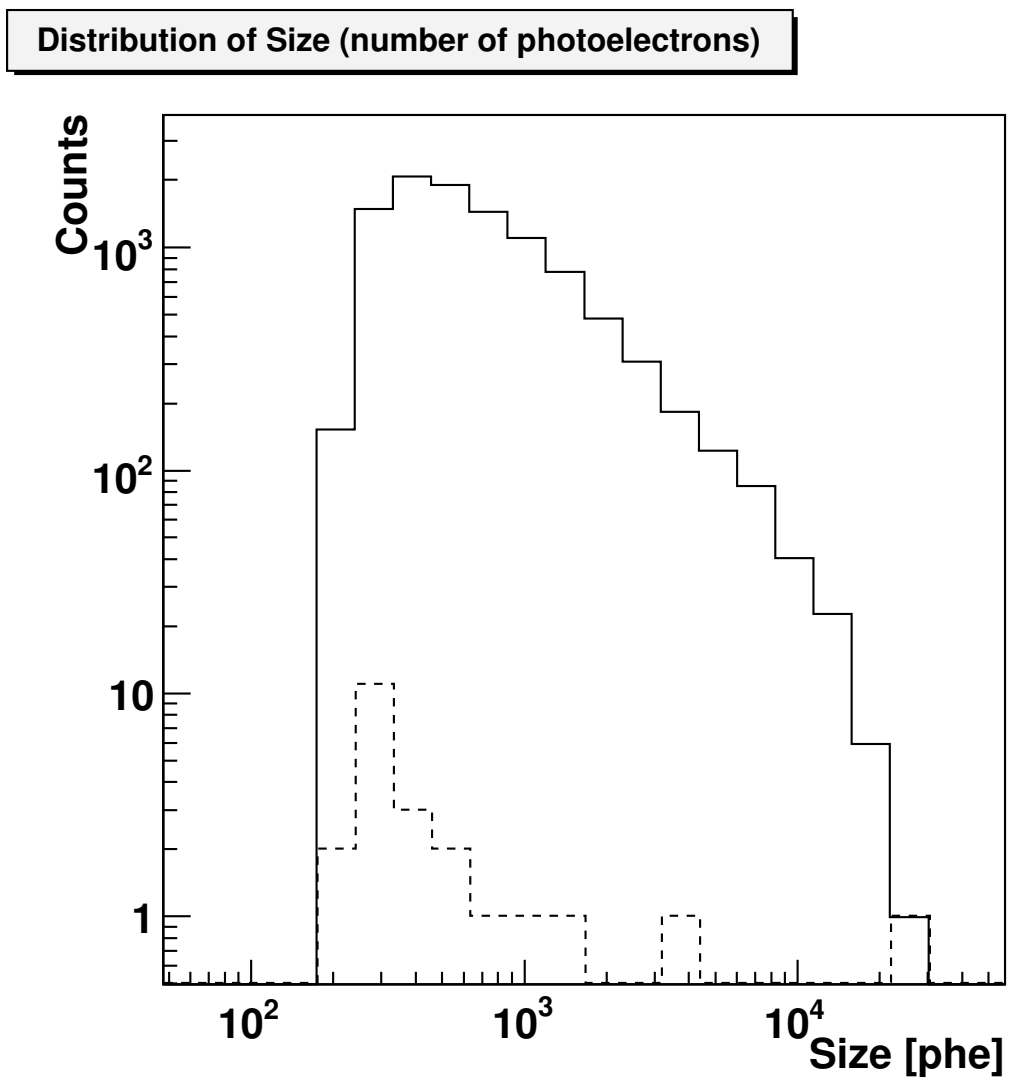

**Figure 2.** Size distribution obtained from Monte Carlo [4] gamma ray events surviving the cuts assigned to the source position (solid) and the anti-source position (dashed).

**Comparison** To avoid systematic errors in on-off observation mode due to wrong scaling, as explained earlier, a background measurement must be done under conditions as similar as possible to the ones taking on-data. Therefore any on-measurement must be followed by a prompt off-measurement. To get enough statistics it is essential to take roughly the same amount of off- and on-data. On the other hand wobble mode observations with the MAGIC telescope result in a loss of sensitivity of less than $20\%$ . This is due to more showers not contained in the trigger area (mainly the inner part of the camera) anymore or leaking out of the camera due their size. Figure 3 shows a comparison of the distribution of the center-of-gravity of Monte Carlo gammas in the camera after cuts f or on-off and wobble observation mode. For this study the position information of the shower in the camera has not been used, it might further improve the sensitivity in wobble mode. To compare both observation modes, data of the Crab Nebula has been taken at similar zenith angles in both modes. The analysis is explained in more details in [2] and [1]. The data analyzed has gone though a strict selection (w.r.t. weather conditions and apparatus performance). Robust and conservative cuts as [2] have been used for both datasets. They were trained dividing the samples each in two subsamples. Gamma-/background-separation has been optimized for maximum significance for both subsamples independently. This, however, raises the analysis threshold, but gives the most reliable output for the a study like this. After being convinced that both subsamples gave similar cut coefficients the full sample was analyzed with an artifical compromise between the two sets of cut coefficients. The energy threshold has been determined as the maximum bin in a dN/dE distribution after cuts of MC gammas produced with a spectral index of -2.6. Table 1 (top) contains the information about the data analyzed for this study and the results. Table 1 (bottom) contains derived values which are more useful for a comparison of both modes.

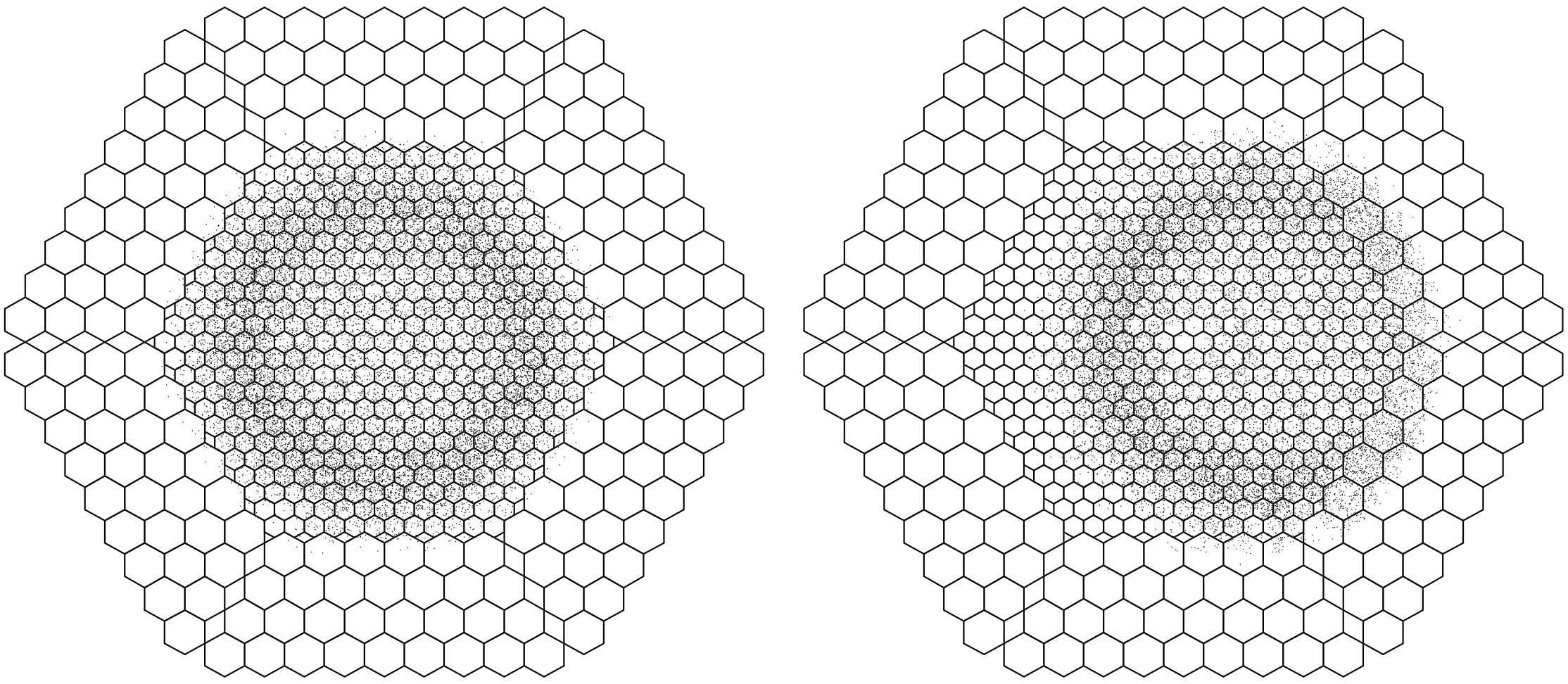

**Figure 3.** Distribution of center-of-gravity of the intensity distribution of Monte Carlo [4] gamma showers. Left: The source centered in the camera. Right: The source 0.4○ off center. The loss into the coarse pixel region of the camera in wobble-mode is small, while there are also showers on the left side which now are fully contained in the camera or the fine pixel region.

| Observation Mode | Data Period | Zd range | Observation Time | | Significance (Li/Ma) | Background Scale Factor |
|---|---|---|---|---|---|---|
| On/Off | Sep'04 - Jan'05 | 5.6°-23.7° | (On) 14837s | 4.12h | 30.5 | 0.94 |
| Wobble | Sep'04 - Jan'05 | 5.6°-17.2° | 13642s | 3.79h | 22.3 | 1.00 |

| Observation Mode | Threshold | Excess Events | Background Events | Excess Rate | Background Rate | Significance per On-Time |
|---|---|---|---|---|---|---|
| On/Off | 205-236GeV | 2233 | 1727 | 9.0evt/min | 7.0evt/min | $15.2\sigma/\sqrt{h}$ |
| Wobble | 205-236GeV | 1621 | 1864 | 7.1evt/min | 8.2evt/min | $11.5\sigma/\sqrt{h}$ |

**Table 1.** Top: Information about the observations and results. The off-data duration only differs from on-data in the order of seconds. Bottom: Information derived from the measurement shown.

**Conclusions** When comparing the on-off and wobble observation modes in the framework of the MAGIC telescope, three aspects deserve particular attention: 1) loss of sensitivity for off-axis observations in the small camera, 2) efficiency of gamma-/background-separation at low energies and 3) the possibility of continuous monitoring of the source on wobble mode. It has been shown that the decrease in sensitivity for wobble observations employing a 0.4° offset from the camera center is compensated by the possibility of observing the source continuously, assuming that in on-off mode half of the time is devoted to the off observations. This can be compensated by defining more than one statistical independent off-region in the same distance to the camera center. The necessity of having non-overlapping on and off regions on the camera makes it difficult to use the wobble mode described here for extended sources ($> 0.1°$). The analysis, specially at low energies, benefits from the precision gained in the background determination in wobble observations. If the background is determined from off observations, scaling and uncontrollable statistical and systematic effects (e.g. due to varying night sky conditions) render the reproducibility of the measurements more difficult, implying generally larger errors of the background estimate in the signal region.

**Acknowledgments** We acknowledge the support of the german BMBF (05 CMOMG1/3).

# The Data Acquisition of the MAGIC II Telescope.

J.A. Coarasa[a], J. Cortina[b], M. Barcelo[b], M. Martinez[b], I. Martinez[b], E. Ona-Wilhelmi[b], R. Paoletti[c], R. Pegna[c], A. Piccioli[c] and N. Turini[c] on behalf of the MAGIC collaboration.
*(a) Max Planck Institut für Physik, Föhringer Ring 6, 80805 München, Germany*
*(b) IFAE, Edifici Cn, Facultat de Ciències, UAB, 08193 Bellaterra (Barcelona), Spain*
*(c) University of Siena and I.N.F.N. Sezione di Pisa, Italy*

Presenter: J.A. Coarasa (tcoarasa@mppmu.mpg.de), ger-coarasa-perez-JA-abs1-og15-poster

The MAGIC telescope is the largest $gamma$-ray Imaging Cherenkov telescope in the world. It is operating since 2004 at the Roque de Los Muchachos observatory, La Palma, Canary islands. MAGIC-II is the upgrade of this project, consisting of a twin telescope frame with innovative features like new photon detectors to lower the threshold energy further, and an ultrafast signal sampling to reduce the effect of the diffuse night sky background. The new acquisition system is based upon a low power analog sampler (Domino Ring Sampler) with frequency ranging from 1.5 to 4.5 GHz and data are digitized with a 12 bits resolution ADC. The analog sampler, originally designed for the MEG experiment, has been successfully tested on site and showed a very good linearity and single photon discrimination capability. Data management is performed by 9U VME digital boards which handle the data compression and reformatting as well. Every board hosts 32 analog channels plus auxiliary digital signals for trigger and monitor purposes. For a 1 kHz trigger rate and a 2 GHz frequency sampling, the data throughput can be as high as 100 MBytes/s, thus being a challenge for modern data transmission and storage solutions. The data are transferred to PCI memory via Gbit optical links using the CERN S-link protocol and to the mass storage system consisting of a RAID system and tape minilibraries. The data acquisition system design and performance will be described in detail.

## 1. Introduction

The MAGIC telescope aims at measuring low energy gamma-initiated showers ($E_{th} \approx 30GeV$) by looking at the Cherenkov light emitted in the atmosphere [1]. This signal is contaminated by background events due to various sources: hadronic showers, local muons, Light of the Night Sky Background (NSB), moon light and bright stars in the field of view of the camera.

The upgrade program of the MAGIC telescope involves the construction of a second telescope (MAGIC-II) which is mechanically identical. Several technological updates are foreseen for MAGIC-II, namely: new photon detectors, either high quantum efficiency photomultipliers or new generation silicon photomultipliers (APD or SiPM); and a new data acquisition system, based on fast (2 GHz) analog samplers, coupled to high bandwidth data transmission. These two changes will allow a more effective suppression of the light of the night sky and a stronger rejection of hadron induced air showers, while efficiently selecting $\gamma$-ray showers through using the more accurate timing information.

The Data taking chain works the following way. The new photon detectors collect the 1-2 nsec full width half maximum Cherenkov light flashes. Their ouput signals are amplified by ultrafast and low-noise transimpedance pre-amplifiers in the camera housing. The amplified analog signals are transmitted over 170 m long optical fibers. In the electronics hut the signals are split. One branch goes to a discriminator with a software adjustable threshold that generates a signal for a 2 level trigger system [2] capable of limiting the trigger rate to $\lesssim 1$ kHz. The signal in the second branch goes to the new analog sampler described in the next section. The digitized signals are stored by a data acquisition system capable to process a sustained data rate of $\approx$ 100 MB/s. Immediately after the data is taken, thanks to an independent network for the storage, an Online Analysis starts on the data.

## 2. Sampling Electronics

The digitizing core of the Data Acquisition for the MAGIC-II telescope is constituted by the analog sampler developed at the Paul Scherrer Institute (PSI, Villigen-Switzerland) [3], called Domino Ring Sampler (DRS) and now at a second version (DRS2). It is an analog sampler designed with 0.25 $\mu$m CMOS technology. An extensive work of characterization has been performed on the Domino chip [4] whose main characteristics are: integration of at least 10 analog channels in one chip; analog bandwidth of around 1 GHz; excellent time resolution, less than 80 picosecond with a sampling frequency of 4 GHz; very low power consumption, around 35 mW per chip; the response of the sampler cells can be calibrated and the useful input voltage can be extended to 2 Volts (twice the range for commercial Flash ADCs).

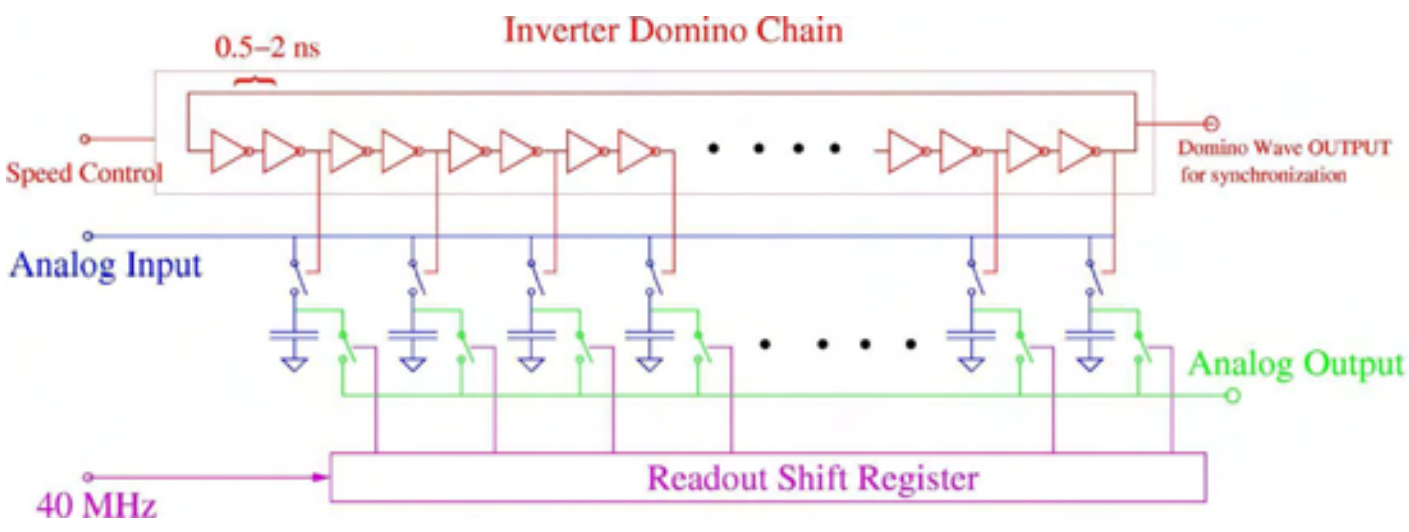

**Figure 1.** Schematic picture of the Domino sampler. Shown on top is the inverters sequence that originates the Domino wave. The shift register on the bottom enables serially the single capacitors connecting them to the output stage.

The sampler consists of two sections, an analog one for the signal sampling and a digital for control and multiplexing. The analog signal is stored in a multi capacitor bank (1024 cell in DRS2) that is organized as a ring buffer, in which the single capacitors are sequentially enabled by a shift register driven by a high frequency clock internally generated (see Figure 1), called Domino wave. At a 2.5 GHz frequency the sampling window is 400 ns wide. The speed of the domino wave is controlled by an external voltage such that the domino inverters can be seen as a Voltage Controlled Oscillator (VCO). Every turn a pulse is produced for synchronization and monitoring. The phase and speed of the domino wave are synchronized to an external common reference clock by an on-board PLL, designed around the domino VCO. The sampling signal jitter is less than 200 ps (see Figure 2a).

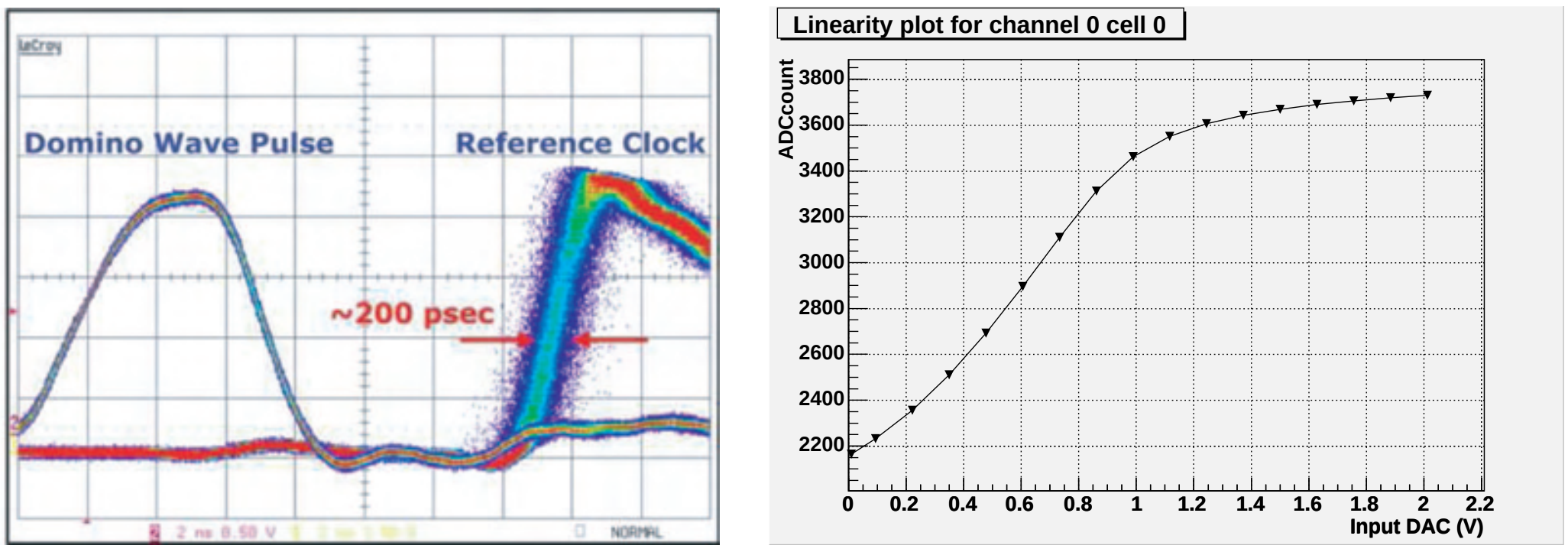

(a) (b)

**Figure 2.** **(a)** Domino wave frequency jitter. **(b)** Domino response as a function of the input voltage.

The chip is housed in a PLCC package and mounted on a mezzanine card, also called CRAB (Capacitor Ring Analog Board). Once the external trigger has been received, the sampled signal in the ring buffer is put on an output stage by a multiplexer and digitized at high resolution (12 bits) at lower frequency (40 MHz). The digital conversion is done by an external ADC. The transfer curve of the output stage is typical of a MOS transistor, with an active region quite linear up to 800 mV, that tends to saturate for larger signals (see Figure 2b). This characteristics turns out to be a useful one since it is a natural signal compression and results in an extension of the response dynamic range. With a calibration procedure that is implemented in the analog mezzanine it is possible to calibrate input signals up to 2 Volt, about twice the range of commercial Flash ADCs.

By using fast sampling frequency and large dynamic range it is possible to reconstruct the shape of individual photons impinging on the photon detector and measure the signal with good precision, allowing for further background rejection.

## 3. Data Readout

The readout system is based on the so called Pulsar board (see Figure 3) that can host up to 4 analog mezzanine boards, for a total of 32 analog channels. Three FPGAs (ALTERA EP20k400) are mounted on each board; they are responsible for interfacing to the CRAB boards and data communication with the transmission board located on the rear VME backplane. Additional digital signals are interfaced via front panel connectors and enter directly in the data acquisition stream. The external trigger signal is sampled together with the analog channels giving a precise time reference, useful for synchronization of different devices or precise time measurement.

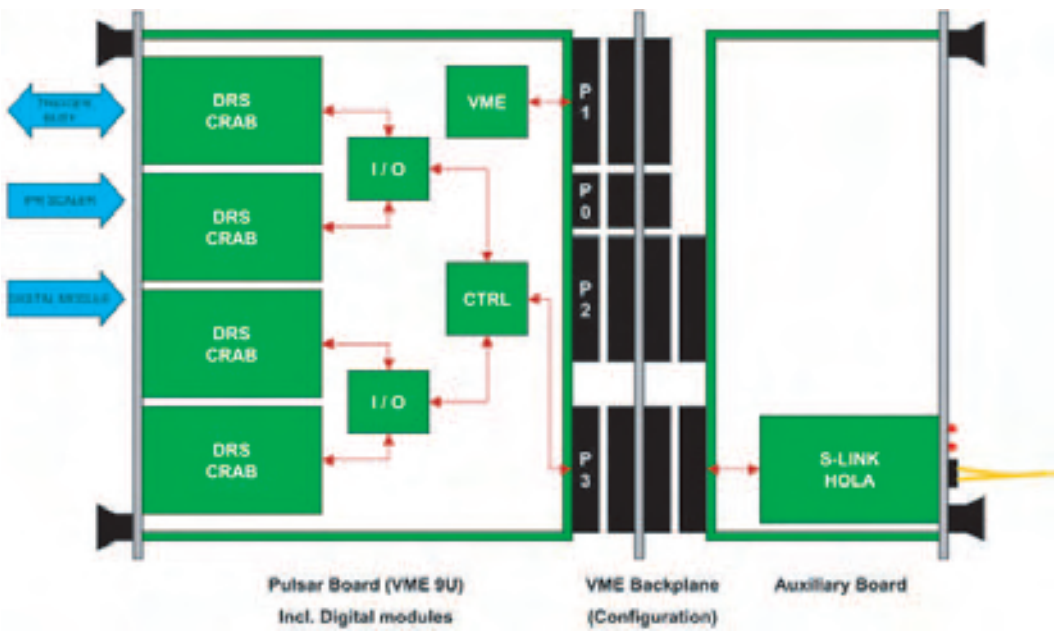

**Figure 3.** Schematic drawing of the data readout board.

The whole readout system consists of 18 Pulsar boards that are accommodated in two VME-9U crates, interfaced to the telescope control system by an embedded CPU. The data transmission is performed by optical link drivers (called HOLA), specifically designed for LHC experiments [5]. The optical fibers are interfaced to a PC memory by dedicated PCI cards named FILAR, also designed for LHC purposes.

The usage of an ultrafast sampler in Cherenkov telescopes with a large number of channels brings to systems that, at 1 kHz trigger rate, produce a data flow of the order of 100 MBytes/s, the data volume to store being of the order of several Terabytes per night. To cope with such large throughput rates and data volumes a RAID system based on a separate network (probably fiberchannel) is being tested. Such a network (allowing up to 4 Gbit/s access) together with a file locking filesystem will allow true Online access to the just recorded data, for analysis and archiving. A cluster of computers will start an Online Analysis immediately after the files are closed.

## 4. Performance test

The readout system has been used to measure the night sky background with standard photomultipliers in the MAGIC telescope framework and a work on the characterization of SiPM has just started.

Tests performed on the MAGIC telescope site have shown that it is possible to disentangle the single photoelectrons (phe) and to measure the diffuse light background (40 phe in a 400 ns window) and to obtain a physics meaningful measurement, provided that the photomultiplier signals are sampled at 2.5 GHz at least. In Figure 4a a typical image of the incoming photons is shown. The poissonian nature of this process is verified by the arrival time distribution of single peaks that shows a typical exponential trend.

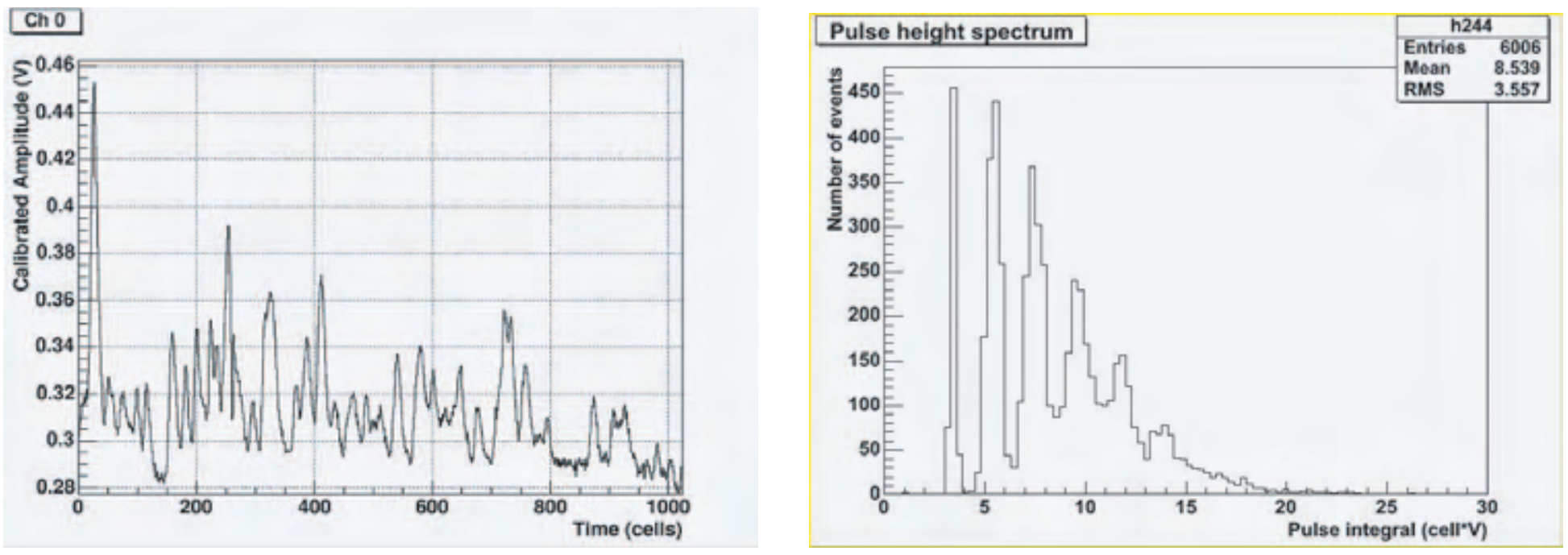

**(a)** **(b)**

**Figure 4.** **(a)** Typical sampling of a MAGIC photomultiplier signal during a 400 nsec interval. **(b)** Charge spectrum on SiPM with simple integrating algorithm.

The Domino readout can be effectively used for the characterization of very fast pulses, such as those produced by Gieger-mode, APD-based, silicon photomultipliers (SiPM) [6]. Figure 4b shows the preliminary charge spectrum obtained by illuminating a SiPM with a fast-pulsed LED at a low light level. The peaks due to the individual photolectrons are clearly resolved.

## 5. Conclusions

The upgrade program of the MAGIC telescope is currently under way. Together with the adoption of new photon detectors, the analog signals are going to be sampled by the Domino ring sampling chip. For a 1 kHz trigger rate and a sampling frequency of 2 GHz, the data throughput is higher than 100 MBytes/s, being a challenge for modern data transmission and storage solutions. The data are transferred to PCI memory via Gbit optical links using the CERN S-link protocol and to the mass storage system.

## 6. Acknowledgements

We would like to thank the IAC for excellent working conditions. The support of the German BMBF and MPG, the Italian INFN and the Spanish CICYT is gratefully acknowledged.

# Technical Performance of the MAGIC Telescope

J. Cortina[a], A. Armada[a], A. Biland[b], O. Blanch[a], M. Garczarczyk[c], F. Goebel[c], P. Majumdar[c], M. Mariotti[d], A. Moralejo[d], D. Paneque[c], R. Paoletti[e], N. Turini[e] and the MAGIC coll.

*(a) IFAE, Edifici Cn, Universitat Autònoma de Barcelona, Bellaterra 08193 Spain*
*(b) ETH Hönggerberg, CH-8093 Zürich, Switzerland*
*(c) MPI für Physik, Föhringer Ring 6, 80805 Munich, Germany*
*(d) Dipart. di Fisica, Univ. di Padova and INFN Padova, Via Marzolo 8, 35131 Padova, Italy*
*(e) Dipart. di Fisica, Univ. di Siena and INFN Pisa, Via F. Buonarroti 2, 56127 Pisa, Italy*
Presenter: J. Cortina (cortina@ifae.es), spa-cortina-J-abs1-og27-oral

The 17m diameter MAGIC Cherenkov telescope was completed in its nominal configuration and started commissioning in summer 2004. MAGIC has emphasized technical innovation in its design. Standard operation regularly involves the largest reflector in the world, active mirror control, analog optical transport of the signal, a fully programmable two-level trigger and 300 MHz signal digitizers. We describe the general performance during the first months of datataking.

## 1. Introduction

The MAGIC (Major Atmospheric Gamma Imaging Cherenkov) Telescope was designed in 1998 [6, 1] with the main goal of being the Imaging Cherenkov Telescope with the lowest possible gamma energy threshold. It was based on the experience acquired with the first generation of Cherenkov telescopes and intended to incorporate a large number of technological improvements. In this paper we present a first review of the performance of the telescope elements.

## 2. Frame and drive system

The 17 m diameter $f/D = 1$ telescope frame is made of light weight carbon fiber tubes (the frame itself weighs $< 20$ tons while the whole structure plus the undercarriage amounts to about 60 ton). The assembly of the whole frame took only one month thanks to a construction based on the so-called tube and knot system of the company MERO.

The azimuth axis of the telescope is equipped with two 11 kW motors, while the elevation axis has a single motor of the same power. The position of the telescope is measured in the mechanical telescope frame by three absolute 14-bit shaft encoders. With this configuration it is possible to measure the telescope position with an accuracy of about $0.02°$. The maximum repositioning time of the telescope is 22 seconds, below the 30 seconds that were set as a target for gamma ray burst follow-ups. By using a CCD camera mounted on the reflector frame we have established that the telescope tracks to better than a 1/10 of a pixel size (see [10] in these proceedings for more details).

## 3. Reflector and Active Mirror Control

The overall reflector shape is parabolic in order to minimize the time spread of the Cherenkov light flashes in the camera plane. The preservation of the time structure of the Cherenkov pulses is important for increasing the signal to noise ratio with respect to the night-sky background light (NSB). The dish is tessellated by 956 0.495 $\times$ 0.495 m$^2$ mirrors covering a total surface of 236 m$^2$. Each mirror is a sandwich of alluminum honeycomb on which a 5 mm plate of AlMgSi1.0 alloy is glued. The alluminum plate is diamond-milled to achieve a spherical reflecting surface with the radius of curvature that is more adequate for its position in the paraboloid. A thin quartz layer protects the mirror surface from aging.

In order to check the light collection efficiency of the reflector we use a direct method based on light measurements with a high-resolution large dynamic range CCD camera. The camera is mounted on the mirror frame in a position that permits simultaneous measurements of part of the focal plane of the camera and of the corresponding portion of the sky. For a measurement, the camera is covered with a diffusely reflective disk [9], of 17 cm diameter. The telescope is directed at a point-like source (star or distant lamp), and the CCD simultaneously records the direct light spot and its reflection from the disk in the focal plane. For strong sources, saturation of the CCD pixels is avoided by defocussing the CCD camera.

The camera used was an STL-1001E from SBIG. The reflectity of the disk is specified to be 99%. The active mirror area at time of measuring (including mirror imperfections, temporary defocussing, and all effects of shadowing) was estimated to be $212m^2$. From several sources with a wavelength spectrum peaking at 500 nm, the average specular reflectivity is measured to be $0.77 \pm 0.04$. This value has been confirmed by independent measurements using PIN diodes. Future measurements at regular intervals are being planned, and will determine the reflectivity with an improved precision of $\pm 0.01$.

A large diameter telescope makes strong requirements on the stiffness of the reflector frame. When directing the telescope to different elevation angles the reflector's surface deviates from its ideal shape under gravitational load. In order to correct these deformations in a light-weight material frame, we equipped the reflector with an "Active Mirror Control" system. Every four mirror facettes are mounted on a single panel. Two of the three mounting points of the panel are equipped with actuators which can be used to adjust its position on the frame. The main elements of each actuator are a two-phase stepping motor (full step 1.8°, holding torque 50 N cm) and a ballspindle (pitch 2 mm, maximum range 37 mm). In the center of the panel a laser module is pointed towards the common focus of the four mirrors. The panels are aligned using the artificial light source, the positions of all the laser spots are recorded and can be used as a reference to re-align them for each elevation angle.

The PSF of the reflector can be extracted from the analysis of the width of muon rings (see [4]) and from the comparison of Hillas parameters in real and MC data. The reflector is focused at a distance of 10 km because this is the typical distance to the shower maximum of low zenith angle 100 GeV $\gamma$-ray showers. After AMC reflector adjustment, a point-like light source at this distance produces a gaussian image at the camera plane with $\sigma$=10.5 mm, which corresponds to 0.035°.

## 4. Camera, signal transmission and readout

The MAGIC camera has 3.5-3.8° FOV. The inner hexagonal area is composed of 397 0.1° FOV hemispherical photomultipliers of 1 inch diameter (Electron Tubes 9116A[7]) surrounded by 180 0.2° FOV PMTs of 1.5 inch diameter (ET 9117A). The time response FWHM is below 1 ns. The photocathode quantum efficiency is enhanced up to 30% and extended to the UV by a special coating of the surface using a wavelength shifter [8]. Each PMT is connected to an ultrafast low-noise transimpedance pre-amplifier, the 6-dynode high voltage system is stabilized with an active load. Dedicated light collectors have been designed to let the photon double-cross the PMT photocathode for large acceptance angles.

The PMT signals are transmitted over 162 m long optical fibers using Vertical Cavity Emitting Laser Drivers (VCSELs, 850 nm wavelength). Transmission over optical links drastically reduces the weight and size of the cables and protects the Cherenkov signal from ambient electromagnetic noise in the line. In the receiver boards located at the electronics room the signal is amplified and split. One branch goes to a software adjustable threshold discriminator that generates a digital signal for the trigger system. The signal in the second branch is stretched to 6 ns FWHM and again split into a high gain line where it is further amplified by a factor $\sim$10 while the low gain line is only delayed by 55 ns. If the signal is above a preset threshold around 50 photoelectrons both lines are combined and digitized by the same FADC channel.

8 bit 300 MHz Flash ADCs continuously digitize the analog signals. If a trigger signal arrives within less than 100 $\mu$s the position of the signal in the ring buffer for each pixel is determined and for each pixel 15 high gain plus 15 low gain samples are written to a FIFO buffer at a maximum rate of 80 Mbyte/s. The readout of the ring buffer results in a dead time $\sim$20 $\mu$s. This corresponds to about 2% dead time at the design trigger rate of 1 kHz. The data are saved to a RAID0 disk system at a rate up to 20 MByte/s which results in up to 800 GByte raw data per night.

The charge in the high gain is intercalibrated for each individual channel with the charge in the low gain by using pulses of about 70 phe that do not saturate the high gain yet but already give a measurable signal at the low gain.

The main source of noise is found to be light of night sky. In average the inner pixels record 0.13 NSB phe/ns. This corresponds to a pedestal RMS of about 4.5 ADC counts in each FADC 3.3 ns time slice, which is more than a factor 2 larger than the pedestal RMS due to electronic noise. The current in the anodes of the PMTs is also proportional to the rate of NSB, so the pedestal RMS is also found out to be proportional to the square root of the anode current. For an average pixel the pedestal RMS in a FADC time slice is 6.5 ADC counts/$\sqrt{\mu A}$.

About 3% of the channels are in average defective during standard operation due to problems in the PMT, electronic base, optical transmission or FADC.

## 5. Trigger

For each channel the above mentioned discriminator issues a digital signal whenever a pulse is above $\sim$10 phe ($\sim$12 phe for galactic sources with an increased light background). The individual pixel rates of the channels included in the trigger are monitored using 100 MHz scalers and used to dynamically regulate the discriminator thresholds. This "Individual Pixel Rate Control" acts only on pixels that are affected by stars brighter than $\sim 4^m$.

The output of the discriminators goes to a second-level trigger system with programmable logic [2]. The first level (L1T) applies tight time coincidence and simple next neighbour logic on the output of the discriminators. The trigger is active in 19 hexagonal overlapping regions of 36 pixels each, to cover 325 of the inner pixels of the camera. The second level (L2T) can be used to perform a rough analysis and apply topological constraints on the event images. Using for instance a fast evaluation of the size of the Cherenkov image it is possible to significantly reduce the NSB, thus allowing a reduction of the discrimination level and the gamma ray threshold.

The global trigger rate is about 250 Hz for extragalactic sources (standard pixel threshold) and about 200 Hz for galactic sources (increased pixel threshold). According to the full MC simulation[5] this rate corresponds to a trigger threshold around 60 GeV.

## 6. Calibration

The calibration system (see [3] for a detailed description) consists of a light pulser and a continuous light source (both situated in the center of the mirror dish), a darkened, single photoelectron counting PMT ("blind pixel"), located in the camera plane and a calibrated PIN-diode 1.5 meter above the light pulser. The pulsed light is emitted by very fast (3-4 ns FWHM) and powerful ($10^8$-$10^{10}$ photons/sr) light emitting diodes in three different wavelenghts (370 nm, 460 nm and 520 nm) and different intensities (up to 2000-3000 photoelectrons per pixel and pulse). It is therefore possible to calibrate the whole readout chain in wavelength and linearity.

The conversion factor from phe to ADC counts is obtained by means of the so-called F-Factor method that estimates the number of photoelectrons from the width of the charge distribution of calibration events. The conversion factor in an individual pixel is known to drift by at most 10% in time scales of several minutes due to instabilites in the optical transmission, and coherently for all pixels in the camera by at most 20% in time scales of hours to days due to thermal effects in the HV regulation and the optical transmission. These fluctuations can be corrected out using "interleaved calibration events": during standard datataking, calibration events with a fixed color (370 nm) and intensity (around 35 phe in the inner pixels) are taken together with cosmic events at a 50 Hz fixed rate. The final precision in the determination of the charge in phe for the individual pixels is about 3%.

## 7. Observation duty cycle

The commissioning phase of the telescope ended in Fall 2004 and MAGIC started regular observations. December 2004, February and March 2005 were practically lost for data taking due to bad weather. During the rest of the months since January 2005 the telescope has recorded an average of 90 h per month. This corresponds to a duty cycle of 13%. On average about 10 hours were taken with moonlight every month.

## 8. Conclusions

The commissioning of MAGIC was successfully completed at the end of 2004. All of the technical innovations have been put to work without major problems. Most of the telescope parameters are within the design specifications. MAGIC is at present taking $\gamma$-ray data regularly.

# The DISP analysis method for point-like or extended $\gamma$ source searches/studies with the MAGIC Telescope

E. Domingo-Santamaría[a], J. Flix[a], V. Scalzotto[b], W. Wittek[c], J. Rico[a]
for the MAGIC collaboration[d]

*(a) Institut de Física d'Altes Energies, Edifici C-n, Campus UAB, 08193 Bellaterra, Spain*
*(b) Dipartimento di Fisica, Università di Padova and INFN Padova, Italy*
*(c) Max-Planck-Institut für Physik, München, Germany*
*(d) Updated collaborators list at http://wwwmagic.mppmu.mpg.de/collaboration/members/*
Presenter: E. Domingo-Santamaría (domingo@ifae.es), spa-domingo-santamaria-E-abs3-og27-poster

Many of the galactic sources of interest for the MAGIC Telescope are expected to be extended. The good telescope angular resolution of about 0.1° allows us to study extended or off-axis sources with the appropriate analysis methods. We have developed an analysis method (called Disp) that uses the information of the shower image shape to reconstruct the position of the source for each detected shower. Starting from the previously successful application by the Whipple Collaboration, the Disp method has been improved and adapted to MAGIC. We report on the performance and present the results achieved when applying it to 5.5 hours of Crab Nebula observed on-axis.

## 1. Motivation for a $\gamma$-ray arrival direction reconstruction

In the standard operation mode, an Imaging Air Cherenkov Telescope (IACT) points to the source under study. It is assumed that the source position is at the center of the camera. However, many observations involve conditions that prevent us from using this standard analysis, like observations of extended sources. This includes Galactic Supernova Remnants, Galactic plane emissions or dark matter searches. Moreover, observations of point-like sources at some offset from the telescope axis, due to important uncertainties in the a priori knowledge of the source position (as the case of unidentified EGRET sources or GRBs), due to serendipitous search for sources (when doing a sky scan [7] or in the field of view of another source), or even because on purpose off-axis observation of a well known source (called 'Wobble' observation mode), require a treatment different from the standard. An analysis method (Disp) which reconstructs the individual $\gamma$-ray arrival direction has been developed to treat all these cases. This contribution reviews the implementation of the method and its performance for the analysis of data recorded with the MAGIC Telescope [6].

## 2. The concept of the Disp method

The Disp method uses the information of the shower image shape to reconstruct the position of the source on an event-by-event basis. The source position lies on the major axis of the Hillas ellipse that fits the shower image in the camera, at a certain distance (DISP) from the image center of gravity. Fomin et al. [2] were the first to propose the use of the 'ellipticity' of the shower images (defined as WIDTH/LENGTH) to infer the position of the source of individual showers using a single IACT. The method was applied by the Whipple Collaboration [5] and provided a good angular resolution for single IACTs (0.12° above 500 GeV). This technique was also adopted by the HEGRA Collaboration when analyzing the data of the stand-alone HEGRA telescope CT1 [4], and by other IACTs.

### 2.1 The DISP parameterization

Lessard et al. [5] proposed a parameterization of DISP using the minor (WIDTH) and major (LENGTH) axes of the Hillas ellipse that characterizes the shower image. Because of the different features of the MAGIC Telescope, such as its parabolic reflecting surface and low energy threshold, we adopted a more general parameterization. This describes better the correlation between the shower elongation and the distance shower/source and improves the angular resolution. Additionally, we have added a dependence of the parameters with the total charge (SIZE) of the shower image:

$$DISP = A(SIZE) + B(SIZE) \cdot \frac{WIDTH}{LENGTH + \eta(SIZE) \cdot LEAKAGE2} \tag{1}$$

We have also included a correction term in LENGTH to account for images truncated at the edge of the camera, similar to the correction introduced by D.Kranich et al. [4] for the CT1 HEGRA telescope. The LEAKAGE2 parameter is defined as the ratio between the light content in the two outermost camera pixel rings and the total light content of the recorded shower image.

The optimal values of the Disp parameters can be determined from Monte Carlo (MC) simulations or real data from a well known point-like source. In this work, we have optimized these values with a MC simulated $\gamma$-ray sample (zenith angle $< 30^\circ$) by minimizing the average angular distance ($\theta^2$) between the real and estimated source position.

The distributions of reconstructed arrival directions are described, in a first approximation, by a bidimensional symmetric Gaussian, so that $\sim 40\%$ of the events lies within a radius of $1\sigma$ and $\sim 85\%$ within $2\sigma$. We adopt $\sigma$ as an angular resolution estimator.

### 2.2 'Head-Tail' information from shower images

The DISP calculation, eq. 1, provides two possible source position solutions along the shower major axis. Therefore, a method to select the correct source position is needed. Images in the telescope camera carry some information about the longitudinal development of the shower in the atmosphere. The 'asymmetry' charge distribution in the images contains the 'head-tail' information of the recorded shower, i.e, which image edge is closer to the source position in the camera plane. Cherenkov photons from the upper part of the shower create a narrower section of the image with a higher photon density ('head') than photons arriving from the shower tail. The photons from the shower tail should normally generate a much more fussy and more spread end of the image.

An image parameter, the so-called ASYMMETRY, is defined as the direction between the center of gravity of the charge distribution image and the position of the maximum signal pixel. It allows one in most cases to determine the 'head-tail' assigment to a shower, providing the selection efficiency for the photon density in the image is high. This is normally the case for high energy showers (>70$\%$ for SIZE>180 photoelectrons [phe]). In addition, we introduced new image asymmetry parameters have been defined to improve the 'head-tail' discrimination, like applying different set of weights to pixel charge contents. By combining them, through a multidimensional events classification algorithm, the achieved ratio of correct 'head-tail' assigment improves to up to $85\%$ for SIZE>180 phe, but the study is not completed. Here, we have used the ASYMMETRY as discriminator, leaving the improved variable combination for further studies.

## 3. Application to real data: Crab Nebula

In order to assess the angular resolution provided by the Disp method, we have analyzed 5.5 hours of Crab Nebula taken on September and October 2004, at zenith angle below 30°. The source was observed on-axis. Also, we took 3 hours OFF data for background estimation.

After data calibration and image cleaning, we have used the Random Forest method [1] to discriminate $\gamma$-ray from hadron events. The Random Forest is trained with MC $\gamma$-rays and a fraction of OFF data as hadron sample. Each event is then tagged with a HADRONNESS parameter which is an estimation of the probability for an event to be a background. As training parameters for the gamma/hadron separation we used those Hillas parameters which are basically 'independent' of the source position in the FOV of the camera, i.e., WIDTH, LENGTH, CONC, and SIZE. We selected the background sample such that its SIZE distribution resembled that of the MC sample in order to avoid dependences on the MC generated spectrum. With a test sample, we optimize the HADRONNESS cuts (maximizing gamma/hadron separation while retaining at least 80% of gammas and sufficient OFF events for background estimation) for the different SIZE bins. The size of the remaining OFF data sample used for the analysis (1.4h) was of about 25% compared to the ON sample. Therefore, in order not to be dominated by the OFF fluctuations we have adopted models to fit the background in the excess region [3].

We approximated the distribution of the reconstructed arrival spots by a 2-dimensional bell-shaped Gaussian function leaving the sigma as a free parameter. The values of $\sigma$, obtained from fits to the MC gamma data and to the Crab Nebula data are shown in Figure 1. The global $\sigma$ for SIZE>180 photo-electrons ($\sim$140 GeV) is $0.102° \pm 0.008°$. The results show a significant improvement in the angular resolution of the MAGIC telescope when compared to the results of Lessard et al [5].

In order to compare our bidimensional analysis to the standard ALPHA-based analysis we have computed the number of excess events and significance for different SIZE obtained using both analysis. To make them comparable, we use for the $\alpha$-analysis those images that points to the center (based on the ASYMMETRY parameter). This adds an additional cut, like the one introduced in the Disp-analysis with the 'head-tail' discrimination, which reduces the background in the excess region by 50% as well as 20% of the excess events. The results are shown in table 1. The $\alpha$-plot and $\theta^2$-plot above 180 phe are shown in figure 2. The used bidimensional analysis gives a better sensitivity compared to the standard $\alpha$-analysis.

**Table 1.** Results for the Disp analysis to Crab Nebula compared to $\alpha$-analysis (numbers in brackets)

| Size Bin [phe] | Excess Counts | Background Counts | Significance | 2-d $\sigma$ [deg] | $\theta^2(\alpha)$ [deg$^2$]([deg]) |
|---|---|---|---|---|---|
| 180 < SIZE < 320 | 316 ± 73 (248 ± 78) | 4158 ± 29 (3688 ± 47) | 4.33$\sigma$ (3.16$\sigma$) | .113 ± .030 | <.115 (<17.5) |
| 320 < SIZE < 570 | 738 ± 48 (794 ± 71) | 1219 ± 19 (2062 ± 46) | 15.31$\sigma$ (11.17$\sigma$) | .086 ± .007 | <.070 (<20.0) |
| 570 < SIZE < 1010 | 801 ± 42 (676 ± 45) | 737 ± 17 (861 ± 22) | 18.72$\sigma$ (14.94$\sigma$) | .101 ± .006 | <.095 (<15.0) |
| 1010 < SIZE < 1800 | 511 ± 27 (432 ± 29) | 198 ± 7 (331 ± 10) | 18.41$\sigma$ (14.62$\sigma$) | .069 ± .005 | <.045 (<10.0) |
| 1800 < SIZE < 3200 | 312 ± 23 (275 ± 23) | 205 ± 6 (218 ± 8) | 13.14$\sigma$ (11.60$\sigma$) | .082 ± .006 | <.065 (<10.0) |
| 3200 < SIZE < 5690 | 72 ± 10 (128 ± 14) | 30 ± 3 (67 ± 4) | 6.79$\sigma$ (8.81$\sigma$) | .035 ± .017 | <.015 (<7.5) |

## 4. Conclusions

The Disp method for the reconstruction of the $\gamma$-ray arrival directions has been successfully used to analyze the MAGIC Telescope data. For energies above 140 GeV both MC and real data measurements yield angular resolutions better than 0.1°. The studies show that the performance does not dramatically degrade for lower energies, but the lack of statistics excluded a possible MC/data comparison. The application of the method to Crab Nebula on-axis data shows that this bidimensional analysis is competitive, respectively slightly superior compared to the standard ALPHA-based analysis for point-like on-axis sources.

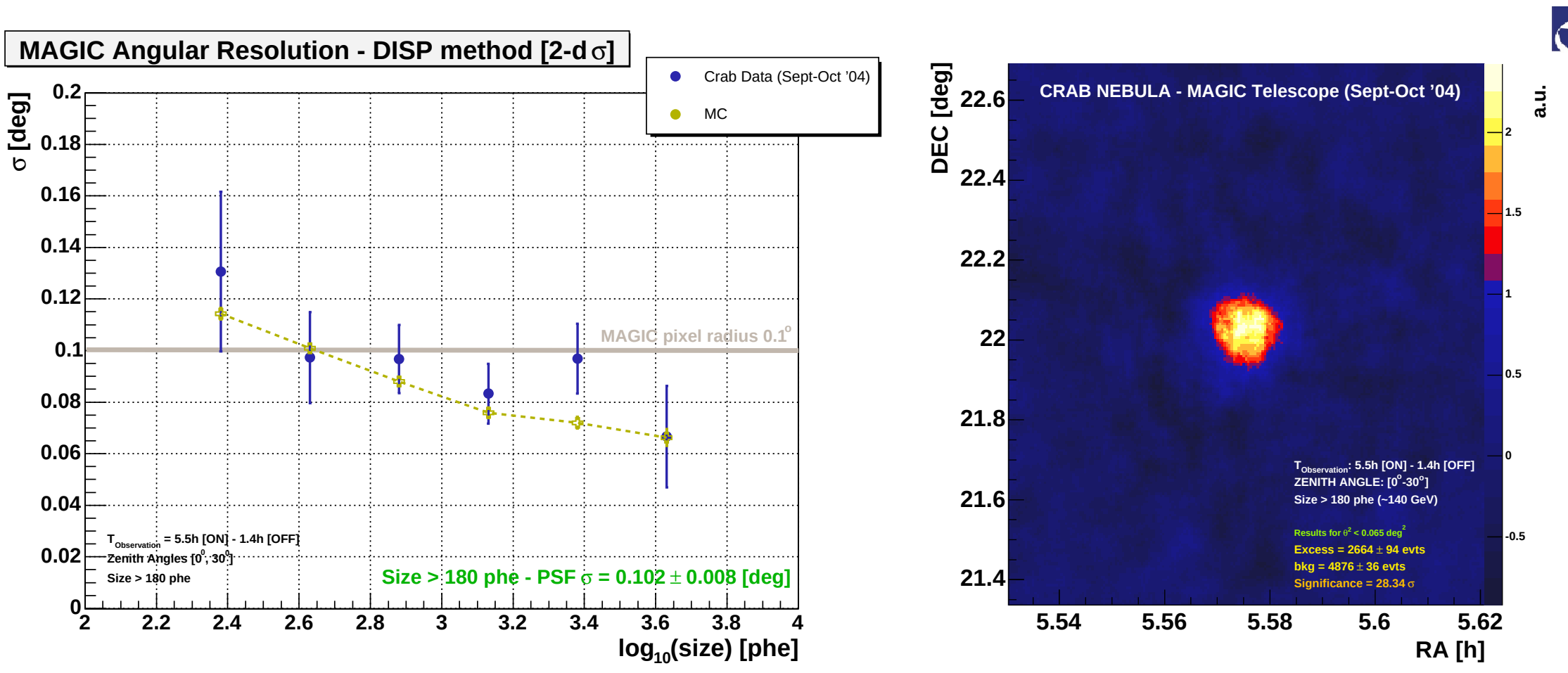

**Figure 1.** (*left*) Results of 2-dimensional Gaussian fi ts to the distribution of reconstructed arrival directions, both for MC and Crab Nebula, for the different SIZE bins considered. The PSF obtained for SIZE $>$ 180 phe is displayed in lower-right text. (*right*) Smoothed skymap for Crab Observations using the DISP analysis method.

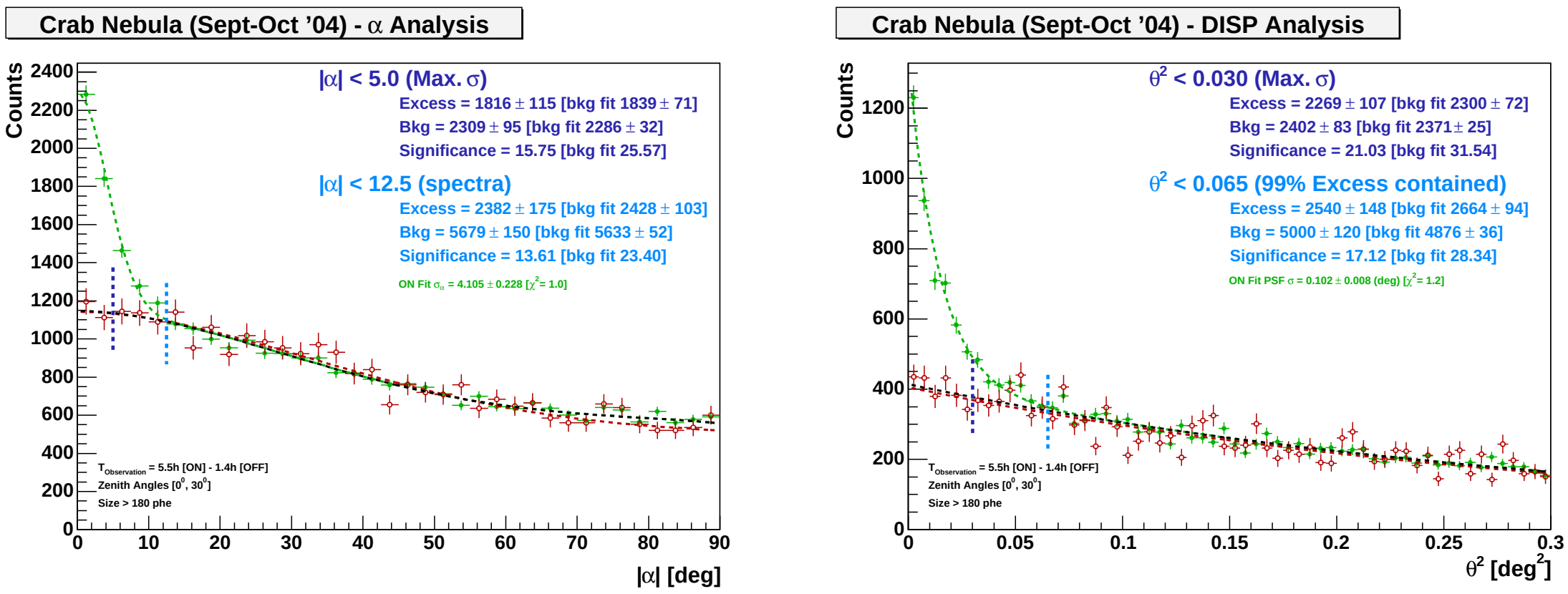

**Figure 2.** Crab Nebula $\alpha$-plot and $\theta^2$-plot for SIZE$>$180 phe. Two cuts applied: the one which maximize signifi cance and the one which retain 99$\%$ of the excess signal. Results for ON-OFF$_{fit}$ are displayed in brackets.

## 5. Acknowledgements

We would like to thank the Instituto de Astrofísica de Canarias (IAC) for excellent working conditions. The support of the German Bundesministerium für Bildung und Forschung (BMBF) and Max-Planck-Gesellschaft (MPG), the Italian Istituto Nazionale di Fisica Nucleare (INFN) and the Spanish Comisión Interministerial de Ciencia y Tecnología (CICYT) is gratefully acknowledged.

# Data Management and Processing for the MAGIC Telescope

D. Dorner[a], K. Berger[a], T. Bretz[a], M. Gaug[b] on behalf of the MAGIC Collaboration[c]

*(a) Institut für Theoretische Physik und Astrophysik, Universität Würzburg, Am Hubland, 97074 Würzburg, Germany*
*(b) Institut de Física d'Altes Engergies (IFAE), 08193 Bellaterra, Barcelona, Spain*
*(c) Updated collaborators list at: http://magic.mppmu.mpg.de/collaboration/members/index.html*

Presenter: D. Dorner (dorner@astro.uni-wuerzburg.de), ger-dorner-D-abs1-og27-poster

Every observing night the MAGIC Telescope is producing up to 170 GB of scientific data and this number will increase with the installation of 2 GHz FADCs. To provide a stable analysis of all the data an automatic data processing is essential.
For MAGIC a concept has been realised, in which an automatic, failsafe preprocessing of all data is ensured. The status of the analysis can be queried at any time from a database. The flexibility of the concept makes it easy to add new steps at any point of the analysis chain.

## 1. Introduction

With its 577 pixel camera and its 30 FADC slices per pixel the MAGIC Telescope produces up to 170 GB of data each night stored in about 500 files. To transfer this large amount of data from the Canary island La Palma, where MAGIC is situated at a height of 2200 m a.s.l., there are two procedures: Small amounts of data like the run information and the output of the online analysis are transferred daily via internet. The large amounts of raw data are then transferred on tape via mail every one or two weeks. Future plans to exchange the current FADCs by 2 GHz FADCs and to store 80 FADC slices promise even larger amounts of data. In addition the data volume will grow with the beginning of operation of the second telescope [1].
To handle such large amounts of data an automatic procedure is mandatory. A concept and its realisation for the MAGIC project will be presented in this paper.

## 2. A Concept of Data Management and Processing

To deal with more than 500 files and more than 150 GB per night the data processing procedure has to fulfil two important needs: It has to be failsafe to guarantee that all data is processed and automatic to make sure that all data is treated the same way and the analysis results are consistent. For MAGIC an automation concept has been developed, which meets not only these requirements but offers also many other usefull features. A systematic scheme of the dataflow and the key elements of the concept are shown in Fig. 1. In the following the concept, which is based on Shell, MySql and C++, will be explained in detail.
One of the core elements of the structure is a MySql database, which consists of different kinds of tables. On the one hand there are tables which contain any kind of information: run information (i.e. information about each run like for example source name, startime, etc.), information from the runbook (i.e. comments from the shiftcrew in the electronic runbook), calibration results (i.e. conversion factors [phe/FADC counts], bad pixels, etc.) and more. On the other hand there are the status tables: for each step in the dataflow and the analysis there is a field in one of the status tables. To meet the given conditions there are several status tables: some steps have to be done per run, therefore a table on run base is needed. Other steps are done per night, others per dataset (to analyse more data of one source together runs are grouped to datasets, as shown in Fig. 3). Adapted to the needs of the experiment there can be several status tables. If a certain step has been executed successfully, the time is entered in the corresponding field. Therefore it is not only known, whether the step has already been done, but also when it has been done. The users thus have the information with which software version the data has been analysed. If a step has to be repeated, for example, when there is an updated software version, this can easily be done, by removing the time from the corresponding field. As the time of execution is known, it is also possible to rerun a step only, if the data was analysed with a certain software version, by sending an appropriate query to the sql server.

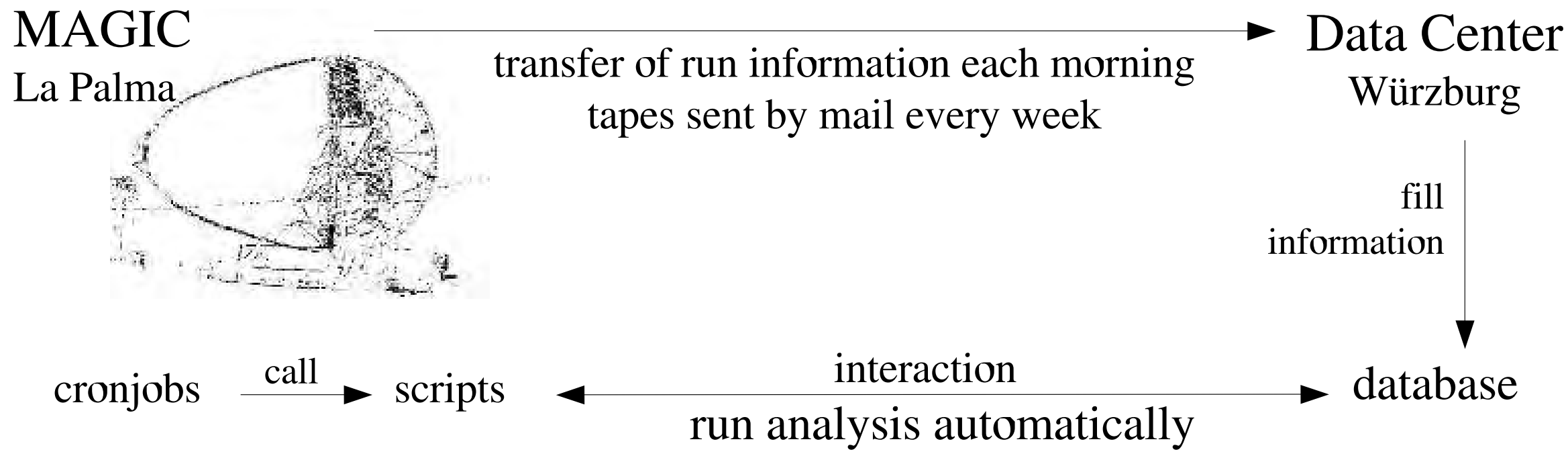

**Figure 1.** Schematic diagram of the automation concept for the MAGIC Telescope. Small files like run information are transferred every day via internet from La Palma to the datacenter. The big amounts of raw data are sent on tape by mail. From the arriving files all useful information is extracted and filled into the database. The automatic analysis is done with scripts that are called by cronjobs. Details on the interaction between the scripts and the database can be seen in Fig. 2.

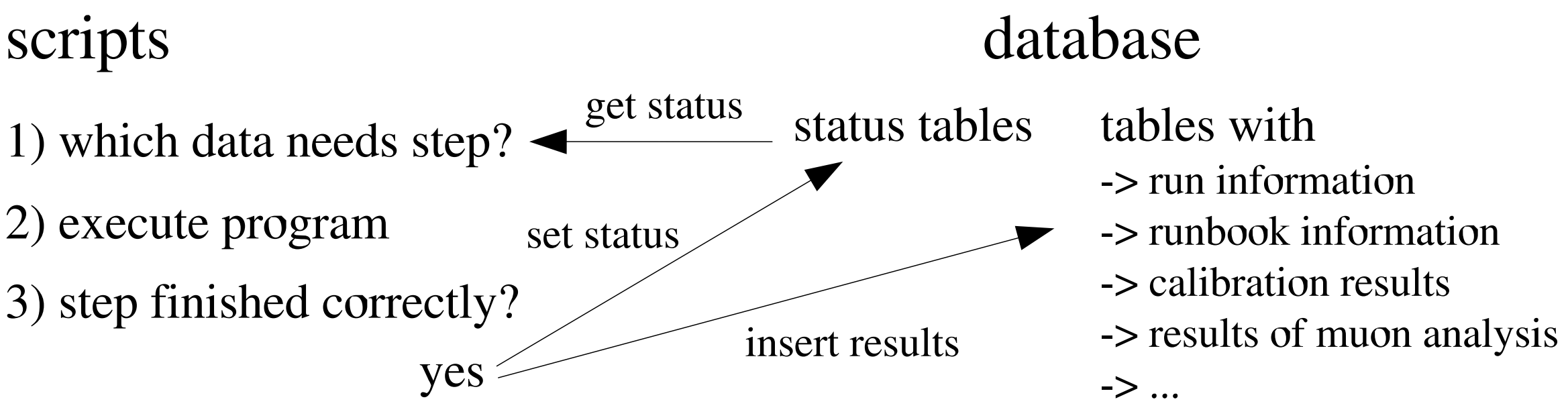

**Figure 2.** Scheme of the interaction between the scripts and the database. For a step, that has to be executed, the script performs the following steps: 1) query from the database, for which data the step is needed, 2) execute the program for this data and 3) if the step has finished without error, insert the new status into the database and insert the results of this analysis step into the appropriate table.

For MAGIC the following implementation has been made (Figs. 1, 3): In a daily transfer run information is copied from the telescope site in La Palma to the datacenter in Würzburg. All usefull information is immediately filled into the database. These steps are done on a nightly base, whereas for example the datacheck is done run wise. The runs are grouped to sequences before calibration [7] and the calculation of image parameters [2] is done. Thus the sequence status table contains for example fields for calibration and image parameter calculation. How the automation works in this case is shown in Fig. 3. To analyse big amounts of data the sequences are grouped to datasets and the remaining steps are done on dataset base. In a structure as explained above the status of each run, sequence or dataset is stored in the database and can be queried either from an automatic script or from a user via web interface so that the status of the analysis can be followed easily.
Based on this database structure the automation can be realised with simple shell scripts, one for each step in the analysis chain. Like shown in Fig. 2 the script queries from the database for which data a certain step has to be done and executes the program for this data. After the step has been completed without returning any error, the new status for the data is inserted into the database by inserting the current time into the corresponding field . Thus it is guaranteed that the data processing is fail-safe even in cases of computer problems, power failure etc. If a step could not be executed completely, it is retried the next time, when the script is started automatically. Only if data cannot be treated with the standard software, manual interaction is needed.
It is important, that all steps are performed in the correct order. For example the calibration can be done only, if all files are available. Therefore the check for disposability of all needed files has to be performed first. But the calibration also influences the next steps: the calculation of the image parameters and the filling of the calibration results into the database need the successful completion of the previous step, the calibration (see Fig. 4). These interdependencies of the steps are stored in a setup file, from where the scripts retrieve them.
By calling the scripts with cronjobs (see Fig. 1) no manual interaction is needed. As soon as the data is on disk, it is processed automatically.
By calling the scripts via cronjob it is very easy to run all analysis programs, that do not need any manual interaction, automatically. Accordingly all data is processed automatically, as soon as the files have been copied from tape. The concept also suits well for the usage with a queuing system like for example condor.
In the implementation for MAGIC the scripts work in interaction with the **M**agic **A**nalysis and **R**econstruction **S**oftware (MARS, [3, 4, 5]). Like this it is easy to implement the steps of the standard analysis chain [2] in the automation concept.
Due to the automation it is possible to obtain first results [2] quickely for all data. With the various tools of MARS the database also gives the possibility to do datachecks and long-term studies of quality parameters. For example the results of the muon analysis are stored in the database and thus the evolution of parameters like the point spread function could be examined [8].

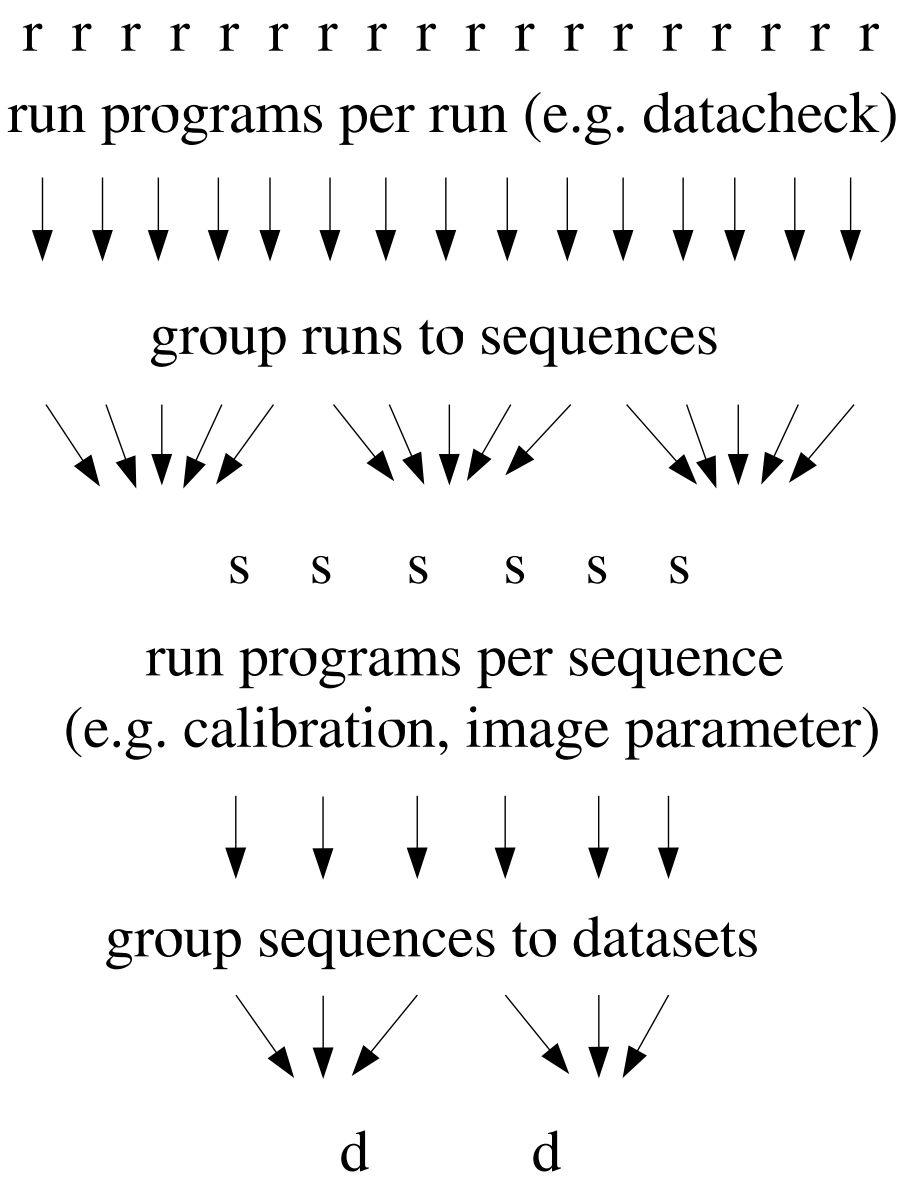

**Figure 3.** Schematic overview of the data structure in MAGIC: After some programs (e.g. datacheck) have been executed run wise, the runs are grouped to sequences. Some steps of the analysis are done sequence wise, like for example the calibration [7] and the calculation of the image parameters [2]. To analyse more data of one source together the sequences are grouped to datasets on which the rest of the analysis is performed.

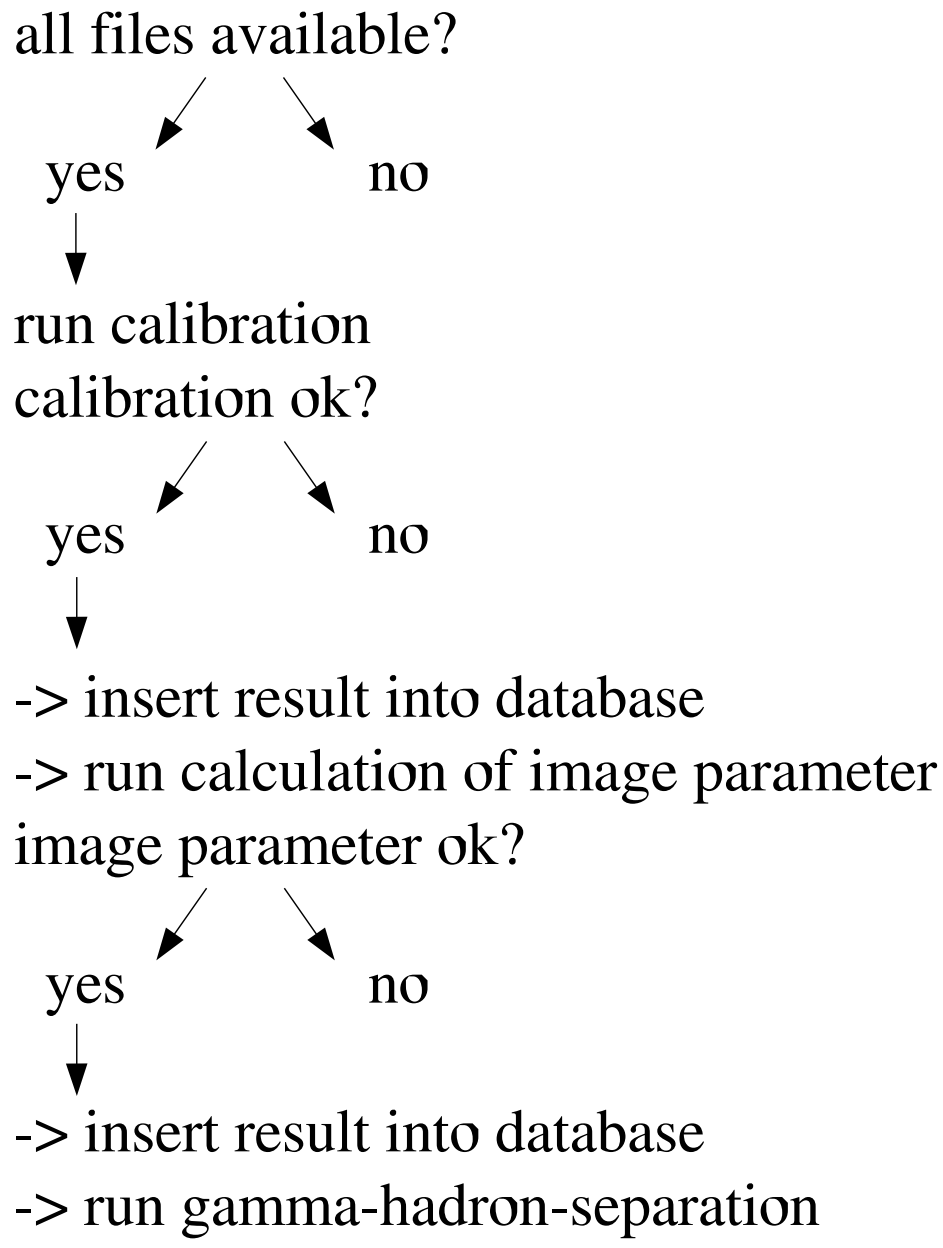

**Figure 4.** Example for the automatic procedure for the analysis of a Cherenkov telescope: If all files of a sequence of runs are available in the datacenter, the calibration is done for this data. If the calibration does not return any error, the result is filled into the database and the calculation of the image parameters is carried out. If this was successful the results are inserted into the appropriate tables in the database and the gamma-hadron separation [6] is performed. How the interaction between the scripts and the database looks like for the single steps, is shown in Fig. 2.

## 3. Flexibility

A very important point is the flexibility of this automation concept. In an evolving and expanding experiment like MAGIC it is important that the concept is adaptable to new software.
The automation for MAGIC has been designed such, that it is possible to add new steps in the analysis to the whole structure easily. Only three small actions are neccessary to implement a new step: 1) add a new field in the corresponding status table in the database, 2) adapt the setupfile with the dependencies of the step, and 3) write new script, that calls the program, and add it in the cronjob.

## 4. Conclusions

Based on the data stream produced by the MAGIC Telescope, we have shown, that it is possible to handle large amounts of data and files automatically with the presented concept. The interaction with a database provides the possibility to check the quality and status of the data at any time. By executing the programs of MARS the implemented structure delivers a quick look into the data with a robust analysis and first results, as soon as the data arrives in the datacenter. Due to its flexibility the concept can be easily expanded and adapted for the next development stage of the MAGIC experiment, but also for other future experiments.

## 5. Acknowledgements

We acknowledge support by the german Federal Ministry of Education and Research (BMBF, 05 CMOMG1/3) and the Astrophysics Institute of the Canary Islands (IAC).

# Calibration of the MAGIC Telescope

M. Gaug[a], H. Bartko[b], J. Cortina[a], J. Rico[a]
*(a) Institut de Física d'Altes Energies (IFAE), 08193 Bellaterra, Barcelona, Spain*
*(b) Max-Planck Institute für Physics, 80805 Munich, Germany*
Presenter: J. Rico (jrico@ifae.es), spa-jrico-J-abs2-og27-poster

The MAGIC Telescope has a 577 pixel photo-multiplier tube (PMT) camera which requires precise and regular calibration over a large dynamic range. A system for the optical calibration consisting of a number of ultra-fast and powerful LED pulsers is used. We calibrate each pixel using the F-Factor Method with signals in three different wavelengths. The light intensity is variable in the range of 4 to 700 photo-electrons per PMT. We achieve an absolute calibration by comparing the signal of the pixels with the one obtained from a 1 cm$^2$ PIN diode. This device is calibrated with the emission lines of two different gamma-emitters ($^{241}$Am and $^{133}$Ba) which produce a precise reference signal in the active region of the PIN diode. The time resolution of the entire MAGIC read-out system has been measured to about 700 ps at intensities of 10 photo-electrons reaching 200 ps at 100 photo-electrons. With an external calibration trigger, it is possible to take calibration events interlaced with normal data at a rate of 50 Hz. The entire system has been used on-site for one year.

## 1. Introduction

The MAGIC Telescope [1] houses a camera of 397 inner pixels (0.1$^\circ\phi$) and 180 outer pixels (0.2$^\circ\phi$), each read out with 300 MSample/s flash-ADCs [2] and an optical link of 260 MHz bandwidth to transfer the electronic signal over 160 m to the counting house. The quantum efficiency (QE) of the MAGIC PMTs strongly depend on the incident wavelength. Moreover, differences in the exact shape of QE($\lambda$) between PMTs had been observed. It is therefore desirable to calibrate the PMT response at different wavelengths.

We use a system of very fast (3–4 ns FWHM) and powerful ($10^8$–$10^{10}$ photons/sr) light emitting diodes [3] (NISHIA, single quantum well) in three different wavelengths (370 nm, 460 nm and 520 nm) and different intensities (up to 700 photo-electrons per inner pixel and pulse) so that we are able to check the linearity of the whole readout chain.

## 2. Excess noise factor method

In the last year, the camera was calibrated using the F-Factor. Assuming that the number of photons impinging on the photo-cathode has a Poisson variance, that the photon detection efficieny is independent from the place where and under which angle the photo-electron is released and that the excess noise introduced by the readout chain does not depend on the signal amplitude, one can derive [4]: $N_{\rm phe} = F^2 \cdot \mu^2/(\sigma_1^2 - \sigma_0^2)$

Here, $\sigma_0$ describes the signal extractor resolution (mostly due to noise of the night sky background), $\sigma_1$ the measured standard deviation of the signal peak and $\mu$ is mean reconstructed (and pedestal-subtracted) signal. $F$ stands for the excess noise factor, previously measured in the lab.

The method yields one value of $N_{\rm phe}$ per pixel the average of which is used to extract an averaged photo-electron fluence per light pulse and inner pixel $<N_{\rm phe}>$. All reconstructed signals from data are then multiplied with a conversion factor $c^i_{\rm phe} = <N_{\rm phe}> * R^i_{\rm area}/\mu^i$ where $i$ stands for the pixel index and $R^i_{\rm area}$ for the ratio of covered areas. $R^i_{\rm area}$ is 1 for all inner pixels and 4 for all outer pixels.

Figure 1 left shows the evolution of the mean number of photo-electrons over one night. One can see that,

despite some long-term dependency, $< N_{\rm phe} >$ remains stable to about 1% on minute scales. More precise investigations revealed a dependency of the light output with ambient temperature of about $2\,\%/K$.

Figure 1 right shows $< N_{\rm phe} >$, calculated at different intensities and with different light colours.

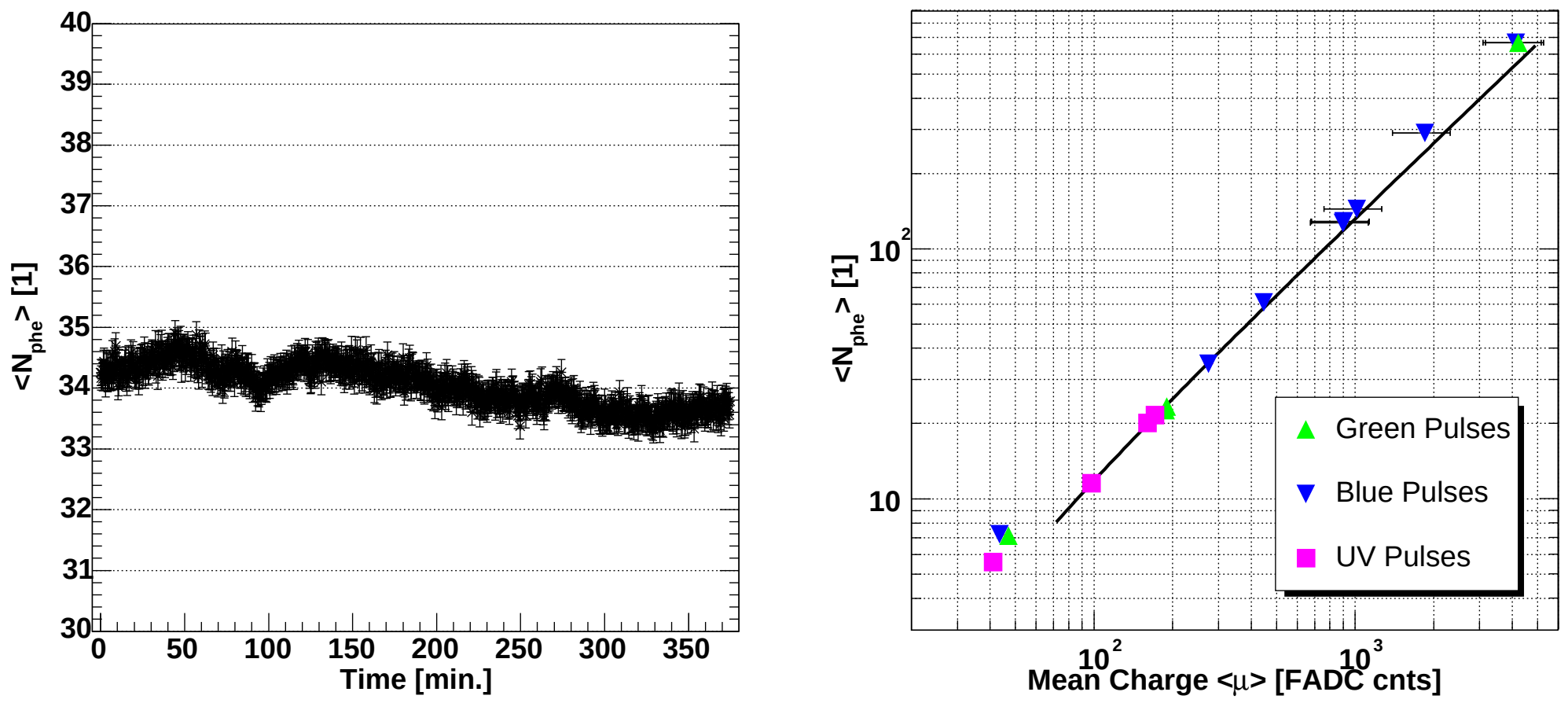

**Figure 1.** Left: Evolution of the mean number of photo-electrons over one night. Right: Mean reconstructed number of photo-electrons vs. mean reconstructed charge at various intensities.

## 3. PIN-Diode method

We measured the absolute light flux with a PIN diode, situated at 105 cm distance from the calibration light pulser and read out with a charge sensitive pre-amplifier (shaping time 1 $\mu$s) [3]. The PIN diode was calibrated with an $^{241}$Am source emitting 59.95 keV photons and a $^{133}$Ba source having an emission line at 81.0 keV [5]. The assumed average energy to create an electron-hole pair by ionization is 3.62 eV at 20°C producing a peak in the recorded PIN Diode spectrum due to photo-absorption. Figure 2 shows the obtained spectra and the signals from two typical calibration light pulses. The quantum efficiency of the diode was obtained by comparison with a calibrated PIN diode. An average QE is obtained by folding the LED spectrum with the QE for each wavelength.

Table 1 shows the result of two dedicated calibration runs where both calibration methods (F-Factor and PIN Diode) have been applied to the same data. One can see a good agreement between the two results, although there is still need to reduce the systematic uncertainties (see caption).

## 4. Time Calibration

The photo-multipliers introduce a time delay, the “transit time (TT)”, in the amplified photo-electrons signal, depending on the applied high-voltage (HV). Together with smaller relative delays due to different lengths of the optical fibers, these delays have to be calibrated relatively to each other in order to obtain a correct timing information for the analysis.

| Number of LEDs fired | $< N_{\rm phe} >$ F-Factor Method | $< N_{\rm phe} >$ PIN Diode Methode |
|---|---|---|
| 10 Leds Blue | 291±3 | 294±3 |
| 20 Leds Blue | 605±6 | 613±6 |

**Table 1.** Average number of photo-electrons for one pixel, calculated with both calibration methods for two different light intensities. For the PIN Diode Method, an additional systematic error of $+8-6\%$ and for the F-Factor Method, ±5% plus possible degradations of the overall PMT quantum effi ciency and transmission coeffi cients have to be added.

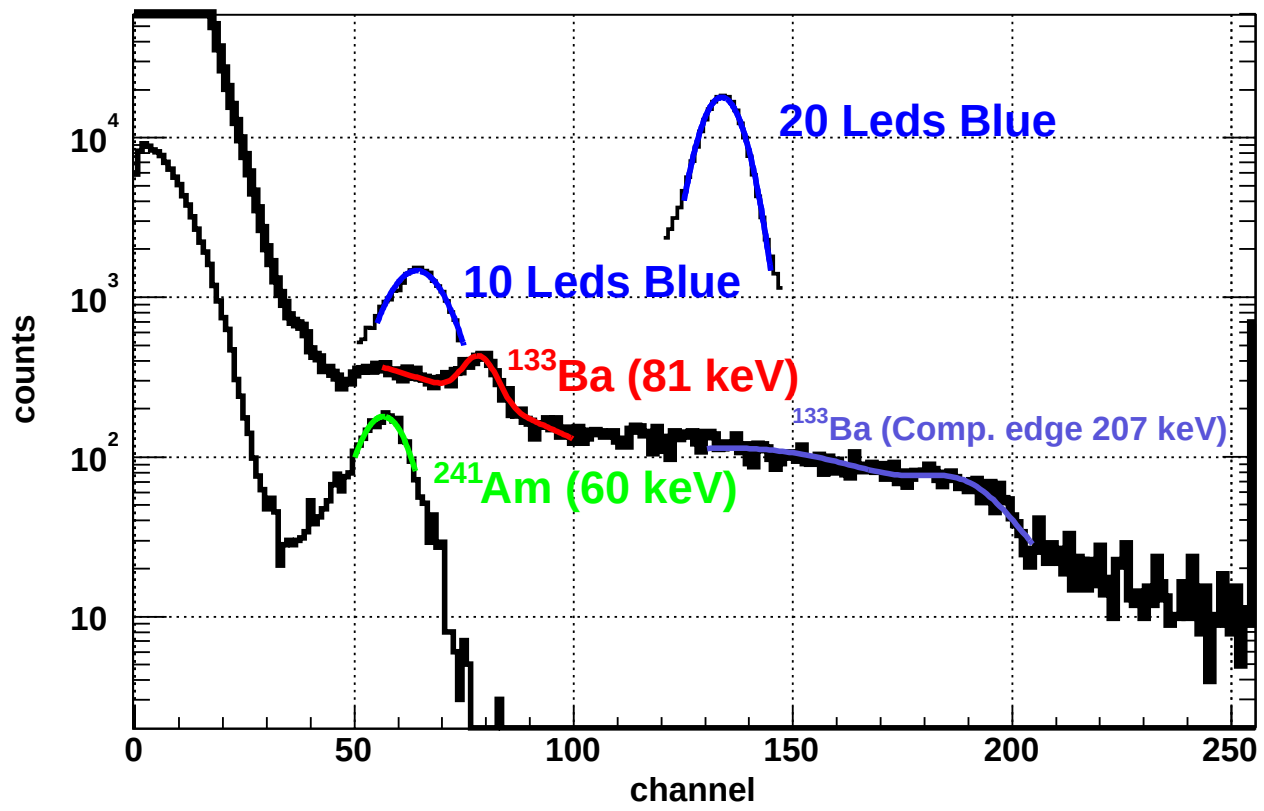

**Figure 2.** Spectra taken with the PIN Diode and various sources: bottom: $^{241}$Am with a 60 keV photon emission line, center: $^{133}$Ba with a photon emission line at 81 keV and a Compton egde at 207 keV, top: the spectrum obtained from light pulses with two combinations of LEDs.

Using the light pulser at different intensities, we measured the time offsets and time spreads of the readout and detection chain. Event by event, the reconstructed arrival time difference of every channel with respect to a reference channel was measured and its mean and RMS calculated. The former yields the measured relative time offset while the latter is the convolution of the arrival time resolution of the measured and the reference channel.

Figure 3 (left) shows the time offset versus the applied HV for each PMT. Like expected, one can see a clear anti-correlation. The smaller the applied HV, the longer the signal takes to travel from the first dynode to the anode.

Figure 3 (right) shows the time resolution (RMS of the arrival time differences histogram, divided by the square root of 2), measured at different intensities. The measurements have been fitted by the following function:

$$\Delta T = \sqrt{\frac{T_1^2}{< Q > /{\rm phe}} + \frac{T_2^2}{< Q >^2 /{\rm phe}^2} + T_0^2}. \quad (1)$$

where $T_1$ parameterizes the combination of intrinsic arrival time spread of the photons from the light pulser and the transit time spread of the PMT, $T_2$ parameterizes the reconstruction error due to background noise and limited extractor resolution and $T_0$ is a constant offset which might be due to the residual FADC clock jitter.

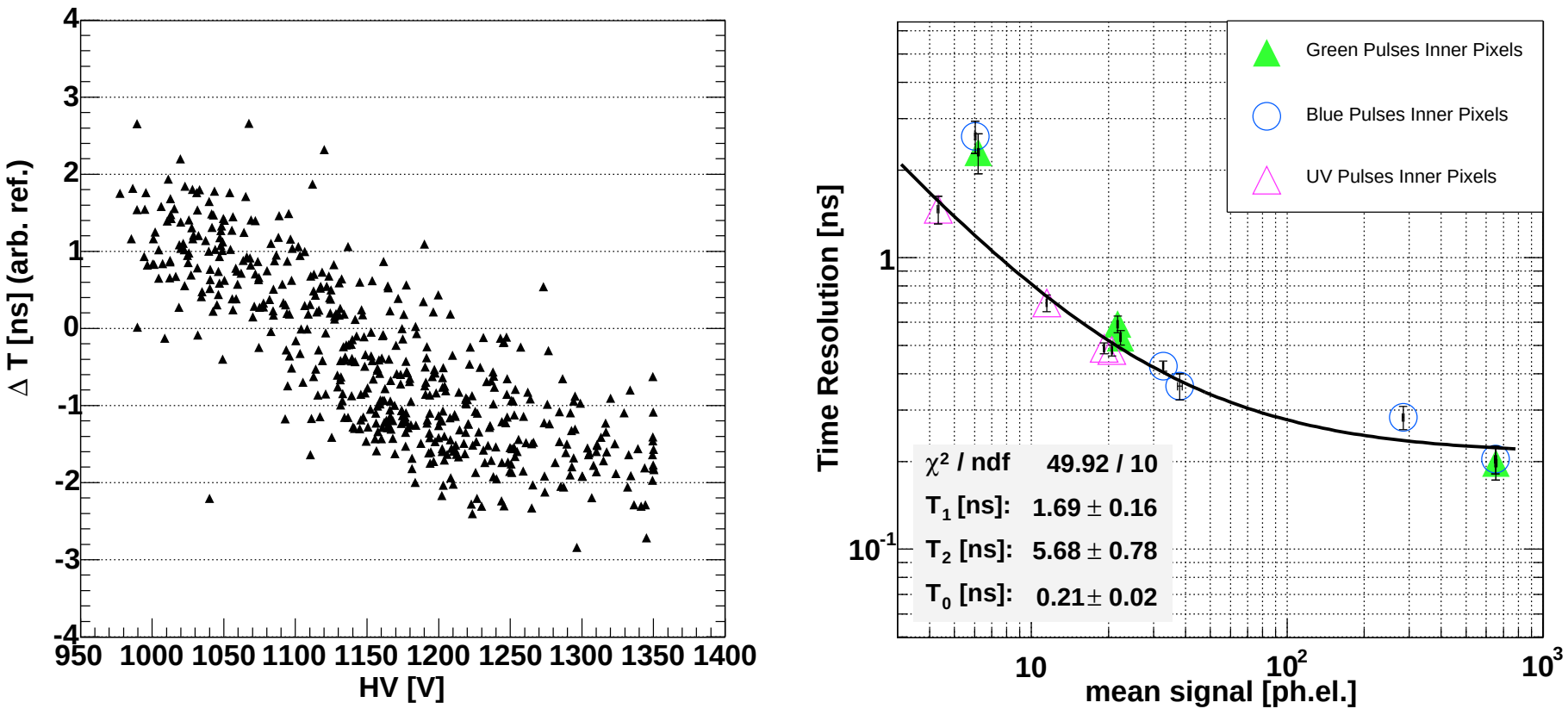

**Figure 3.** Left: Calibrated arrival time offsets $\Delta T$ vs. applied HV. Right: Calibrated time resolution for various intensities.

## 5. Conclusions

The LEDs pulser calibration system has been used to calibrate the MAGIC camera for about one year. If run in standard mode, it fires UV-pulses of 370 nm in dedicated calibration runs and additionally as interlaced calibration events with a frequency of 50 Hz. This frequency allows to accumulate enough statistics before reaching the typical time scales of residual short-term fluctuations of the optical transmission gains. The camera is such continuously monitored and re-calibrated [6] using the F-Factor method.

In dedicated calibration runs with different colours and intensities, the response of the system and signal extraction methods [7] have been tested.

Using a calibrated PIN Diode, the light flux of the pulser has been measured with an independent method yielding consistent results with the F-Factor method. The systematic error of both methods is still above 5% (8% for the PIN Diode method) and will be reduced in the future.

Using the light pulses to calibrate the arrival times extracted from each channel, we find an upper limit to the time resolution of the MAGIC telescope for cosmics pulses of:

$$\Delta T_{\mathrm{cosmics}} \approx \sqrt{\frac{4\,\mathrm{ns}^2}{<Q>/\mathrm{phe}} + \frac{20\,\mathrm{ns}^2}{<Q>^2/\mathrm{phe}^2} + 0.04\,\mathrm{ns}^2}. \qquad (2)$$

# Absolute energy scale calibration of the MAGIC telescope using muon images

F.Goebel[a], K. Mase[a,b], M. Meyer[c], R. Mirzoyan[a], M. Shayduk[d], and M. Teshima[a]
*on behalf of the MAGIC collaboration*
*(a) Max-Planck-Institut für Physik (Werner Heisenberg Institut), Föhringer Ring 6, 80805 Munich, Germany*
*(b) Chiba University, Yayoi-cho 1-33, Inage-ku, Chiba-shi 263-8522, Japan*
*(c) Institut für Theoretische Physik und Astrophysik, Universität Würzburg, Am Hubland, 97074 Würzburg, Germany*
*(d) Humboldt Universität zu Berlin, Institut für Physik, Newtonstraße 15, 12489 Berlin, Germany*
Presenter: F. Goebel (fgoebel@mppmu.mpg.de), ger-goebel-F-abs1-og27-poster

The absolute overall light collection efficiency of the MAGIC telescope can be calibrated using isolated muons hitting the reflector. The geometry and the energy of the muons are reconstructed from the measured ring images and compared with Monte Carlo predictions. The amount of Cherenkov light produced by muons can be modeled with small systematic uncertainties. Muon images are recorded during normal observation with a rate of about 2 Hz. A continuous calibration can therefore be performed with no need for dedicated calibration runs. In addition the width of the muon ring images can be used to monitor the spot size of the reflector during normal data taking.

## 1. Introduction

The MAGIC telescope [1] for $\gamma$-ray astronomy in the energy range between 30 GeV and several TeV is situated on the Roque de los Muchachos on the Canary Island La Palma (28.8$^o$ N, 17.8$^o$ W) at 2200 m altitude. The 17 m diameter parabolic tessellated mirror is mounted on a light weight carbon fiber structure. The 3.5$^o$ field of view (FOV) camera is equipped with 576 high quantum efficiency photo-multiplier tubes (PMTs). The inner area is composed of 396 PMTs of $0.10^o$ FOV surrounded by 180 $0.20^o$ FOV PMTs. The analog signals are transferred via optical fibers to the trigger electronics and the 300 MHz flash analog to digital converter (FADC). The telescope has been fully operational since August, 2004.

## 2. Calibration Principle

The standard calibration of the MAGIC telescope [2] uses a light pulse generator which illuminates the PMT camera uniformly. This procedure provides an absolute calibration of the camera and the signal processing chain of the MAGIC telescope. In order to calibrate the overall light collection of the whole telescope including e.g. the reflectivity of the mirror dish additional information is required. A useful tool for the overall absolute calibration is provided by ring images generated by muons hitting the reflector [3, 4].

Muons hitting the reflector produce ring images on the camera plane, fully or partially contained inside the camera depending on the incident angle $\xi$. For muons hitting the mirror with known energy and geometry the number of photo electrons N collected per azimuth angle $\Phi$ by a mirror dish of radius R (see Fig. 1) is given by [5]:

$$\frac{dN}{d\phi} = \frac{\alpha I}{2}\sin(2\theta_c)D(\phi) \qquad (1)$$

$$I \equiv \int_{\lambda 1}^{\lambda 2} \frac{\psi(\lambda)}{\lambda^2}\, d\lambda, \qquad D(\phi) = R\ \left[\sqrt{1-(\rho/R)^2\sin^2\phi} + (\rho/R)\cos\phi\right],$$

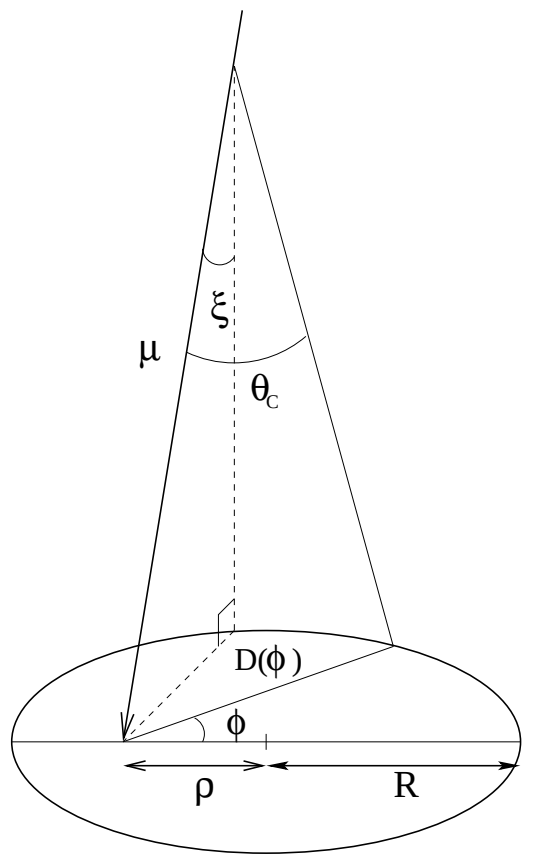

**Figure 1.** Geometry of Cherenkov light emitted by a muon hitting the reflector dish (R: mirror radius, $\rho$: impact parameter, $\xi$: incident angle, $\theta_C$: Cherenkov angle

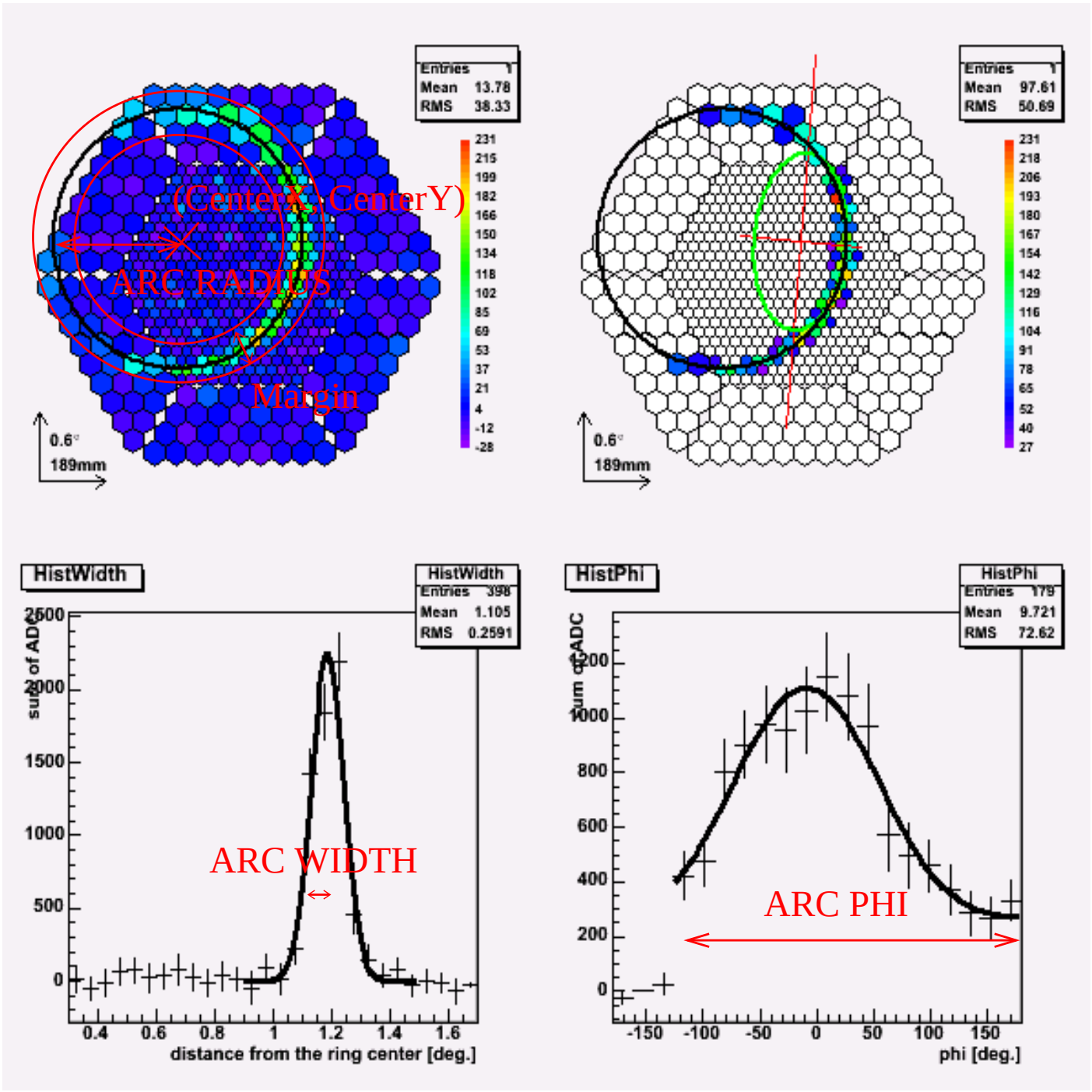

**Figure 2.** An example of a muon ring image in the MAGIC camera. The lower plots show the corresponding ARC WIDTH and ARC PHI distributions.

where $\alpha$ is the fine structure constant, $\psi(\lambda)$ is the overall photon to photo electron conversion efficiency (CE) and $\lambda$ is the wavelength of the Cherenkov light. The impact parameter $\rho$ and the energy (or $\theta_C(E)$) of the incident muon can be reconstructed from the observed muon ring radius and the $\Phi$-distribution ($dN/d\Phi$) of the light intensity along the ring. The calibration then consists in adjusting $\psi$ in order to match the observed and the predicted number of photo electrons.

## 3. Data Analysis

The reconstruction of the muon ring images starts with the standard signal reconstruction for each camera pixel. The resulting image is then fitted with a circle of radius $R_{arc}$. The integrated signal of all pixels inside a donut $\pm 0.2^o$ around the fitted circle ($SIZE_{muon}$) is computed. The width of the muon ring ($WIDTH_{arc}$) is determined as the $\sigma$ of a Gaussian fit to the signal distribution projected onto the radial distance from the center of the circle. The signal within the $\pm 0.2^o$ donut is plotted as a function of $\Phi$. $ARC_\Phi$ is the $\Phi$ range above a fixed threshold and the impact parameter can be estimated by fitting the $\Phi$-distribution with eq. 1.

A clean sample of muons can then be obtained using cuts on the quality of the muon ring fit, the parameters described above and the leakage parameter which is defined as the ratio of the signal in the outer pixels of the camera to the total signal. After these cuts a muon rate of 2.3 Hz is obtained. This is enough to calibrate the energy scale every 10 minutes with an accuracy of better than 3%.

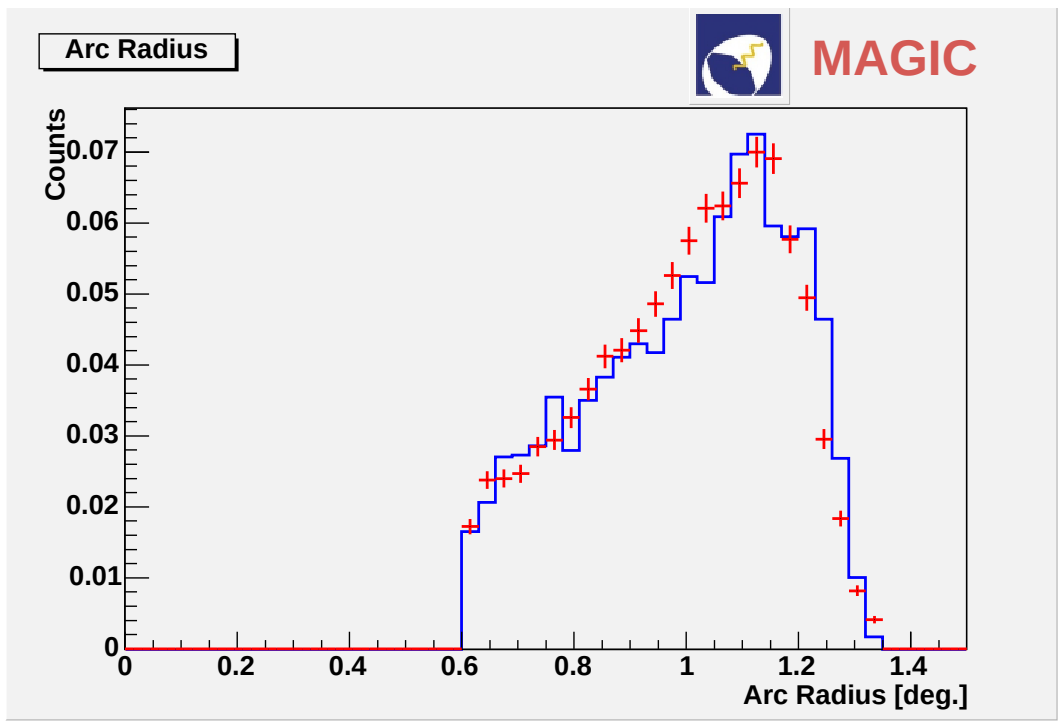

**Figure 3.** The distribution of the radius $R_{arc}$ of the fitted muon rings is shown for data (red dots) and MC (blue histogram)

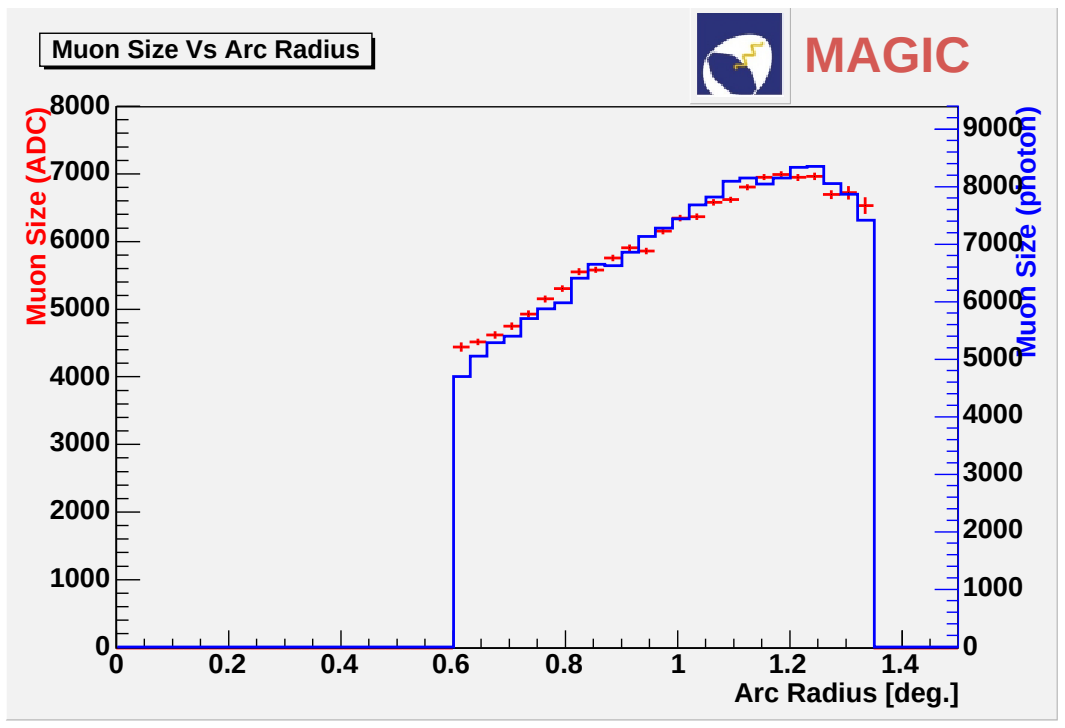

**Figure 4.** The dependence of $SIZE_{muon}$ on $R_{arc}$ is shown for data (red dots) and MC (blue histogram). After overall CE calibration good agreement is achieved.

## 4. MC simulation

The results of the telescope data analysis are compared with MC simulations. This takes into account effects like multiple scattering of muons, the refractive index of the air as well as detector and reconstruction inefficiencies. In the first step of the simulation muons with an energy spectrum according to [6] are generated with the CORSIKA air shower program [7]. A high energy (10-1000 GeV) and a low energy (5-10 GeV) sample have been generated with different differential flux indices (-2.69 and -1.39 respectively). An impact parameter range of up to 15 m, a starting altitude of 600 g/cm$^2$ (corresponds to 2 km above the MAGIC site) and an opening angle of up to 1.2$^o$ were simulated. In the second step the standard MAGIC detector simulation [8] is used.

## 5. Light collection efficiency

The $R_{arc}$ distribution shown in Fig. 3 shows good agreement between data and MC. Since $R_{arc}$ is related to the muon energy, this indicates that the energy distribution of the muons is sufficiently well simulated. Also the shapes of the $SIZE_{muon}$ distributions as a function of $R_{arc}$ agree well between data and MC. The conversion factor from ADC counts to photons hitting the reflector can thus be extracted by normalizing the distributions. A value of Conv$_{ADC\rightarrow photon}$ = $1.38 \pm 0.01(stat.)$ has been obtained. Using Conv$_{ADC\rightarrow phe}$ = $0.149 \pm 0.005(stat.) \pm 0.017(sys.)$ as obtained from the standard MAGIC light calibration [2], one obtains the photon to photo electron conversion efficiency: Conv$_{photons\rightarrow phe}$ = $0.108 \pm 0.003(stat.) \pm 0.012(sys.)$.

The Cherenkov light from the muons in this analysis is produced very near to the telescope. The light from muons is therefore less affected by Rayleigh scattering than the light from typical $\gamma$-ray showers. This leads to a small difference in the spectral distribution of the Cherenkov light. Since also the photon collection efficiency is wavelength dependent, a small correction of 3% has to be applied to the conversion factor for $\gamma$-ray showers. The resulting conversion factor is:

$$Conv_{photons\rightarrow phe} = 0.105 \pm 0.003(stat.) \pm 0.012(sys.)$$

The above results have been obtained using a 2 hours data sample of September '04. The analysis has been automatized and applied during the standard data reconstruction [9]. In Table 1 the the average CE per month is shown for 3 selected months. A small degradation of $\sim 2\%$ has been observed for January '05 which is com-patible with the number of mirror panels excluded from the AMC (see next section) due to faulty connectors during that time.

| Month | CE ratio [%] | PSF [deg] |
|---|---|---|
| Sept ’04 | 101.2±1.3 | 0.037±0.006 |
| Jan ’05 | 99.5±1.7 | 0.046±0.003 |
| May ’05 | 102.5±2.0 | 0.035±0.003 |

**Table 1.** The CE and the PSF for September ’04, January ’05 amd May ’05 are shown. The CE is given in % of the value implemented in the MC used for this analysis and the PSF is given in degree FOV.

## 6. Point Spread Function

The MAGIC telescope uses an active mirror control (AMC) which adjusts the orientation of each of the 247 mirror panels in order to optimize the mirror focusing for every position of the telescope. Muon rings provide a powerful tool to measure the point spread function (PSF) of the reflector continuously during data taking.

The PSF can be estimated from the $WIDTH_{arc}$ distribution of the muon rings. Several effects contribute to the broadening of the muon ring such as multiple scattering of the muons, mirror aberrations and the effect that the telescope focuses at 10 km for best imaging of $\gamma$-ray showers (the muon rings are smallest when focusing to infinity). The $WIDTH_{arc}$ distribution in data is thus compared to the corresponding distribution obtained from MC with different PSFs. For the September ’04 data set best agreement has been achieved for PSF = 0.037$^o$ ($\sigma$ of Gaussian fit). This is a satisfactory value when compared to the 0.1$^o$ inner pixel diameter. For January ’05 a slightly worse PSF has been found (see table 1) which is consistent with a known temporary degradation of the AMC accuracy during winter ’05.

## 7. Conclusions

Ring images of muons hitting the telescope reflector have been used to calibrate the overall light collection efficiency. A conversion factor of $Conv_{photons \rightarrow phe} = 0.105 \pm 0.003(stat.) \pm 0.012(sys.)$ has been obtained. The width of the muon rings is used to monitor the PSF of the mirror of the MAGIC telescope.

**Acknowledgments:** We would like to thank the IAC for excellent working conditions. The support of the Italian INFN, German BMBF and MPG, Spanish CICYT and Japanese JSPS is gratefully acknowledged.

# Monte Carlo simulation for the MAGIC telescope

P. Majumdar[a], A. Moralejo[b], C. Bigongiari[b], O. Blanch[c], D. Sobczynska[d]
*on behalf of the MAGIC collaboration*
*(a) Max Planck Institut für Physik, Munich (Germany)*
*(b) Department of Physics and INFN, University of Padua (Italy)*
*(c) Institut de Física d'altes energies, Barcelona (Spain)*
*(d) Division of Experimental Physics, University of Lodz (Poland)*
Presenter: P. Majumdar (pratik@mppmu.mpg.de), ger-majumdar-P-abs1-og27-oral

The operation of ground based Imaging Cherenkov telescopes requires a detailed Monte Carlo (MC) simulation of $\gamma-$ ray and hadron initiated air showers, as well as of the detector response to them. An overview of the MAGIC telescope MC simulation is presented, showing comparisons between the features of the simulated showers and those of real data taken during the first year of operation.

## 1. Introduction

The Monte Carlo simulation for the MAGIC telescope is divided into three stages. The development of $\gamma$ and hadron-initiated air showers is simulated with CORSIKA 6.019 [1], using VENUS for hadronic interactions and the US standard atmosphere. Cherenkov photons arriving around the telescope location are stored in binary files containing all the relevant parameters (including wavelength). The second stage of the simulation, the so called *Reflector* program, accounts for the Cherenkov light absorption and scattering in the atmosphere (using the US standard atmosphere to compute the Rayleigh scattering plus the Elterman model [2, 3] for the distribution of aerosols and ozone), and then performs the reflection of the surviving photons on the mirror dish (composed of 964 tiles) to obtain their location and arrival time on the camera plane. Finally, the *camera* program simulates the behaviour of the MAGIC photomultiplier camera, trigger system and data acquisition electronics. Realistic pulse shapes, noise levels and gain fluctuations obtained from the real MAGIC data have been implemented in the simulation. The overall light collection efficiency of the telescope has been tuned at the camera simulation level, using data from the comparison of the intensity of observed and simulated ring images from single muons at low impact parameters [4], resulting in an effective aperture of around 26 $m^2$.

For the present study a total of 2.6 $\times$ $10^7$ protons and 1.3 $\times$ $10^7$ Helium nuclei between 30 GeV and 30 TeV, have been produced, as well as 7.8 $\times$ $10^6$ gammas between 10 GeV and 30 TeV. The energy distribution of primary $\gamma$ rays is a pure power law with index -2.6, whereas charged primaries follow the known cosmic ray spectra [5, 6]. The telescope pointing directions range from 0 to $\simeq$ 30° in zenith angle $\theta$ (flatly distributed in $\cos\theta$), with the directions of protons and Helium nuclei scattered isotropically within a $5^\circ$ semiaperture cone around the telescope axis. Gammas were simulated as coming from a point source $0.4^\circ$ off the center of the camera, in order to compare them to a special sample of real data taken in similar conditions. Maximum impact parameters of 300 and 400 m have been simulated for gammas and nuclei respectively.

## 2. Comparison of Monte Carlo events and real MAGIC data

The calibration and image reconstruction of both the Monte Carlo events and data have been done following standard procedures of MARS (Magic Analysis and Reconstruction Software [7]). Signal intensity in each pixel is obtained by interpolating the pulse in the FADC with a cubic spline which is then integrated in a range of 3.3 ns around the peak. Tail cuts select pixels with a signal of at least 7 photoelectrons above the pedestal (*core pixels*), and those with 5 or more photoelectrons which are neighbours to any of the former (this applies to the fine pixels within $\simeq 1^\circ$ of the center of the camera; for the larger outer pixels, tail cuts were scaled up such that they correspond to the same light density as those for inner pixels). A classical (Hillas) image parametrization is then performed, including the moments up to third order of the photoelectron distribution on the camera.

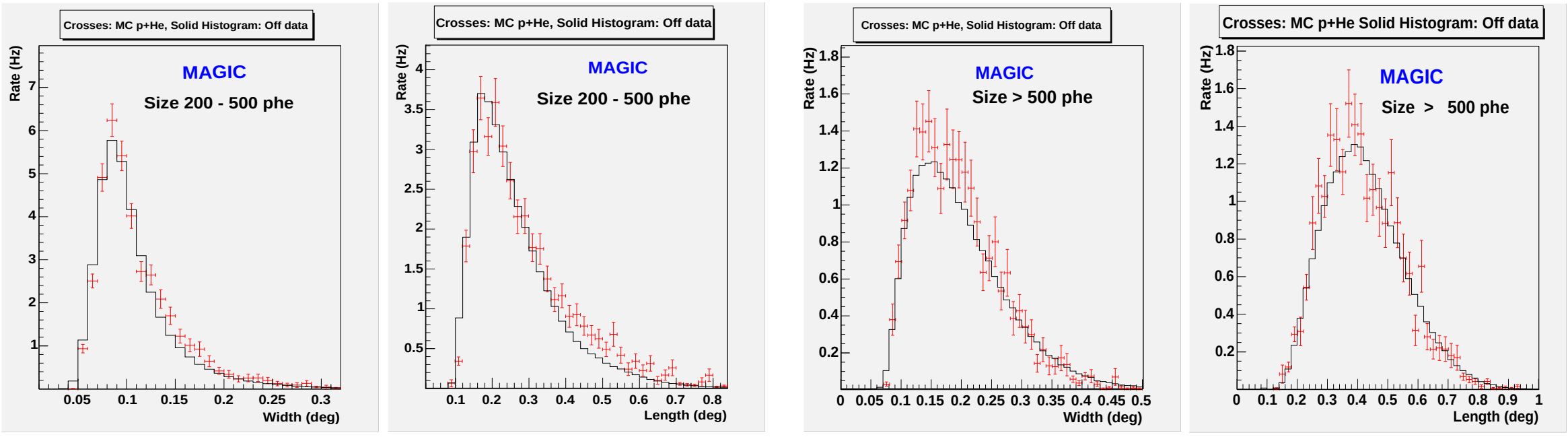

**Figure 1.** Distribution of Width and Length parameters for Monte Carlo hadrons (crosses) and real OFF data (solid line) for size bins $200-500$ photoelectrons and $>500$ photoelectrons respectively.

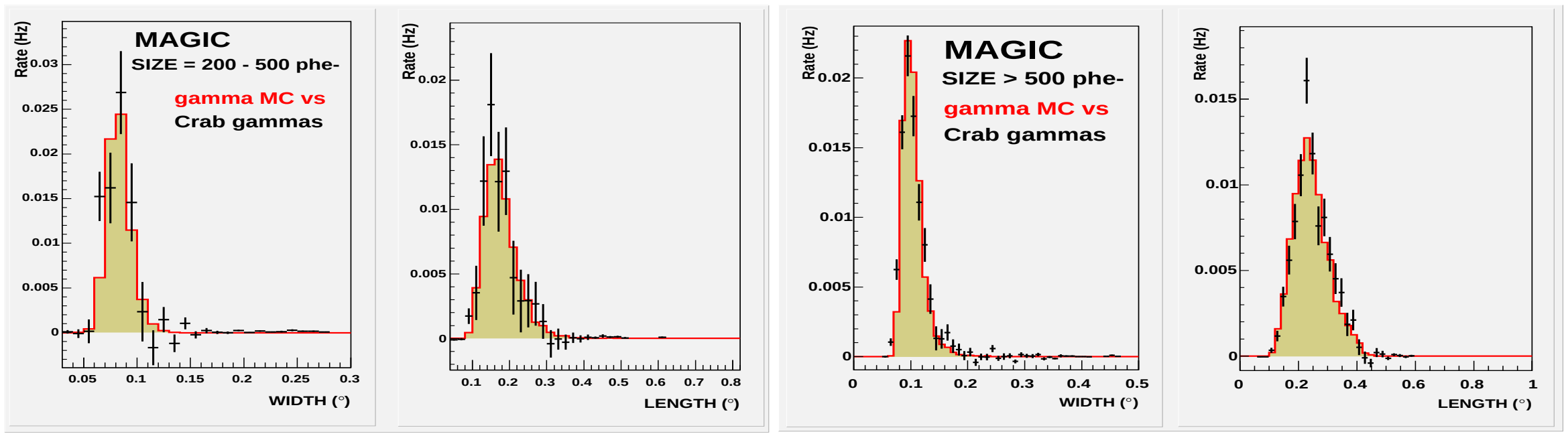

**Figure 2.** Distribution of the WIDTH and LENGTH parameters of shower images for Monte Carlo gammas (shaded histograms) and real Crab gammas (crosses). Left: SIZE between 200 and 500 phe$^-$ (corresponding to a gamma peak energy of 140 GeV); right: SIZE $> 500$ phe$^-$ (peak energy 270 GeV).

The real data used for comparing with MC samples were taken in January 2005. They consist of 156 minutes of observations of an *OFF* sky region (containing no known $\gamma$ ray source), and 309 minutes of observations of the Crab Nebula in good weather conditions. The Crab data were taken off-axis, with the center of the Nebula located $0.4^\circ$ off the camera center, by tracking two different sky locations alternatively (wobble mode). Tight quality cuts were applied to the events surviving the image cleaning, namely: $>5$ core pixels, size $>200$ photoelectrons and less than 10% of the image light contained in the outermost ring of pixels. As much as $74\%$ of the events are rejected by these cuts, which raise the typical energy of the remaining events well above the trigger threshold of the telescope (which we estimate to be currently between 50 and 60 GeV). The remaining sample of MC gammas has an energy peak of 140 GeV (assuming a -2.6 differential energy spectrum). The tight cuts were dictated by the need of having a good angular resolution (see below) and a significant excess of gamma events from Crab Nebula even before applying strong hadron discrimination cuts. The total background trigger rates after imposing these cuts for MC and data are about 65 Hz and 60 Hz respectively. Figure 1 shows the comparison of the WIDTH and LENGTH distributions for Monte Carlo and real hadrons (from the *OFF* runs) in two different size bins: $200-500$ photelectrons and $>500$ photoelectrons. The SIZE spectrum of hadrons is displayed on the right pad of Fig 3. The observed distributions agree well with expectations (it must be noted that the applied cuts and the limited trigger region of the MAGIC camera result in a negligible contribution from hadrons beyond the maximum simulated impact parameter of 400 m).

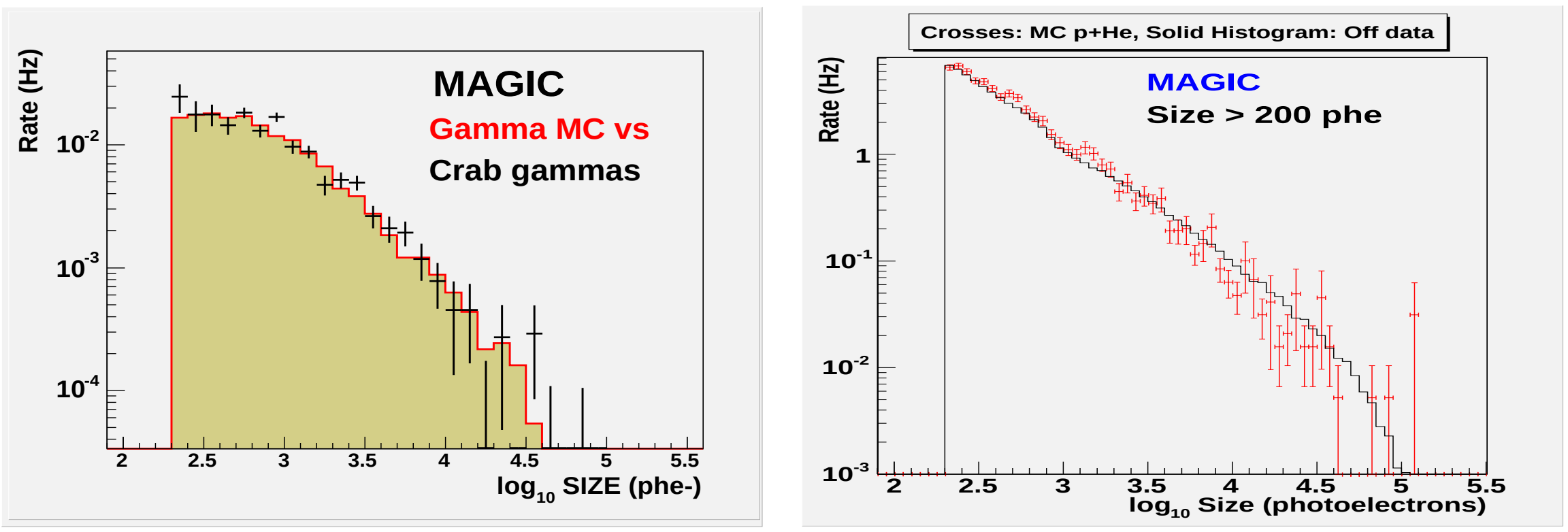

**Figure 3.** Left: distribution of the SIZE parameter for Monte Carlo gamma (shaded histogram) and real excess events from the Crab Nebula. Right: SIZE for Monte Carlo hadrons (crosses) vs. real OFF data.

For the comparison of simulated and real gamma-initiated showers, only the Crab wobble runs were used (with no additional OFF runs needed). With a simple DISP method [9] the incident direction of each event is estimated using the ratio of WIDTH and LENGTH as a measure of the distance between the center of gravity of the phe$^-$ distribution and the gamma-ray source position on the camera. The third moment along the major image axis is used to resolve the head-tail ambiguity. On the resulting event map a circular ON region of $0.25°$ radius is defined around the nominal position of the Crab Nebula, together with three identical non-overlapping OFF regions at the same distance from the camera center. By plotting the distribution of any image parameter both for the events in the ON and the OFF regions, and subtracting the latter from the former (after suitable normalization), one obtains the distribution of the parameter for the excess events, which are gamma rays from the Crab Nebula. With the source just 0.4 degree away from the center, tight cuts were necessary in order to achieve a good enough angular resolution such that OFF regions were not significantly contaminated by gamma rays. Apart from the quality cuts, an extremely loose hadron suppression cut (rejecting just $2\%$ of gammas while halving the background) has also been applied to reduce the number of background events in the subtracted histograms and hence the fluctuations in the resulting distributions. This cut, based only on "shape" parameters (no source-dependent parameter was used), and performed with the *Random Forest* classification method [10], has to be loose to avoid biasing the image parameter distributions of gammas. The MC gamma sample undergoes the same treatment. In filling the histograms, the Monte

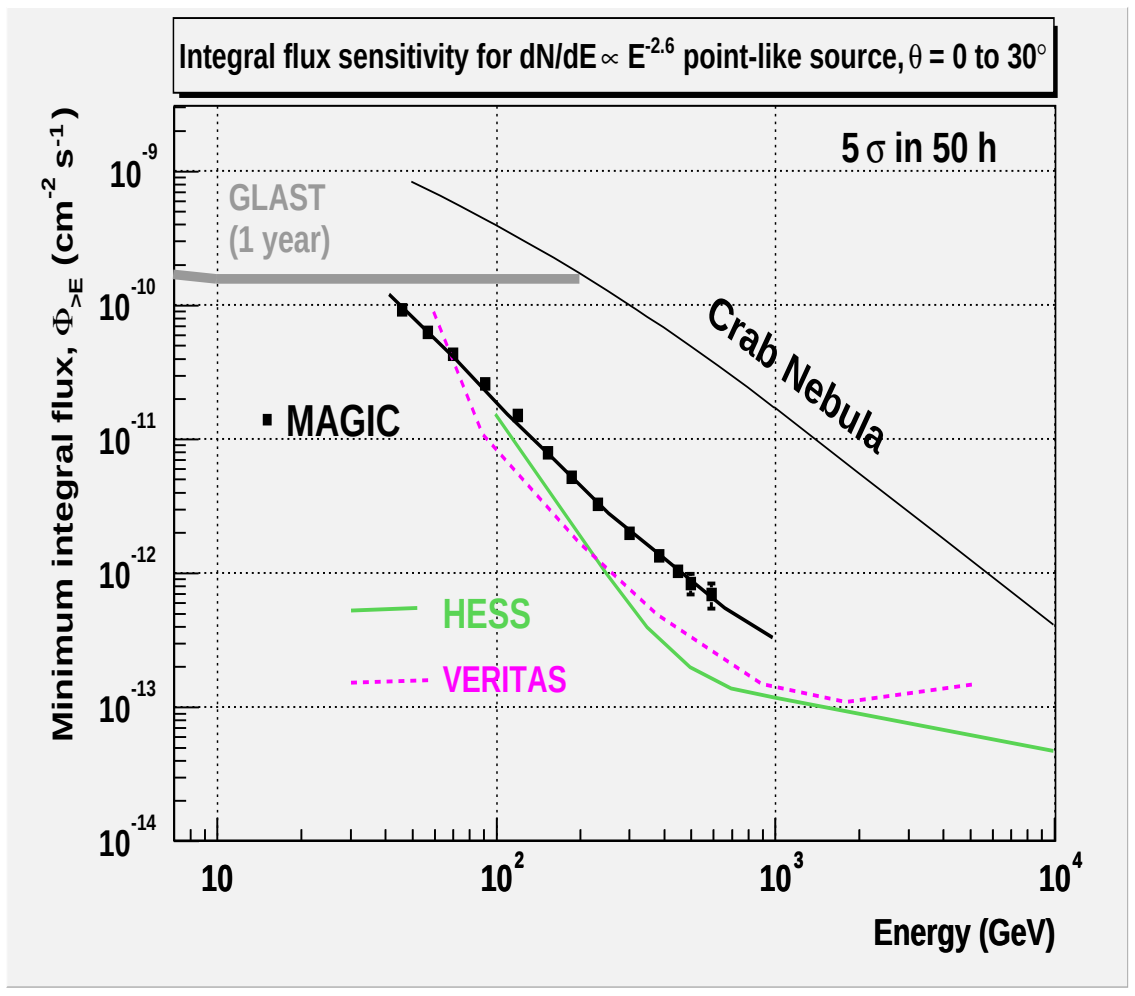

**Figure 4.** Integral point-source flux sensitivity of MAGIC as obtained from the MC simulation. The curve does extend beyond 1 TeV, but values have not been computed due to low hadron MC statistics.

Carlo gammas have been weighted to account for the deviation of the Crab spectrum from a pure power law towards low energies [11].

The distributions of WIDTH and LENGTH for MC gammas are shown in figure 2, and the SIZE spectrum in Fig. 3, compared to those of real Crab Nebula data. The real data are found to be in good agreement with Monte Carlo expectations. No arbitrary normalization factors have been applied: the observed total rate of gammas from Crab for this sample (the integral of the shown histograms) is of 9.9 events per minute, in reasonable agreement with the MC prediction of 9.5 (taking into account that the assumed Crab spectrum [11] is just a parametrization from data taken at higher energies). These are the rates of events reconstructed in the defined $0.25°$ radius region around Crab. From MC we can estimate that the rate of Crab gammas reconstructed *outside* this circle is of about 6 per minute, mostly close to the low energy end of the sample. Hence the degradation of the angular resolution (for the particular DISP method here used) is the reason for the flattening of the SIZE spectrum of gammas below $10^3$ phe$^-$ (Fig. 3, left). The SIZE distribution of hadrons in the same figure does not show such a feature, since no selection on shower direction was applied for them.

## 3. Conclusions

The Monte Carlo simulation of the MAGIC telescope has been shown to reproduce the experimental data, both for hadron and gamma primaries. The method used for the validation of the gamma MC could be applied only to a sample well above the trigger energy threshold of the experiment, which was nevertheless enough to confirm the validity of some of the main performance parameters of the telescope used in the simulation, like the overall light collection efficiency. The on-axis sensitivity of MAGIC for point sources obtained from MC is shown in Figure 4, and has likewise been confirmed in observations of Crab down to 100 GeV. Below 100 GeV, background discrimination becomes more difficult: WIDTH and LENGTH for hadrons and gammas become more and more similar for decreasing SIZE (left pads of Figs. 1 and 2), and a better tuning of the simulation will be needed to reach the optimal performance of the telescope as shown in Figure 4 (both in terms of analysis energy threshold and flux sensitivity). The performance of MAGIC from the point of view of energy resolution, as obtained from the same MC simulation, is discussed elsewhere in these proceedings [8].

**Acknowledgement** : The authors gratefully acknowledge the support of MPG and BMBF (Germany), the INFN (Italy), the CICYT (Spain) and the IAC for this work.

# Status Report on the 17m Diameter MAGIC Telescope Project

Razmick Mirzoyan
On behalf of the MAGIC Collaboration*
*Max-Planck-Institute for Physics (Werner-Heisenberg-Institute), Foehringer Ring 6, 80805 Munich, Germany*
Presenter: R. Mirzoyan (razmik@mppmu.mpg.de), ger-mirzoyan-R-abs1-og21-oral

In this report we will present the current status of the MAGIC Telescope Project. Since the early fall 2004 the first MAGIC telescope is regularly collecting data from a long list of astrophysical objects. The main parameters of the telescope are experimentally evaluated and compared to the expectations and to the Monte Carlo simulations. For the time being we are able to analyze gamma energies as low as 60 GeV. The high sensitivity of the telescope could be demonstrated by the unprecedented measured rate of gamma rays from Crab Nebula of ~1Hz on the trigger level. Gamma ray signals from Crab Nebula, Mkn-421, Mkn-501, 1ES1959 and the Galactic Center were observed. Signals from a few more sources will be reported during this conference. Construction of the second MAGIC telescope has started. It will be located at 85m distance from the first MAGIC telescope. In the design of the second telescope we are planning to introduce several improvements compared to the first MAGIC telescope, especially in the light sensor part of the imaging camera. We are scheduling to operate the second telescope in the beginning of 2007.

## 1. Introduction

The window of ground-based γ astronomy was opened in 1989 by the observation of a strong signal from the first TeV γ source, the Crab Nebula by the Whipple collaboration [1]. As instrument the 10 m diameter Whipple atmospheric air Cherenkov imaging telescope on mount Hopkins in Arizona has been used. The breakthrough in the technique was achieved by means of the image parameterization suggested by Hillas [2] allowing one to separate the rare γ showers from the orders of magnitude more intense background from showers induced by the charged cosmic rays (CR). Since then the new field of astronomy was progressing very rapidly and all new source discoveries have been made by means of this new type of telescopes, the so-called imaging air Cherenkov telescopes (IACT).

Currently, a new generation of very large IACTs [3] have either started to operate or are in the final phase of completion. One hopes to exploit the energy window from a few tens of GeV (30-200 GeV, depending on the instrument) up to the multi-TeV energy range. Already the discussion and planning for the future IACTs with a threshold close to 5-10 GeV have begun.

In 1995 worldwide there were just a few instruments operational the largest one being the 10m diameter Whipple telescope with a lower threshold setting of > 300 GeV (in fact it was the largest telescope from 1968 till 2002). On the other hand there existed no instruments to measure in the energy range below 300 GeV and above 10 GeV (below 10 GeV recent satellite born EGRET instrument onboard of Compton Gamma Ray Observatory have performed numerous measurements of sources). The above mentioned energy gap seemed to be very interesting from the observational point of view because there was a discrepancy of more than an order of magnitude in the number of sources: the satellites have discovered and measured few hundred sources while the IACT telescopes have measured just few. It was very clear that some important astrophysical processes were happening in the energy gap not accessible to none of the known techniques. It was in this time that we came along with a project to build a 17m diameter imaging air Cherenkov telescope MAGIC [4]. Since then a large international collaboration, mostly from European countries, has been collected around the core groups of MAGIC. The design report of the project has appeared in spring 1998 [5] and the financial support was allocated in late 2000 and early 2001. In late 2001 the mechanical structure of the telescope was installed and in 2002-2003 it was equipped with

* Please find enclosed the full list of the collaboration memebres on a separate page

all the different components and put into operation. From fall 2003 till the early fall 2004 the telescope was in the commissioning phase. After that MAGIC started to regularly observe gamma ray source candidates and collect data.

## 1. The MAGIC-I Telescope

When starting to design MAGIC it was clear for us that one cannot build a large telescope just by up-scaling the size of the smaller, ~4m diameter telescopes of HEGRA (few members of MAGIC collected their experience in HEGRA). Moreover, in order to be able to re-position the telescope within ~20 seconds to the coordinates of a Gamma Ray Burst provided by detectors flown on satellites and thus to contribute in the understanding of the origin of those enigmatic sources, we have designed the telescope to be light-weight. This had immediately introduced lot of constraints in the design. We have suggested few new techniques and technologies to implemented in MAGIC-I (see Fig.1 below) in order to enable the fast re-positioning. Below we list the important innovations:

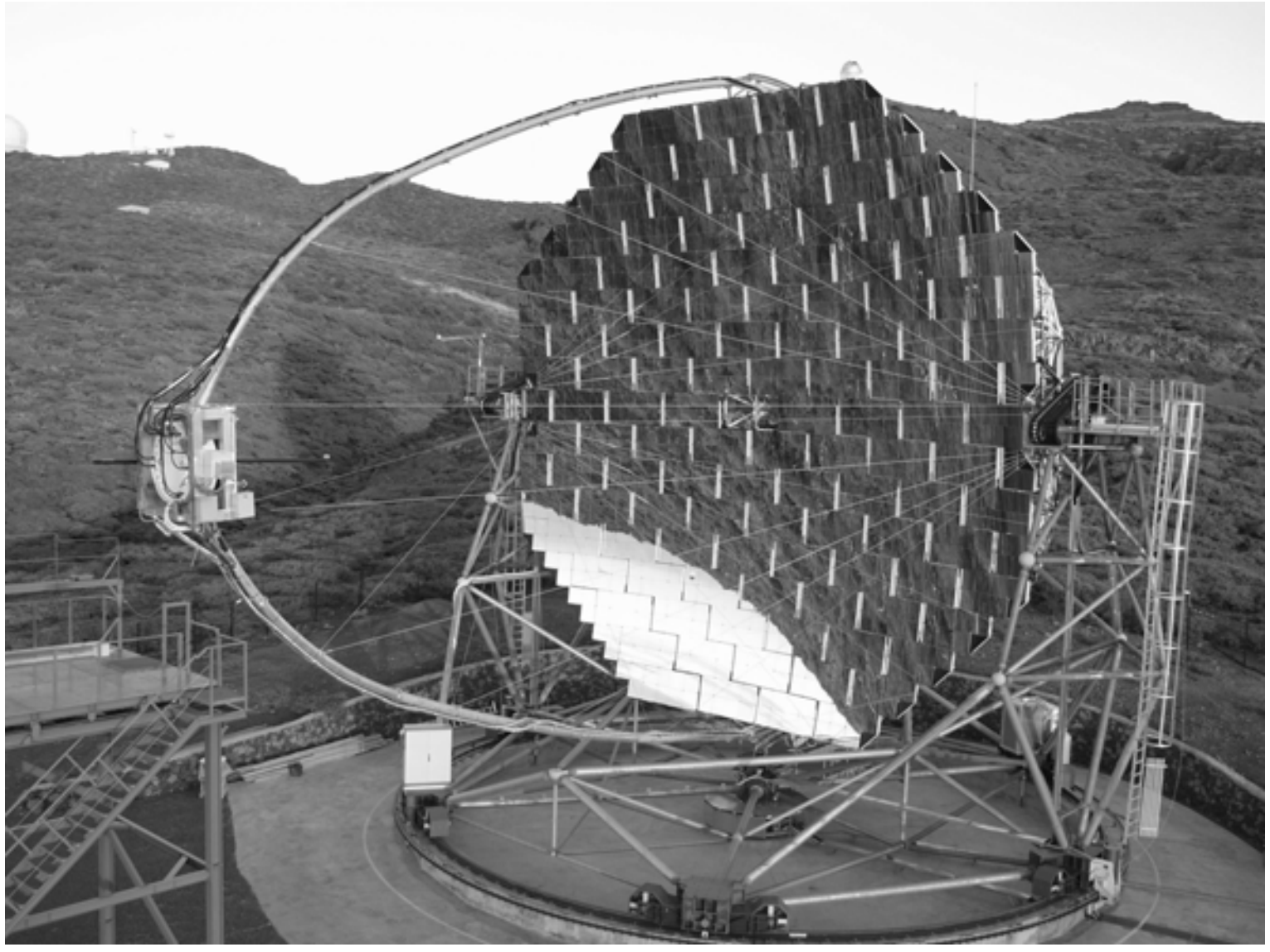

**Figure 1.** Photo of the MAGIC telescope. Location: 2200m a.s.l., Canary island of La Palma, Roque de los Muchachos Observatory

- Reflector frame made from carbon-fiber tubes
- All Al diamond milled light-weight mirrors with quartz protection and with internal heating
- Active control of the reflector shape because of varying gravitational loads while moving the elevation axis

- Transfer of ultra-fast analog signals from the camera to the experimental control room via optical fibers
- 300 Msample/s FADC system for the readout of data
- 6 dynode ultra-fast hemispherical photo multiplier tubes from *Electron Tubes*, England
- Special milky coating of PMT input window for enhanced by about 20 % quantum efficiency
- 10 layer printed circuit board in the imaging camera both for carrying the PMT pixels and for providing necessary power and connections to the computer readout system
- Temperature controlled closed-loop water cooling system of the camera

All of the above listed techniques were thoroughly tested in several iterations over ~5 years and only after that implemented in the construction. Note please that the other leading gamma observatories are already using some of the innovations pioneered by MAGIC in the design of their telescopes.
Obviously implementation of so many innovations simultaneously made it necessary to allow for some time for understanding and when necessary improving the telescope's performance in real operating conditions. Moreover, hence from the beginning of our project it was clear for us that we have chosen a difficult way: we were consciously and consecutively working on opening of a new path with a new instrument in a not yet studied energy domain where we were intending to be the pioneering explorers.

## 2. Where We Are With MAGIC-I

Today we can say that to large extent we understand the performance of MAGIC. Our measurements of sources at energies above ~100GeV showed that we succeeded to built a very sensitive instrument. Also, we could measure gamma ray signals from Crab Nebula at energies as low as 50-60 GeV. Demonstration of the achieved unprecedented high sensitivity is the measured integral rate of ~1Hz of gamma rays from Crab Nebula (above the currently achieved analysis threshold, on the trigger level). Nevertheless we shall mention that we are still working on improving the sensitivity of the telescope for energies below 100GeV. Not all of the problems are yet fully understood at sub-10GeV energies and we are currently working to significantly improve the response of the telescope in that important energy domain.
The best sensitivity of the telescope in terms of Signal/Noise ratio of the measured signal from the standard candle, the Crab Nebula, is ~20 $\sigma \cdot h^{0.5}$ and is achieved for energies (150-180)GeV. The telescope has very high hadron rejection power for energies above (500-600)GeV; for sources strong as Crab Nebula the response is practically background free. Below ~100GeV energy regime the image shapes for gammas and hadrons are becoming more similar; still gammas could be selected but only at the expense of relatively high left-over background after the image analysis. Our studies indicate that the main reason for the high background should be the gamma ray showers induced by the decay of $\pi^0$ from hadron showers. Those gamma rays develop quite "normal" electromagnetic air showers that except for angular orientation can pass the genuine gamma selection criteria.
Since several months the telescope is running smooth without any major technical problems and we reached an efficiency of ~95 % in possible observation time. With MAGIC-I we succeeded to measure gamma ray signals from Crab Nebula (R. Wagner, et al., these proc.), from AGNs Mkn-421 (D. Mazin, et al., these proc.) and 1ES1959 (N. Tonello, et al., these proc.), from the Galactic Center (H. Bartko, et al., these proc). The measured signals allowed us to evaluate the performance of the telescope (J. Cortina, et al., these proc.) and to compare it with Monte-Carlo simulations. We found a satisfactory agreement with the predictions of extensive simulations, especially for energies above ~100GeV (P. Majumdar, et al., these proc.).
Also, we have measured gamma signals from a few more sources (or source candidates) but these data are still in extensive analysis and checking stage. We are hoping to present more about new results during the conference.

## 3. Construction of MAGIC-II

Recently we have started the construction of the second telescope, MAGIC-II on 85m distance from MAGIC-I that will help to stronger suppress the backgrounds and to lower the threshold setting down towards 30 GeV. Operating two telescopes in coincidence will not only allow us to lower the threshold but also to essentially double the sensitivity. The second telescope will be similar to the first one with some improvements, especially in the camera design: we are intending to use very high quantum efficiency hybrid photo diodes with GaAsP photo cathodes that will double the light sensitivity (please see for details the talks Teshima, et al., and Hayashida, et al., these proc.).

## 4. Conclusions

Since early fall 2004 MAGIC-I is fully operational and is regularly observing sources or possible source candidates. Since several months the telescope is running smooth and we reached ~95 % efficiency in time for observations. Gamma signals from sources Crab Nebula, Mkn-421, Mkn-501, 1ES1959 and the Galactic Center have been measured and are presented at this conference. Gamma signals from few more sources or source candidates are also measured but these data are still under extensive checks. We are hoping to show more results during this conference. MAGIC-II telescope is under construction at 85m distance from MAGIC-I. Operation of both telescopes in coincidence is scheduled for early 2007 and will provide double sensitivity. Also, it will allow us to lower the threshold setting down towards the 30GeV energy regime.

## 4. Acknowledgements

We thank the Max-Planck-Society, the German BMBF, the Italian INFN, the Spanish CICYT as well as Ministries of Education and of Science of Italy, Switzerland, Poland and few other MAGIC Collaboration member countries for the support. Also we thank Instituto Astrophisico de Canarias and CCI for providing infrastructure and supporting our project at the Roque de Los Muchachos Observatory.

# Reconstruction methods of energy spectra for high redshift sources with the MAGIC Telescope

S. Mizobuchi[ab], E. Aliu[c], D. Mazin[a], M. Teshima[a], R.M. Wagner[a], W. Wittek[a], H. Yoshii[b]
*(a) Max-Planck-Institut für Physik, Föhringer Ring 6, 80805 München, Germany*
*(b) Department of Physics, Ehime University, Matsuyama, Ehime 790-8577, Japan*
*(c) Institut de Fisica d'Altes Energies, Universidad Autonoma de Barcelona, 08193 Bellaterra, Spain*
Presenter: S. Mizobuchi (satoko@mppmu.mpg.de), jap-mizobuchi-S-abs1-og27-poster

Very high energy gamma-rays are absorbed in the propagation through the intergalactic space by interaction with infrared and visible background photons. Cut-off energies are dependent on the redshift of the emitting source, and they are expected in the energy range from several tens of GeV to 100 GeV for high redshift sources like GRBs and distant AGNs. Measurement of these cut-off energies give us a unique opportunity to put constraints on the star formation rate in the history of the universe, and on cosmological parameters. MAGIC is a unique detector to explore this important energy range. The real energy spectrum at the top of the atmosphere needs to be derived from the observed one using the unfolding method or by model fitting (the forward folding method), to remove the experimental systematics. In this paper, we discuss the forward folding method in reconstructing the high-energy gamma-ray spectra observed by MAGIC.

## 1. Introduction

The MAGIC (Major Atmospheric Gamma-ray Imaging Cherenkov) Telescope is located on the Canary island of La Palma (28.8°N, 17.9°W) at 2200m above sea level. With its 17m diameter mirror, it is the largest Imaging Atmospheric Cherenkov Telescope (IACT) world-wide. The main aim of the experiment is the cosmic gamma-ray observation in the energy gap between satellite experiments and current IACTs. By lowering the energy threshold (eventually down to $\sim$ 30 GeV) with the large reflector, we can access high redshift sources like GRBs and distant AGNs because of the high flux of gamma-rays and the less significant absorption in the propagation. We expect a break in the energy spectrum in the energy region around 100 GeV, due to the gamma-ray absorption process in the propagation. A measurement of these cut-off energies is very important in order to understand the gamma-ray horizon. The accuracy of determining such energy spectra depends on the statistics and the reliably observable energy range below and above the cut-off energies. In this respect, MAGIC is well suited to study spectral breaks around 100 GeV. In this report, we would like to discuss how reliably we can reconstruct the gamma-ray energy spectrum with a cut-off using the forward-folding method in the MAGIC observation.

## 2. Energy spectra

High energy gamma-rays are absorbed in the propagation through the interaction with infrared and visible background photons. For the detection of gamma-rays, this is a somewhat negative aspect because of the reduction of the flux at high energies. However, by studying this attenuation effect in dependence of the distance of sources, we can obtain information on the star formation rate in the history of the Universe. Currently, evidence for a cut-off has been observed for relatively nearby TeV-blazars (Mkn501, Mkn421 and 1H 1426+428) by Whipple, HEGRA and CAT. These sources have low redshifts, z= 0.03, 0.03 and 0.129, respectively, and the expected cut-off energies range from sub TeV to a few TeV. For high redshift sources (z$\sim$1), gamma-rays start to be absorbed around 100 GeV [1].
We assume a spectral shape for high redshift AGNs and GRBs like

$$\frac{dN}{dE} = F_{10GeV} \times \left(\frac{E}{10GeV}\right)^{\alpha} \times exp\left(\frac{-E}{E_{cut}}\right) \qquad (E\ in\ GeV). \tag{1}$$

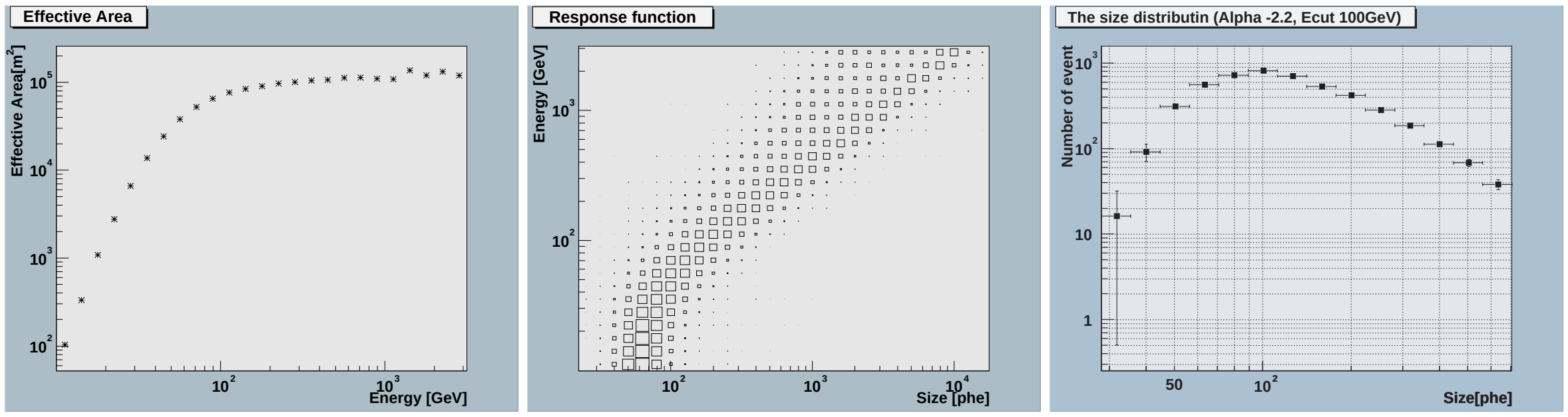

**Figure 1. Left:** Effective collection area as a function of the true energy. **Center:** Response function from the true energy to the observable size. **Right:** Expected size distribution from the energy spectrum with physics parameters ($F_{10GeV} = 8.8 \times 10^{-4} m^{-2} s^{-1} GeV^{-1}, \alpha = -2.2, E_{cut} = 100GeV$).

## 3. Spectrum reconstruction method (Forward folding method)

There are two kinds of methods for reconstructing the energy spectrum, the so-called ”Unfolding” [2, 3] and ”Forward-folding” methods. The Unfolding method is used to determine a true function by deconvolution, multiplying the experimental data by a deconvolution matrix determined from MC data. Instability of the solutions, which often occur in unfolding methods are removed by regularization procedures. On the other hand, the concept of the Forward-folding method is simpler. It aims at finding the true distribution by maximizing the agreement probability between the experimental data distribution and the one expected from Monte Carlo. In this paper, we only discuss the forward-folding method. In general,the distribution of observable parameters is related to the physical parameters through the response function, because the observation (experimental measurement) can be done only through the physical process and the detector response. It is important to obtain the matrix relating observables to physics parameters as precise as possible.
We proceeded as follows: First we prepared the matrix which relates the true energy to the measurable size distribution, including the shower fluctuation and the detector response, by MC simulation. The definition of the size is the total number of photo electrons. Then we assumed the energy spectrum with parameters (the power index, and the cut-off energy), and then multiplied the energy spectrum by the effective collection area and by the response function, to obtain the expected size distribution. This expected size distribution can be compared with the observed one. Until getting a reasonable agreement, we vary the parameters of the energy spectrum and iterate the procedure.
In Figure 1, the effective collection area as a function of the energy, the response function from the energy to the size, and the expected size distribution from the energy spectrum with physics parameters ($F_{10GeV} = 8.8 \times 10^{-4} m^{-2} s^{-1} GeV^{-1}, \alpha = -2.2, E_{cut} = 100GeV$) are shown. The integrated value of elements in the energy column is normalized to 1. Elements in each energy column represent the response function of the detector for triggered events, but do not include the collection efficiency. This collection efficiency in each energy column can be represented by the effective collection area.

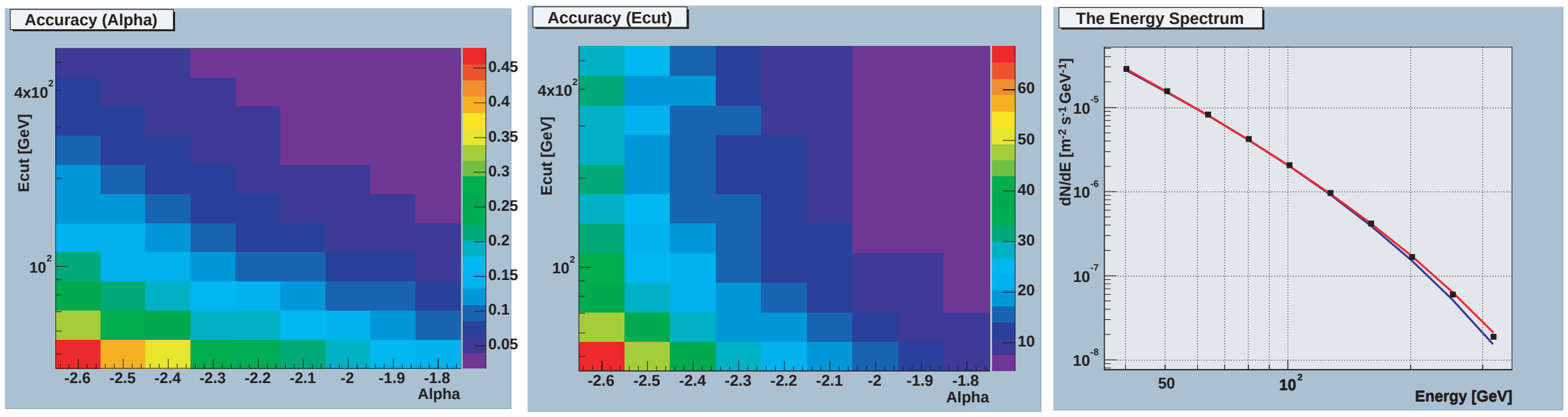

**Figure 2.** Result of the Forward-folding method for a GRB. **Left:** The error calculated for the physics parameter ”$\alpha$”. The color code represents $\Delta\ \alpha$. x- and y-axes are ”$\alpha$” parameter and ”$E_{cut}$” parameter of the sample data sets, respectively. **Center:** The error calculated for ”$E_{cut}$”. The color code shows $\Delta\ E_{cut}$ in percent. **Right:** The closed circles represent the true energy spectrum. The two lines show the reconstructed spectra with 1$\sigma$ deviations in $\Delta\ \alpha$ and $\Delta\ E_{cut}$.

## 4. Results

The errors in the spectrum reconstruction are expressed by the errors calculated for the physics parameters, ”$\alpha$” and ”$E_{cut}$”. Here we assumed an AGN- and GRB-like energy spectrum, a power law spectrum with the absorption feature. Our simulation procedures were as follows: We normalized the spectrum of AGNs at 10 GeV to the flux of $4.9 \times 10^{-5} m^{-2} s^{-1} GeV^{-1}$. We assumed the gamma-ray flux of $8.8 \times 10^{-4} m^{-2} s^{-1} GeV^{-1}$ at 10 GeV for GRBs. Then we assumed observation times of 300 seconds and 10 hours for GRBs and AGNs, respectively. This assumption of the energy spectrum might be reasonable for GRBs, if the energy spectrum is extending with the power law index of $-2.2$ up to several tens of GeV. However, there is no clear idea about the observation time of a 10$-$100 GeV gamma-ray emission from GRBs. Here, the assumed 300 seconds’ observation time might be optimistic for some GRBs. The observation time of 10 hours can be a typical observation time for a certain AGN flare observation in one week. We made 1000 data sets in each physics parameter set ”$\alpha$” and ”$E_{cut}$”, varying from $-2.6$ to $-1.8$ and from 50 GeV to 500 GeV, respectively. We individually calculated the determination accuracy of these parameters in each physics parameter set. The errors of the parameters ”$\alpha$” and ”$E_{cut}$” with a 68 % confidence level are shown in the left and center of Figure 2, respectively.
By assuming a typical GRB with a power law index of $-2.2$ and a cut-off energy of 100 GeV (assumed redshift z $\sim$ 1) [1], the energy spectrum is derived as shown in the right of Figure 2. The parameter ”$\alpha$” is determined with an error of $+0.09 - 0.11$ in the power law index value, the energy cutoff position ”$E_{cut}$” is determined with an accuracy of $+14$% and $-11.5$% of a confidence level of 68 %. In the same plot, we additionally show two size distributions which are allowed within a 68% confidence level. For distant AGNs, we also estimated the error of the ”$\alpha$” and the ”$E_{cut}$” determination as shown in Table 1, for two different power law cases.

## 5. Discussion and Conclusions

A summary of the determination accuracies of the physical parameters is shown in Table 1 for various energy spectra. The cut-off energies are well determined in all cases, as long as these energies are inside the observable energy range of MAGIC.
The energy spectrum and the flux of GRBs observed by satellite experiments vary considerably from event to event, and there is some uncertainty as to extrapolating the energy spectrum up to several tens of GeV from a theoretical point of view (internal absorption processes, hadronic model or SSC model, etc.). The MAGIC Telescope will have a reasonable sensitivity to detect GRBs and to determine the physics parameters even for mean size GRBs, if we do not have strong internal absorption process in the source.

| source | observation time | assumed $F_{10GeV}$ | assumed $\alpha$ | assumed $E_{cut}$[GeV] | determined $\alpha$ | determined $E_{cut}$[GeV] |
|---|---|---|---|---|---|---|
| GRBs | 300 sec | $8.8 \times 10^{-4} m^{-2} s^{-1} GeV^{-1}$ | $-2.2$ | 100 | $-2.2^{+0.09}_{-0.11}$ | $100^{+14.0}_{-11.5}$ |
| AGNs | 10 hrs | $4.9 \times 10^{-5} m^{-2} s^{-1} GeV^{-1}$ | $-2.2$ | 100 | $-2.2^{+0.09}_{-0.13}$ | $100^{+16.9}_{-11.3}$ |
| AGNs | 10 hrs | $4.9 \times 10^{-5} m^{-2} s^{-1} GeV^{-1}$ | $-2.6$ | 100 | $-2.6^{+0.19}_{-0.32}$ | $100^{+83.6}_{-23.6}$ |

**Table 1.** The determination accuracy dependence on the spectrum shape.

## 6. Acknowledgements

We are grateful to the IAC for their valuable help. We feel privileged to work with the support of the German BMBF and MPG, the Italian INFN and the Spanish CICYT.

# A tracking monitor for the MAGIC Telescope

B.Riegel[a], T.Bretz[a], D.Dorner[a], R.M.Wagner[b] for the MAGIC Collaboration[c]

*(a) Institut für Theoretische Physik und Astrophysik, Universität Würzburg, Am Hubland, 97074 Würzburg, Germany*
*(b) Max-Planck-Institut für Physik (Werner-Heisenberg-Institut), München, Germany*
*(c) Updated collaborator list at: http://magic.mppmu.mpg.de/collaboration/members/index.html*

Presenter: B.Riegel (riegel@astro.uni-wuerzburg.de), ger-riegel-B-abs2-og27-poster

The 17m diameter MAGIC Cherenkov telescope on the Canary island La Palma measures cosmic rays by detecting Cherenkov light from atmospheric air showers. To increase the accuracy of the tracking to the arcsec level a guiding system is used. Since gamma ray sources typically are not visible in the optical, we rely on bright "guiding stars", so a starfield guidance is mandatory. As a result we get the positional offset of the telescope, which is determined by comparing the positions of the detected stars with the positions calculated from a star catalog. As an additional feature, the CCD camera of the starguiding system can be used to check the point spread function and to monitor the sky brightness.

## 1. Introduction

The tracking system of the MAGIC telescope [1], located on the Canary Island La Palma, must meet different demands: During normal operations the 60 ton telescope has to be repositioned accurately and maintain a given position during tracking with high precision, i.e. better than the PSF of the $\gamma$-ray signal ($\sim 0.1°$). For GRB follow-up observations it is necessary to reposition the telescope to an arbitrary point in the sky within 20 seconds.
Up to now the 14 bit-shaftencoders measure the angular position of the telescope with an accuracy of about 0.022°. The star guiding system provides yet higher precision to monitor the performance of the tracking accuracy.

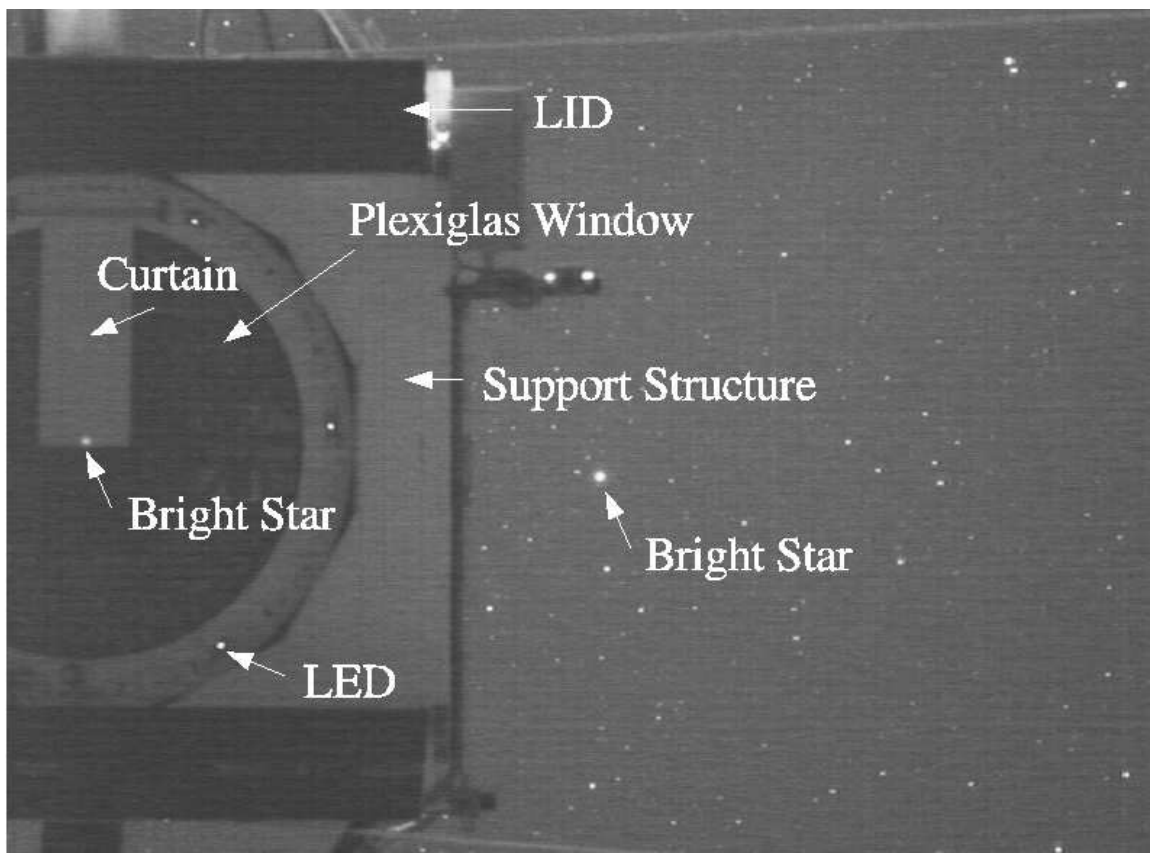

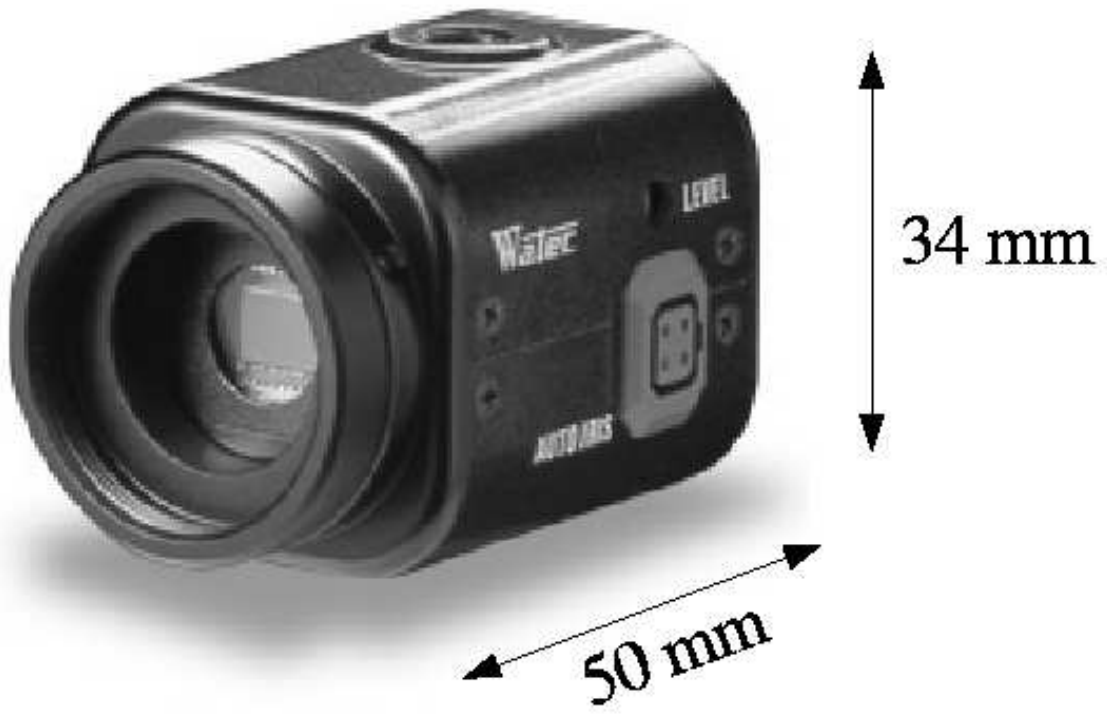

**Figure 1.** Left: Display window of the tracking monitor. With the three LEDs mounted on the PMT-camera frame a calibration is possible. To visualize the image of the bright star reflected from the main mirror to the PMT-camera a curtain was attached. Right: CCD-camera Watec WAT 902H

In contrast to optical astronomy, gamma-ray sources are typically too dim in the optical to be used for centering and monitoring the source position. Guiding Cherenkov telescopes therefore relies on tracking stars with their well-known sky positions. A low-prize, off-the-shelf video camera (Fig. 1) was attached to the center of the main dish of the MAGIC telescope monitoring LEDs mounted on the PMT-camera frame and stars from the celestial background.

The CCD-camera features a very high sensitivity of 0.0003 lux at the cost of a poor signal to noise ratio. By averaging over 125 frames (five seconds) the high noise level can be reduced and full use of the high sensitivity can be made (Fig. 2).

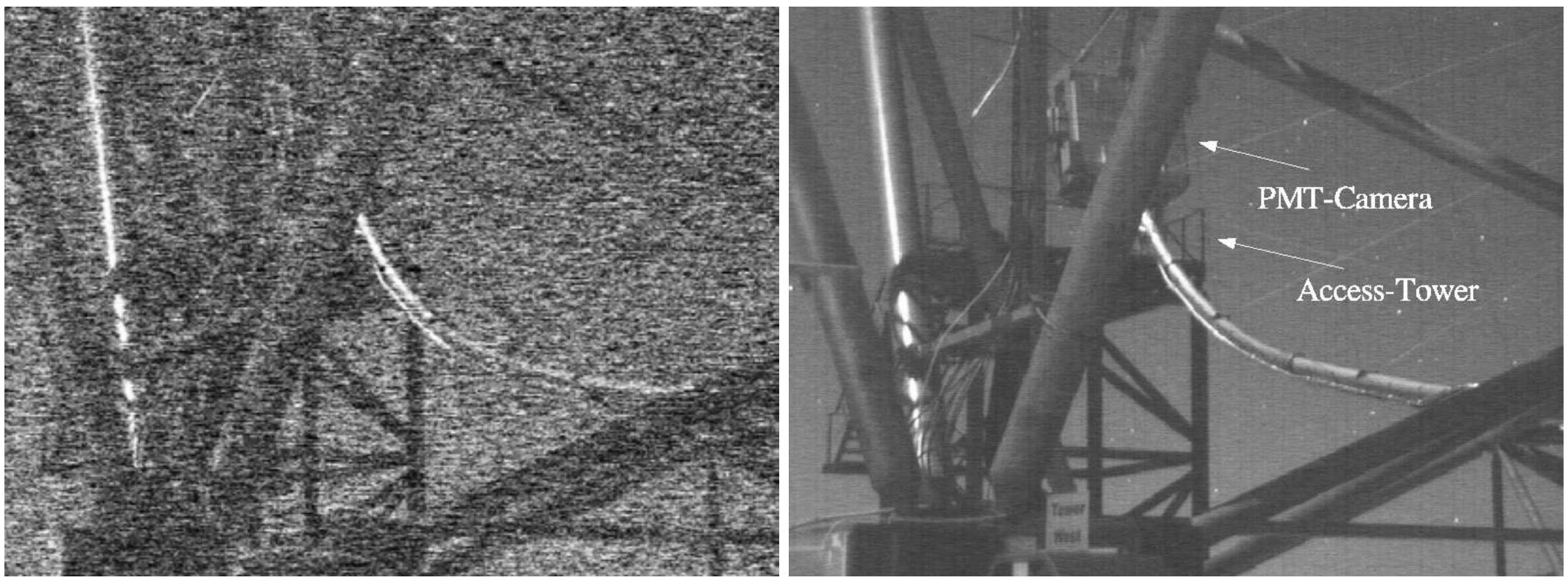

**Figure 2.** Comparison of a single frame and a picture integrated over 125 frames (five seconds)

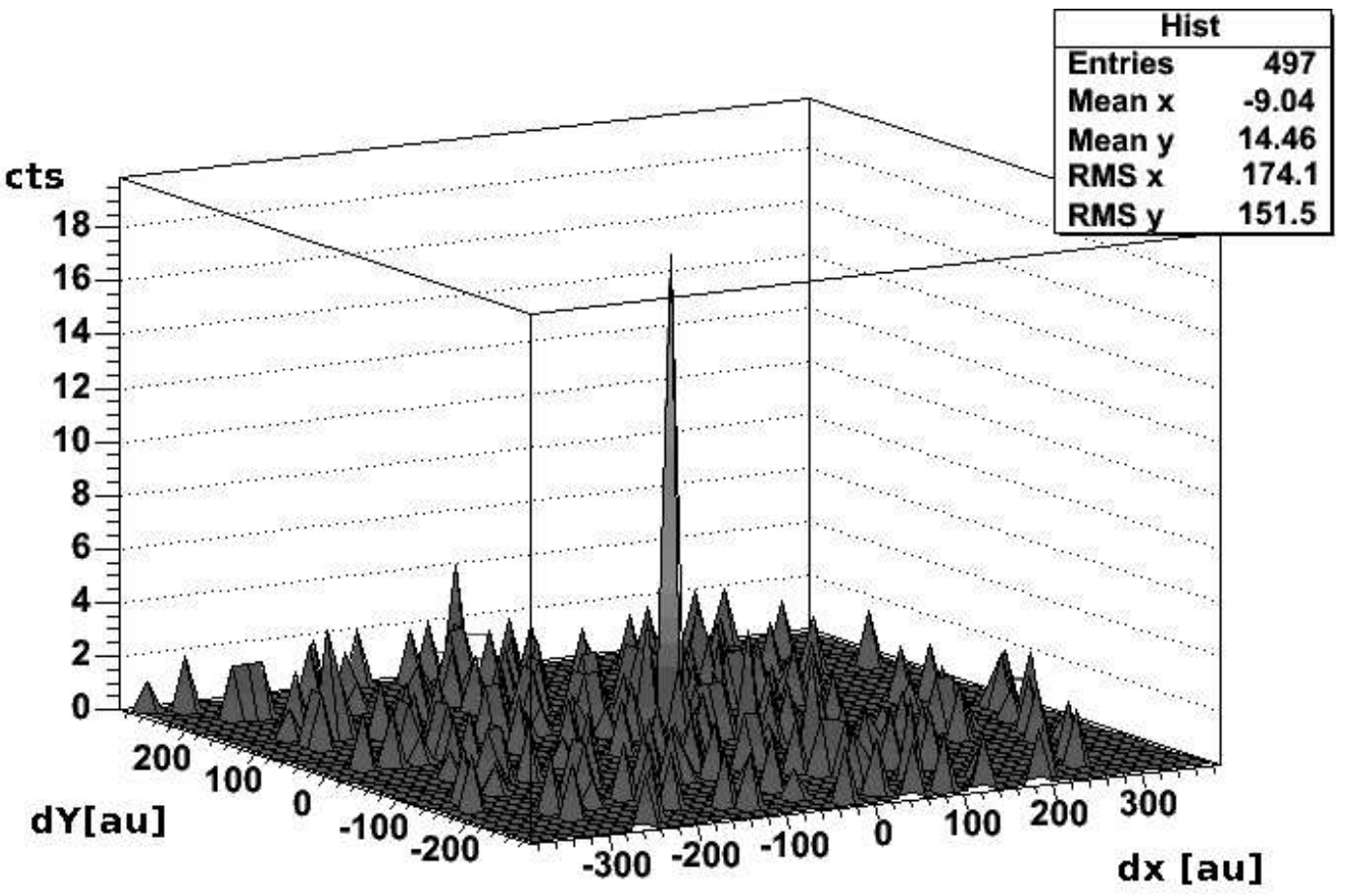

**Figure 3.** This figure shows the two dimensional distribution of the x- and y-components of the differences between the nominal and the measured star positions. Due to the strong correlation between the CCD-image and the catalog starfield a clear maximum peak is visible providing the mispointing of the telescope

## 2. Tracking Monitor

In the display monitor with 460×460 pixels (6.2°× 6.2°) typically 40-50 stars are visible (limiting magnitude 8.6) of which the brightest 10 (mag 7.2) are readily recognized.

The difference between nominal (after astrometrical and misalignment corrections) and actual pointing direction of the telescope, the mispointing, is calculated by a comparison between the starfield around the pointing position and cataloged stars. A two-dimensional histogram (dx/dy) is filled with the differences between the catalog positions of each star and all measured star positions. Because of the randomness of the dx/dy-distribution of the uncorrelated data points a clear maximum can be identified as shown in Fig. 3. The maximum in the histogram characterizes the mispointing of the telescope. By correlating also their magnitudines the algorithm can be improved further. This offset is reported to the run control system for on-site inspection and is read out to the datastream.

To demonstrate the performance of the tracking control system, we show in Fig. 4 the effect of a sudden stop during a regular tracking maneuver, which enforces an artificial mispointing.

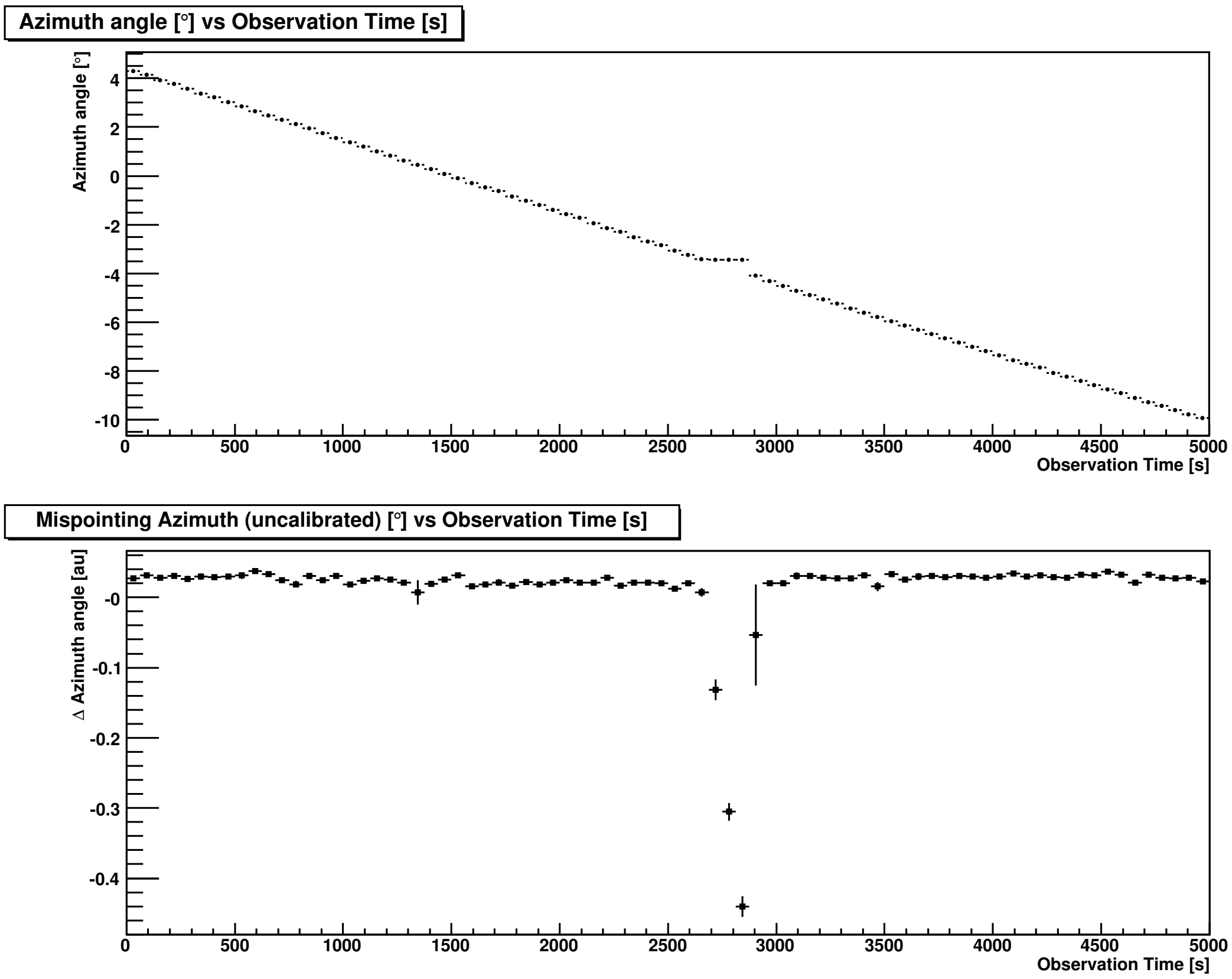

**Figure 4.** Upper panel: Azimuth angle during tracking with a short interruption at ∼ 2600 s - 2900 s. Lower panel: Artificial Mispointing enforced by stopping to track

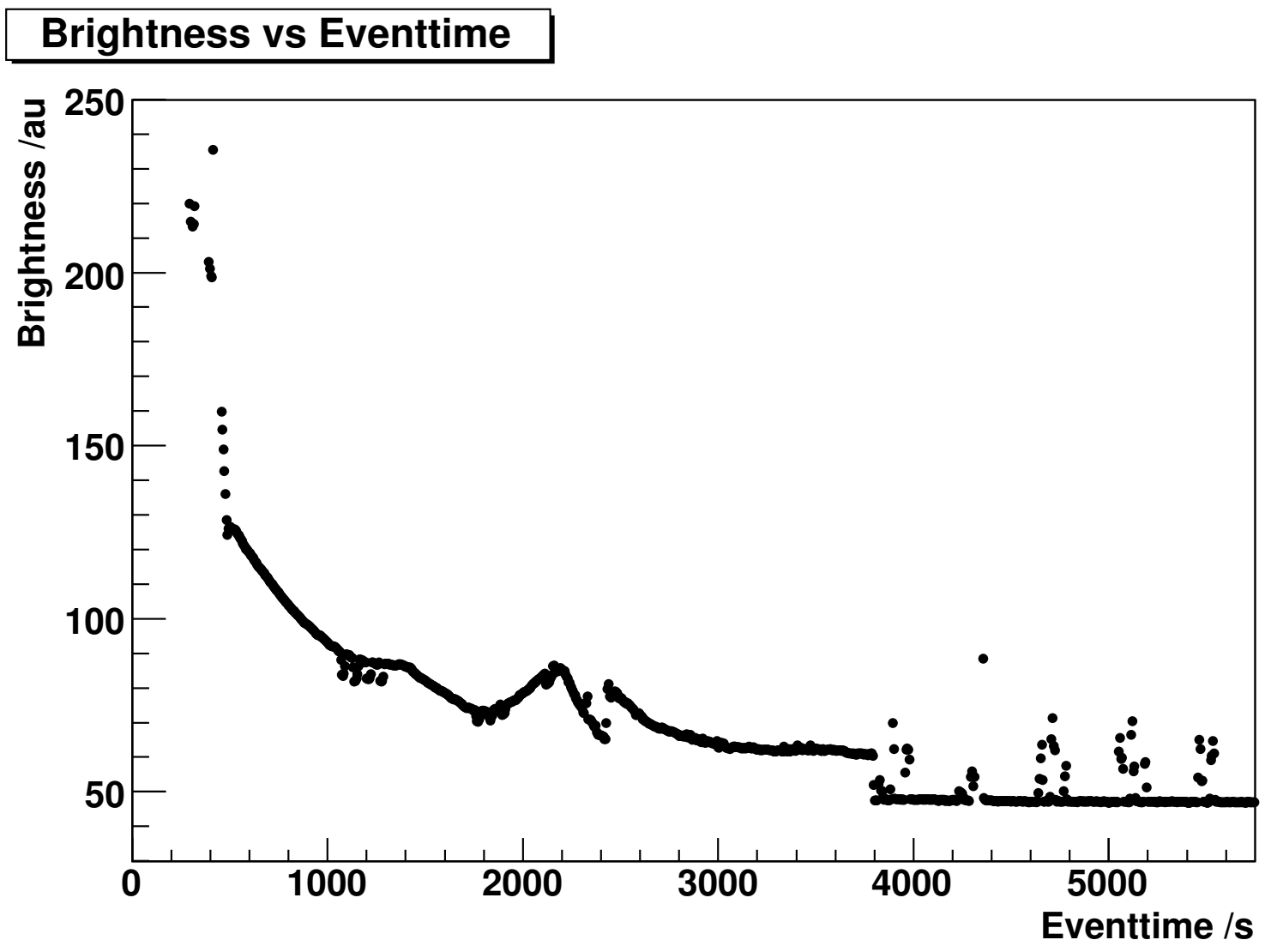

**Figure 5.** Sky brightness vs observation time. Starting at dawn into the astronomical night. Brightness fluctuations reflects the image of bright stars through the FOV

## 3. Additional Features

There are diverse possibilities for applications of the tracking monitor of the MAGIC telescope. First it is used to check the accuracy of the position adjustment both on location and afterwards during data analysis. For example, a problem which became apparent during commissioning, viz. a short mispointing during culmination of the tracked targets, could be solved. Also the sagging of the PMT-camera with increasing zenith angle could be visualized and used to obtain accurate pointing models correcting the effect.

Other interesting features are: (i) Measuring the point spread function of the telescope by attaching a curtain on the PMT-camera (see Fig. 1), and (ii) measuring the sky brightness by analyzing the noise content of the video camera (Fig. 5).

## 4. Acknowledgments

We acknowledge the support of the German Ministry of Education and Research BMBF (05 CMOMG1/3). We are also grateful to the Instituto de Astrofísica de Canarias (IAC) for the use of the MAGIC site at the Observatorio del Roque de los Muchachos (ORM) and for the excellent working conditions on La Palma.